\documentclass[11pt]{article}
\usepackage{amssymb, amsmath, bbm}
\usepackage{mathrsfs}
\usepackage{dsfont}
\usepackage{authblk}

\allowdisplaybreaks \numberwithin{equation}{section}
\renewcommand\arraystretch{1.2}
\newtheorem{thm}{Theorem}[section]
\newtheorem{prp}[thm]{Proposition}
\newtheorem{lem}[thm]{Lemma}
\newtheorem{dfn}[thm]{Definition}
\newenvironment{defn}{\begin{dfn} \rm }{\end{dfn}}
\newtheorem{cor}[thm]{Corollary}
\newtheorem{example}[thm]{Example}
\newenvironment{exa}{\begin{example} \rm }{ \end{example}}

\newtheorem{rmk}[thm]{Remark}
\newenvironment{prf}{\noindent {\it Proof:}}{\hfill $\Box$}

\newcommand\od{\mathrm{d}}
\newcommand\ad{\mathrm{ad}}

\newcommand{\nn}{\nonumber}
\newcommand\pd{\partial}
\newcommand{\ld}{\lambda} \newcommand{\Ld}{\Lambda}
\newcommand{\al}{\alpha}
\newcommand{\gm}{\gamma} \newcommand{\Gm}{\Gamma}
\newcommand{\sg}{\sigma}
\newcommand{\om}{\omega} \newcommand{\Om}{\Omega}
\newcommand{\ep}{\epsilon} \newcommand{\ve}{\varepsilon}
\newcommand{\dt}{\delta} 
\newcommand{\ta}{\theta} 
\newcommand{\ka}{\kappa}  \newcommand{\vp}{\varphi}

 \newcommand{\one}{\mathds{1}}

\newcommand{\im}{\mathrm{Im}}

\newcommand\sL{\mathscr{L}} 
\newcommand\sR{\mathscr{R}}
\newcommand\fg{\mathfrak{g}}
\newcommand\mH{\mathcal{H}}  \newcommand\mV{\mathcal{V}}
\newcommand\Z{\mathbb{Z}}  \newcommand\Zop{\mathbb{Z^{\mathrm{odd}}_+}}
\newcommand\C{\mathbb{C}}
\newcommand\R{\mathbb{R}}

\newcommand{\p}{\partial}

\newcommand{\bt}{\mathbf{t}}  \newcommand{\bu}{\mathbf{u}}
  
\newcommand{\rs}{\mathbf{s}}

\def\bea{\begin{eqnarray}}
\def\eea{\end{eqnarray}}

\allowdisplaybreaks

\begin{document}
\title{Virasoro Constraints for
Drinfeld-Sokolov Hierarchies and Equations of Painlev\'e Type}

\author[1]{Si-Qi Liu}
\author[2]{ Chao-Zhong Wu}
\author[1]{Youjin Zhang}
\affil[1]{Department of Mathematics, Tsinghua University, Beijing
100084, P.\,R. China \authorcr
Email: liusq@tsinghua.edu.cn; youjin@tsinghua.edu.cn}

\affil[2]{School of Mathematics, Sun Yat-Sen University,
   Guangzhou 510275, P.\,R. China \authorcr
   Email: wuchaozhong@sysu.edu.cn}

\date{}
\maketitle

\begin{abstract}
We construct a tau cover of the generalized Drinfeld-Sokolov hierarchy associated to an arbitrary affine Kac-Moody algebra with gradations $\mathrm{s}\le\mathds{1}$ and derive its Virasoro symmetries. By imposing the Virasoro constraints we obtain solutions of the Drinfeld-Sokolov hierarchy of Witten-Kontsevich and of Brezin-Gross-Witten types, and of those characterized by certain ordinary differential equations of Painlev\'e type. We also show the existence of
affine Weyl group actions on solutions of such ordinary differential equations, which generalizes the theory of Noumi and Yamada on affine Weyl group symmetries of the Painlev\'e type equations.
\end{abstract}

\tableofcontents

\section{Introduction}

{In the seminal paper \cite{DS} of Drinfeld and Sokolov, an integrable hierarchy of Korteweg-de Vries (KdV) type was constructed from any given affine Kac-Moody algebra $\fg$ and a
vertex of its Dynkin diagram. The construction and properties of these integrable hierarchies together with their generalizations \cite{BGHM,FHM,dGHM} constitute an important part of the theory of integrable systems. They also have close relationships with several different research areas of mathematics and physics, such as conformal and cohomological field  theories, see \cite{FJR,FF2,FF1,FBZ,LRZ} and references therein. In particular, it was proved in \cite{FJR} that the total descendant
potential (or the partition function) of the Fan-Jarvis-Ruan-Witten (FJRW) invariants of ADE-singularities are tau functions of the Drinfeld-Sokolov hierarchies associated to the untwisted affine Kac-Moody algebras of ADE type, which generalizes the Witten-Kontsevich theorem on the relationship between the topological 2d gravity and the KdV hierarchy \cite{Wi}. Such relationships were also studied for the FJRW theory and the Drinfeld-Sokolov hierarchies associated to the boundary singularities and untwisted affine Kac-Moody algebras of BCFG type respectively \cite{LRZ}. In establishing these relationships the Virasoro symmetries and constraints to the integrable hierarchies play an important role. More exactly, they are used to select the solutions of the Drinfeld-Sokolov hierarchies whose tau functions coincide with the total descendant potential of the FJRW invariants. }

{In this paper we consider the generalized Drinfeld-Sokolov hierarchies associated to an arbitrary affine Kac-Moody algebra, of either untwisted or twisted type.
Recall that in \cite{BGHM,dGHM}, the construction of the generalized Drinfeld-Sokolov hierarchies depends on two gradations $\rs \le \rs'$. When $\rs$ is the gradation $\rs^m$ and $\rs'$ is the principal gradation $\mathds{1}=(1,1,\dots,1)$ (see their definitions given below), the corresponding generalized Drinfeld-Sokolov hierarchies coincide with the original Drinfeld-Sokolov hierarchies.
The generalized Drinfeld-Sokolov hierarchies we consider here is for $\rs'=\mathds{1}$,
and we will omit the word ``generalized'' henceforth.
For such integrable hierarchies, we defined their tau functions in \cite{LWZZ} by using the approach of \cite{Wu}, and now we continue to study their Virasoro symmetries represented via the tau functions and then solve the Virasoro constraints. As to be seen, the solutions of Drinfeld-Sokolov hierarchies together with certain Virasoro constraints are characterized by some ordinary differential equations (ODEs) of Painlev\'e type, on which there are affine Weyl group actions.  }
For this purpose, we need to consider a certain extension, called the tau cover, of the Drinfeld-Sokolov hierarchy to avoid certain nonlocal terms in the Virasoro symmetries (see, for example, \cite{GO} for the case of the KdV hierarchy).

We proceed to state the main results of the present paper.
Let $\fg$ be an arbitrary affine Kac-Moody algebra of rank $\ell$. Denote by $\rs=(s_0,s_1,\dots,s_\ell)$ an arbitrary gradation satisfying $0\le s_i\le 1$ ($\rs\le\one$ for short; see Subection~\ref{sec-gVir} below). For instance, the
gradation $\rs^m$ is defined as $s_i=\dt_{i m}$. In particular,
$\rs^0=(1,0,\dots,0)$ is called the
homogeneous gradation.
It is known that the Drinfeld-Sokolov hierarchy associated to the triple $(\fg, \rs, \one)$ (see \cite{DS,dGHM,LWZZ}) can be represented as the following system of evolutionary equations of an unknown vector function $\mathbf{u}=(u_1, u_2, \dots, u_\ell)$ as
\begin{equation}\label{ut0}
\frac{\p u_i}{\p t_j} =X^i_{j}(\mathbf{u}, \mathbf{u}', \mathbf{u}'', \dots), \quad i=1,2,\dots,\ell; ~~ j\in J_+.
\end{equation}
Here $J_+$ stands for the set of positive exponents \cite{Kac} of $\fg$, and $X^i_j$ are differential polynomials of $\bu$. Note that in this paper we identify $t_1=x$ and write $\bu'=\p\bu/\p x$.
Given a solution of the hierarchy \eqref{ut0}, we define its tau function $\tau^\rs$ such that \cite{LWZZ}
\begin{align}\label{taucover0}
 \frac{\p \log\tau^\rs}{\p t_j} &=\omega_j, \quad
  \frac{\p \om_j}{\p t_k} =\Om^\rs_{k j}(\mathbf{u}, \mathbf{u}', \mathbf{u}'', \dots), \quad j,k\in J_+,
\end{align}
where $\Om^\rs_{k j}$, symmetric with respect to the indices $k$ and $j$, are certain differential
polynomials of $\mathbf{u}$ (cf. \cite{EF,Mi,Wu}). Following the notions in \cite{DZ}, the system
consists of \eqref{ut0} and \eqref{taucover0} is called the \emph{tau cover} of the Drinfeld-Sokolov hierarchy \eqref{ut0}.
In fact, let us denote by $m_1, m_2, \dots, m_\ell\in J_+$ the lowest $\ell$ positive exponents, then the unknown functions $u_1,\dots,u_\ell$ can be represented by $\p_x\om_{m_1}, \dots, \p_x\om_{m_\ell}$ via a Miura-type transformation \cite{BGHM,DS}, hence the Drinfeld-Sokolov hierarchy can be represented as a system of evolutionary equations of a single tau function.

Our first main result is a reformulation of the tau cover of the Drinfeld-Sokolov hierarchy.
\begin{thm}\label{thm-taucover0}
The tau cover \eqref{ut0}, \eqref{taucover0} of the Drinfeld-Sokolov hierarchy is equivalent to the following system of evolutionary equations of an unknown function $V$ taking value in $\fg_{<0\,[\rs]}$:
\begin{equation}\label{nablaV0}
\sum_{m\ge0}\frac1{(m+1)!}(\ad_V)^m\frac{\pd
V}{\pd t_j}=\left(e^{\ad_V}\Ld_j\right)_{<0}, \quad j\in J_+.
\end{equation}
Here $\Ld_j$ are certain generators for the principal Heisenberg subalgebra of $\fg$, and the subscript ``$<0$'' means the projection to the negative component of the decomposition $\fg=\fg_{<0\,[\rs]}\oplus\fg_{\ge0\,[\rs]}$ with respect to the gradation $\rs$.
\end{thm}

Note that if we introduce the Kac-Moody group associated to $\fg$ and the exponential map from $\fg$ to this group, then $\Theta=e^V$ has been introduced in \cite{HM} to study tau functions of the generalized Drinfeld-Sokolov hierarchies.
However, the notion of Kac-Moody group is sophisticated \cite{Kumar}. The above theorem enables us to avoid the use of this notion by representing the Virasoro symmetries in terms of the elements of $\fg$ only.
The above theorem also implies that the components of $V$ with respect to a certain basis can be represented as differential polynomials of $u_i$ and $\omega_j$,
which will be used to construct Virasoro symmetries of the Drinfeld-Sokolov hierarchy. As to be seen, such a property of $V$ ensures that these Virasoro symmetries are indeed local symmetries for the tau cover \eqref{ut0}, \eqref{taucover0}, i.e. they can be represented via differential polynomials of $u_i$ and $\omega_j$.

We recall that the Virasoro symmetries of the Drinfeld-Sokolov hierarchies were studied in \cite{HMG,Wu} and references therein. In \cite{HMG} Hollowood {\it et al} constructed the Virasoro symmetries of the generalized Drinfeld-Sokolov hierarchies associated to untwisted affine Kac-Moody algebras, based on a zero-curvature formalism of these integrable hierarchies that involves certain functions taking values in the corresponding Lie groups. In particular, when the affine Kac-Moody algebra is of ADE type, such symmetries can be represented as infinitesimal transformations of the form
\begin{equation}\label{Lktau0}
\tau\mapsto \tilde{\tau}=\tau+\ep L_k\tau, \quad k=-1,0,1,2,\dots.
\end{equation}
Here the tau function $\tau$ was introduced via the representation theory of affine Kac-Moody algebras \cite{HM} and the linear operators $L_k$, independent of $\tau$, obey the
Virasoro commutation relations. For the Drinfeld-Sokolov hierarchy associated to an
arbitrary affine Kac-Moody algebra and the zeroth vertex of its Dynkin diagram, the
Virasoro symmetries acting on the tau function were studied in \cite{Wu}, in which the tau function was defined by choosing a
special class of Hamiltonian densities (see Remark~\ref{rmk-tauham} below).

In this paper, we consider the Virasoro symmetries of the
tau cover \eqref{nablaV0} of the Drinfeld-Sokolov hierarchy \eqref{ut0} associated to a general triple $(\fg, \rs, \one)$.
Inspired by the approach of \cite{HMG,Wu}, we first extend $\fg$ to the Kac-Moody-Virasoro algebra $\mathfrak{d}^{\rs}\ltimes\fg$, where $\mathfrak{d}^{\rs}$ is a Virasoro algebra generated by a set of operators $\{d_k^\rs\mid k\in\Z\}$ (see e.g. \cite{Wa}). We remark that these operators $d_k^\rs$ are constructed such that $d_k^\rs-d_k^{\rs'}\in\fg$ for any two gradations $\rs$ and $\rs'$ of $\fg$.  Then we introduce the following evolutionary equations
in terms of the unknown function $V$:
\begin{equation}\label{nablabtV0}
\sum_{m\ge0}\frac1{(m+1)!}(\ad_V)^m\frac{\pd
V}{\pd \beta_k}=-\left(e^{\ad_V}e^{-\sum_{j\in J_+}t_j\ad_{\Ld_{j}} } d_k^\mathds{1}-d_k^\rs\right)_{<0}.
\end{equation}
Here the index $k$ is chosen in the following way:
\begin{itemize}
\item[(I)] $k=-1,0,1,2,\dots$ when $\fg$ is untwisted and $\rs$ equals to $\rs^0$ up to a diagram automorphism of $\fg$;
\item[(II)] $k=0,1,2,\dots$ for other case.
\end{itemize}
The range of the index $k$ will be explained in the proof of Lemma~\ref{zh-4}.
Based on \eqref{nablaV0} and \eqref{nablabtV0}, we will show that the flows $\p/\p\beta_k$ commute with $\p/\p t_j$ for all $k$ and $j$ in their ranges. In other words, the flows $\p/\p\beta_k$ are symmetries for the tau cover of the Drinfeld-Sokolov hierarchy.
Furthermore, we can represent these symmetries in terms of the tau function $\tau^\rs$ as follows (see Theorem~\ref{thm-vir} below for the definition of the operators $S_k$):
\begin{equation}\label{vir0}
\frac{\pd\log\tau^\rs}{\pd\beta_k}=S_k(\log\tau^\rs),
\end{equation}
and prove the following Virasoro commutation relations
\begin{equation}\label{taubtkl0}
\left[\frac{\p}{\p \beta_l}, \frac{\p}{\p \beta_k}\right]\log\tau^\rs=(k-l)  \frac{\p\log\tau^\rs}{\p \beta_{k+l}}
\end{equation}
with $k$ and $l$ given in the cases (I) or (II) above.
In particular, when $\rs=\rs^0$ we recover the corresponding results given in \cite{Wu}.
Here we note that there is a typo in the equation (4.25) of \cite{Wu} for the twisted case, where the index should be $k\ge0$ rather than $k\ge-1$.

To select the tau function of the Drinfeld-Sokolov hierarchy for case (I)
which coincides with the partition function of the FJRW theory for an ADE singularity or its BCFG-type generalization, one can
impose the string equation, i.e. the $(-1)$-th Virasoro constraint
\[\frac{\p \tau^{\rs}}{\p\beta_{-1}}=\frac{\p \tau^{\rs}}{\p t_1},\]
then show that there is a unique tau function (up to multiplication of a constant) satisfying this condition.
Furthermore, this tau function also satisfies the other Virasoro constraints
\[\frac{\p \tau^{\rs}}{\p\beta_k}=\frac{\p \tau^{\rs}}{\p t_{1+h(k+1)}},\quad k\ge 0,\]
where $h$ is the Coxeter number.
For example, the topological solution $\tau^{\rs^0}$ for $\fg=A_1^{(1)}$ is just the well-known Witten-Kontsevich tau function \cite{Ko,Wi} up to rescaling the time variables.
We can also consider more general constraints of the following form:
\begin{equation}\label{str00}
\frac{\p \tau^{\rs}}{\p\beta_{-1}}=
\sum_{p\in J_+}a_p\frac{\pd \tau^{\rs}}{\pd
t_{p} },
\end{equation}
where $a_p$ are constants that vanish except finitely many of them. We call it the generalized string equation.

On the other hand, there is no $(-1)$-th Virasoro constraint for case (II), so we can not select a particular tau function in this case by using the string equation.
In particular, the connection between the Drinfeld-Sokolov hierarchy for a twisted affine Lie algebra and the cohomological field theory is still unknown.
Nonetheless, we can still impose the following Virasoro constraints
associated to the zeroth Virasoro symmetry:
\begin{equation}\label{sim00}
\frac{\p \tau^{\rs}}{\p\beta_0}=
\sum_{p\in J_+}b_p\frac{\pd \tau^{\rs}}{\pd
t_{p} },
\end{equation}
where $b_p$ are constants that vanish except finitely many of them, and show that this constraint also implies further Virasoro constraints (see Theorem~\ref{thm-Virconstraint} for details).
We will call the equation \eqref{sim00} the
similarity equation, for it is related to the so-called similarity
reductions of integrable hierarchies in the literature (see e.g.
\cite{CK,FS-D,FS-E,FS-A}). Note that the constraint \eqref{sim00} can also be imposed on the tau function of the Drinfeld-Sokolov hierarchy of case (I).

If we take  $b_p=\dt_{p 1}$ in the similarity equation \eqref{sim00}, then {\em the solution $\log\tau^\rs$ of the Drinfeld-Sokolov hierarchy is determined up to $\ell-1$ free
parameters (see Proposition~\ref{thm-simred} below)}. For example, when $\fg=A_1^{(1)}$ and
$\rs=\rs^0$, the solution $\tau^{\rs^0}$ is called the Brezin-Gross-Witten tau
function of the KdV hierarchy \cite{BG,GW}, and it
gives (after rescaling the time variables) a generating function for the intersection numbers on the moduli spaces $\overline{\mathcal{M}}_{g,n}$ of stable curves with certain Theta cohomology classes involved \cite{Nor}.
For this reason, such kind of solutions of the Drinfeld-Sokolov hierarchy will be also called of Brezin-Gross-Witten type.
In general, we have the following theorem (see Theorems~\ref{thm-Lax} and \ref{thm-weylaction} below for more details).
\begin{thm}\label{thm-sim0}
For the Drinfeld-Sokolov hierarchy associated to $(\fg,\rs,\one)$ together with the similarity equation \eqref{sim00}, the following assertions hold true:
\begin{itemize}
\item[(i)]
The solution space is characterized by a system of ODEs given by the compatibility condition of a Lax pair for a function $\Psi(x; z)$ as follows (see \eqref{simred} and \eqref{Laxpair} for more details):
\begin{equation}\label{Laxpair0}
z\frac{\pd\Psi}{\pd z}=M\Psi, \quad \frac{\pd\Psi}{\pd x}=-L\Psi.
\end{equation}
\item[(ii)] When $\rs=\one$,  the compatibility condition of \eqref{Laxpair0} yields a system of ODEs of the form:
\begin{equation}\label{vpta0}
\vp_i'+\ta_i\vp_i+\chi_i=0, \quad i=0,1,2,\dots,\ell,
\end{equation}
where $\ta_i$ are unknown functions of $x$ and $\chi_i$ are constants, with the conditions \eqref{tachi} being fulfilled, and $\vp_i=\vp_i\left(x,\ta_j, \ta_j', \ta_j'', \dots\right)$ are polynomials of their arguments. Moreover, the system of ODEs \eqref{vpta0} admits a class of rational B\"acklund transformations $\sR_j$ with $j=0,1,2,\dots,\ell$, which give a realization of the affine Weyl group corresponding to $\fg$. Namely, these  B\"acklund transformations satisfy
\begin{equation}
{\sR_j}^2=\mathrm{Id}, \quad (\sR_i \sR_j)^{m_{i j}}=\mathrm{Id}\quad \hbox{for} \quad i\ne j,
\end{equation}
where $m_{i j}=2, 3, 4, 6$ or $\infty$ when $a_{i j}a_{j i}=0, 1,
2, 3$ or $\ge4$ respectively,
with  $A=(a_{i j})_{0\le i,j\le\ell}$ being the generalized Cartan matrix of affine type for $\fg$.
\end{itemize}
\end{thm}

If $b_p\ne0$ for some exponents $p\in J_{>1}$ in the similarity equation \eqref{sim00}, then the ODEs given by the compatibility condition of \eqref{Laxpair0} are of Painlev\'e type. For instance, if one take $\fg=A_1^{(1)}$ and $b_p=\dt_{p 3}$, then the equation \eqref{vpta0} gives the second Painlev\'e equation P2 for $\rs=\one$, and the thirty-fourth Painlev\'e equation P34 (or P4$'$ in the appendix of \cite{CM}) for $\rs=\rs^0$. We will also give some other examples, including the ODEs for $\fg=A_2^{(1)}$ and $b_p=\dt_{p 2}$ that are related to the fourth Painlev\'e equation P4 (see also \cite{CK}).

The study of the relationship between (generalized) Drinfeld-Sokolov hierarchies and higher-order ODEs of Painlev\'e type may date back to  Noumi and Yamada \cite{Nou,NY-A,NY}. For such ODEs of Painlev\'e type, by representing them in a certain symmetric form, Noumi and Yamada constructed a class of birational B\"acklund transformations, whose commutation relations admit the generating relations for affine Weyl groups \cite{NY-W,NY}. This approach was developed by a series of work, for example, \cite{FS-P6,FS-D,FS-E,FS-A,KK,Na}, most of which rely on matrix realizations of affine Kac-Moody algebras of some particular types. Our Theorem~\ref{thm-sim0} gives a unified construction of the birational B\"acklund transformations related to the Drinfeld-Sokolov hierarchy associated to $(\fg, \rs, \one)$. In particular, for $\fg=A_\ell^{(1)}$ with $\ell\ge2$ and $b_p=\dt_{p 2}$, our formulae for $\sR_j$ coincide with the results in \cite{NY-A,SHC} obtained in a different way.

The paper is arranged as follows. In Section~2 we present some properties
of affine Kac-Moody algebras. In Section~3 we first recall the definition of Drinfeld-Sokolov hierarchies and their tau-covers, then prove Theorem~\ref{thm-taucover0} and propose an algorithm to solve the Cauchy problem of Drinfeld-Sokolov hierarchies. In Section~4 we construct the Virasoro symmetries of the tau cover of  Drinfeld-Sokolov hierarchies, and study their solutions satisfying the Virasoro constraints. In Sectoin~5, we derive ODEs of Painlev\'e type from the similarity reductions of Drinfeld-Sokolov hierarchies, and study their discrete B\"acklund transformations. The final section is devoted to some concluding remarks.

\section{Preliminaries}

Let us first recall, mainly following \cite{Kac,Wa}, some properties of affine Kac-Moody algebras.

\subsection{Affine Kac-Moody algebras and their principal Heisenberg
subalgebras} \label{sec-g}

Let $A=(a_{i j})_{0\le i, j\le \ell}$ be a generalized Cartan
matrix of affine type $X_{\ell'}^{(r)}$ with $r=1,2,3$.
The corresponding Kac labels  and the dual Kac labels are denoted by $\{k_i\}_{i=0}^\ell$ and $\{k_i^\vee\}_{i=0}^\ell$ respectively, which satisfy the relations:
\begin{equation}\label{kkvee}
\sum_{m=0}^\ell a_{i m} k_m=\sum_{m=0}^\ell k_m^\vee a_{m j}=0, \quad k_i^\vee a_{i
j}k_j=k_j^\vee a_{j i} k_i, \qquad \forall\, i, j=0,1, \dots, \ell.
\end{equation}
Denote by $\fg(A)$ the complex affine Kac-Moody algebra associated to
$A$. Let $\mathfrak{h}$ be a fixed Cartan subalgebra of $\fg(A)$, $\Pi=\{\al_0,
\al_1, \dots, \al_\ell\}$ and $\Pi^\vee=\{\al_0^\vee, \al_1^\vee,
\dots, \al_\ell^\vee\}$ be the corresponding sets of simple roots and simple
coroots respectively, and $\Delta$ be the root system. Then the algebra $\fg(A)$ admits the following root space decomposition:
\begin{equation}\label{rootdec}
\fg(A)=\mathfrak{h}\oplus\left(\bigoplus_{\al\in\Delta}\fg_\al
\right).
\end{equation}
There is a set $\{e_i\in\fg_{\al_i}, f_i\in\fg_{-\al_i} \mid i=0, 1, \dots,
\ell\}$ of Chevalley generators satisfying the following Serre relations:
\begin{align}\label{efal}
& [e_i, f_j]=\dt_{i j}\al^\vee_i, \quad [\al^\vee 	_i, \al^\vee_j]=0; \\
& [\al^\vee_i, e_j]=a_{i j}e_j, \quad [\al^\vee_i, f_j]=-a_{i j}f_j; \\
& (\ad_{e_i})^{1-a_{i j}}e_j=0, \quad (\ad_{f_i})^{1-a_{i j}}f_j=0, \quad i\ne j,
\label{adee}
\end{align}
where $0\le i, j\le \ell$, and $\dt_{ij}$ is the Kronecker symbol. The Cartan subalgebra $\mathfrak{h}$ can be decomposed as
\[
\mathfrak{h}=\C\al_0^\vee\oplus\C\al_1^\vee\oplus\dots\oplus\C\al_\ell^\vee\oplus\C
d,
\]
with a scaling element $d$ that satisfies:
\begin{equation}\label{d}
[d, e_i]=e_i, \quad [d, f_i]=-f_i, \quad i=0,1,\dots, \ell.
\end{equation}
The canonical central element of $\fg(A)$ is given by
\begin{equation}\label{c}
c=\sum_{i=0}^\ell k_i^\vee \al^\vee_i.
\end{equation}

On the Cartan subalgebra $\mathfrak{h}$ there is a nondegenerate symmetric bilinear form
defined by
\begin{align}\label{blf}
(\al_i^\vee\mid\al_j^\vee)=a_{i j}\frac{k_j}{k_j^\vee}, \quad
(d\mid\al_j^\vee)=\frac{k_j}{k_j^\vee}, \quad (d\mid d)=0, \qquad \forall\, i, j=0,1, \dots, \ell.
\end{align}
It is easy to see that
\begin{equation}\label{Coxeter}
(\al_i^\vee\mid c)=0 \hbox{ for } 1\le i\le\ell, \quad (c\mid c)=0, \quad (d\mid c)=\sum_{i=0}^\ell k_i=:h,
\end{equation}
here $h$ is the Coxeter number.
The bilinear form on $\mathfrak{h}$ can be uniquely extended to the normalized invariant symmetric bilinear form $(\cdot\mid\cdot)$ on $\fg(A)$.

Let $\fg=[\fg(A),\fg(A)]$ be the derived algebra of $\fg(A)$. Namely, the Lie algebra $\fg$ is
generated by the above Chevalley generators, and it satisfies $\fg(A)=\fg\oplus\C d$. We will also call $\fg$
the affine Kac-Moody algebra associated to $A$ below in case there is no confusion.
According to \eqref{d}, the adjoint action of $d$ induces on $\fg$ the {\em principal gradation}
\begin{equation}\label{prin}
\fg=\bigoplus_{k\in\Z}\fg^k, \quad \fg^k=\left\{X\in\fg\mid [d,X]=k
X\right\}.
\end{equation}
We fix a cyclic element
\[
\Ld=\sum_{i=0}^\ell  e_i\in\fg^1
\]
and consider its adjoin action on $\fg$. It is known that
\begin{equation}\label{dec}
\fg=\im\,\ad_{\Ld}+\mathcal{H}, \quad
\im\,\ad_{\Ld}\cap\mathcal{H}=\C c
\end{equation}
with $\mathcal{H}=\{X\in\fg\mid \ad_\Lambda X\in \C c\}$ being the so-called
principal Heisenberg subalgebra of $\fg$. In more details, let $J$ be the set of exponents given by \begin{equation}\label{mirh}
J=\{m_1, m_2, \dots, m_{\ell'}\}+r h\Z, \quad 1= m_1< m_2\le m_3\le \dots \le m_{\ell'-1}
< m_{\ell'}=r h-1,
\end{equation}
then there exists a class of elements
$\Ld_j\in\fg^j$ such that
\begin{equation}\label{zh-1}
\mathcal{H}=\bigoplus_{j\in J}\C\Ld_j\oplus\C c,
\end{equation}
and these elements obey the commutation relations:
\begin{align}
[\Ld_i, \Ld_j]=i \dt_{i,-j} c, \quad i, j\in J. \label{Ldij}
\end{align}
Note that $\dim \left(\mathcal{H}\cap \fg^1\right)=1$,
so there is a constant $\nu$ such that
\begin{equation}\label{Ld1Ld}
\Ld_1=\nu\Ld.
\end{equation}

\subsection{The Kac-Moody-Virasoro algebras}\label{sec-gVir}

Besides the principal gradation \eqref{prin}, let us consider gradations on $\fg$ that are indexed by integer vectors of the set
\begin{equation}\label{} \mathrm{S}=\{\rs=(s_0, s_1, \dots,
s_\ell)\in\Z^{\ell+1}\mid s_i\ge0, s_0+s_1+\cdots+s_\ell>0\}.
\end{equation}
For any given vector $\rs=(s_0, s_1, \dots, s_\ell)\in\mathrm{S}$, by using the nondegenerate bilinear form on $\mathfrak{h}$ there is an element $d^{\rs}\in\mathfrak{h}$ defined by the conditions:
\begin{equation}\label{dsal}
(d^{\rs}\mid \al_i^\vee)=\frac{k_i}{k_i^{\vee}}s_i \ (0\le i \le \ell), \quad (d^{\rs}\mid d^{\rs})=0.
\end{equation}
Clearly, if an element $X\in\fg$ has restriction
$X|_{\fg^0}=\sum_{i=0}^\ell x_i\al_i^\vee$ with respect to the principal gradation \eqref{prin}, then
\begin{equation}\label{blfdsX}
(d^\rs\mid X)=\sum_{i=0}^\ell\frac{  x_i k_i s_i}{ k_i^\vee}.
\end{equation}
In particular, the representation \eqref{c} of the central element gives
\begin{align}\label{dsc}
&(d^{\rs}\mid c)=k_0 s_0+k_1 s_1+\dots+k_\ell
s_\ell=:h^\rs.
\end{align}
Here $h^\rs$ is called the Coxeter number of $\fg$ with respect to the gradation $\rs$. One can verify that
\[
[d^{\rs}, e_i]=s_i e_i, \quad [d^{\rs}, f_i]=-s_i f_i, \quad
i=0,1,\dots,\ell,
\]
so the element $d^\rs$ induces a gradation on $\fg$ as
\begin{equation}\label{fgks}
\fg=\bigoplus_{k\in\Z}\fg_{k\,[\rs]}, \quad \fg_{k\,[\rs]}=\{X\in\fg\mid [d^\rs, X]=k X\}.
\end{equation}

\begin{exa}
The vector $\mathds{1}:=(1,1,\dots,1)$ gives the principal
gradation \eqref{prin} on $\fg$, with
$d^{\mathds{1}}=d$ and $h^\mathds{1}=h$ given in \eqref{d} and \eqref{Coxeter} respectively. In contrast, the vector $\rs^0:=(1,0,0,\dots,0)$ induces the {\em homogeneous
gradation} on $\fg$, with $h^{\rs^0}=k_0$ being the zero-th Kac label.
\end{exa}

Let us recall the realization of $\fg$ of type $X_{\ell'}^{(r)}$ graded by some vector $\rs=(s_0,s_1,\dots,s_\ell)\in\mathrm{S}$ (see \S\,7 and \S\,8 of \cite{Kac}). We start with a simple Lie algebra $\mathcal{G}$ of type $X_{\ell'}$, on which there is a diagram automorphism $\sigma$ of order $r$.
Let $\{E_i, F_i, H_i \mid i=0, 1, \dots, \ell\}$ be a set of elements of $\mathcal{G}$ that is defined in \S\,8.3 of \cite{Kac}.
It is known that $E_i$ ($i=0,1,2,\dots,\ell$) generate the Lie algebra $\mathcal{G}$, and so do $F_i$ ($i=0,1,2,\dots,\ell$).
The assignment
\[\deg E_i=-\deg F_i=s_i,\quad i=0,1, \dots, \ell\]
induces a $\Z/r h^\rs\Z$-gradation of $\mathcal{G}$ as
\[\mathcal{G}=\bigoplus_{k=0}^{r h^\rs-1}\mathcal{G}_k.\]
Then we have the following infinite dimensional Lie algebra:
\begin{equation}\label{gAs}
\fg^\rs=\bigoplus_{k\in\Z}\left( z^k\otimes\mathcal{G}_{k\, \mathrm{mod}\, r
h^{\rs}}\right)\oplus \C\,c'
\end{equation}
with $z$ being a parameter and $c'$ a central element. More precisely, if we denote by $X(k)$ an element $z^k\otimes X\in z^k\otimes\mathcal{G}_{k\, \mathrm{mod}\, r h^\rs}$, then the Lie bracket and the normalized invariant bilinear form on $\fg^\rs$ are defined by
\begin{align}\label{XYbr}
&[X(k)+\xi  c', Y(l)+\eta  c']=[X, Y](k+l)+\dt_{k,-l}\frac{k}{r
h^\rs}(X\mid Y)_\mathcal{G} c', \\
&(X(k)+\xi  c'\mid Y(l)+\eta c')=\frac{\dt_{k,-l}}{r}(X\mid
Y)_\mathcal{G},
\end{align}
where $\xi, \eta\in\C$ and $k, l\in\Z$, and $(\,\cdot\mid\cdot\,)_\mathcal{G}$ is the normalized bilinear form on $\mathcal{G}$. As it is shown in \S\,8.7 of \cite{Kac}, the Lie algebra $\fg^\rs$ gives a faithful realization of $\fg$. In other words, there is an isomorphism
\begin{equation}\label{Rs}
R^\rs: ~\fg^\rs \longrightarrow \fg
\end{equation}
such that the following elements are mapped to the Chevalley generators and the simple coroots of $\fg$:
\begin{equation}\label{weyls}
 E_i(s_i)\mapsto e_i, \quad  F_i(-s_i)\mapsto f_i, \quad
H_i(0)+\frac{k_i s_i}{k^\vee_i h^\rs} c'\mapsto \al_i^{\vee}.
\end{equation}
Clearly, one has $R^\rs(c')=c$.

\begin{lem}\label{thm-HEFind}
For $i=0,1,2,\dots,\ell$, the following elements
\[
R^\rs\left( E_i(r h^\rs k+s_i)\right), ~ R^\rs\left( F_i(r h^\rs k-s_i)\right), ~ k\in\Z; \quad R^\rs\left( H_i(r h^\rs k )\right), ~ k\in\Z\setminus\{0\}
\]
of $\fg$ are independent of the gradation $\rs$.
\end{lem}
\begin{prf}
The statement is trivial for $R^\rs\left(E_i(s_i)\right)$ and $R^\rs\left(F_i(-s_i)\right)$. According to the definition of $E_i$, $F_i$ and $H_i$ and the
root-space decomposition of $\mathcal{G}$, one can represent $H_i$ in the form
\[
H_i=\sum a_{i_1 i_2 \dots  i_m}[E_{i_1},[E_{i_2}, \dots, [E_{i_{m-1}}, E_{i_m}]\dots] ]
\]
with $i_1$, $i_2$, $\dots$ $i_m$ contain exactly $r k_j$ times of $j\in\{0,1,\dots,\ell\}$. So
\[
R^\rs\left(H_i(r h^\rs)\right)=\sum a_{i_1 i_2 \dots i_m}[ R^\rs\left(E_{i_1}(s_{i_1})\right), \dots, [R^\rs\left(E_{i_{m-1}}(s_{i_{m-1}})\right), R^\rs\left(E_{i_m}(s_{i_m})\right)]\dots]
\]
is independent of $\rs$. For $k\ge2$, the independence of $R^\rs\left( H_i(r h^\rs k )\right)$ on $\rs$ can be derived recursively by using the following relations:
\begin{align*}
 R^\rs\left(H_i(r h^\rs k))\right)=&\frac{1}{2}\sum a_{i_1 i_2\dots  i_m}[ R^\rs\left(E_{i_1}(s_{i_1})\right), \dots,\\
&\quad [R^\rs\left(E_{i_{m-1}}(s_{i_{m-1}})\right),
  [R^\rs\left(H_{i_m}(r h^\rs (k-1)))\right), R^\rs\left(E_{i_m}(s_{i_m})\right)] ]\dots].
\end{align*}
In the same way, when $k\le-1$ we can show the validity of the statement for $R^\rs\left( H_i(r h^\rs k )\right)$ with $E_i(s_i)$ replaced by $F_i(-s_i)$.
Finally, we complete the proof by using the relations:
\begin{align*}
R^\rs\left(E_{i}(r h^\rs k+s_{i})\right)=&\frac{1}{2}[R^\rs\left(H_i(r h^\rs k)\right), R^\rs\left(E_{i}(s_{i})\right)],\\
R^\rs\left(F_{i}(r h^\rs k-s_{i})\right)=&-\frac{1}{2}[R^\rs\left(H_i(r h^\rs k)\right), R^\rs\left(F_{i}(-s_{i})\right)]
\end{align*}
for $k\ne0$.
\end{prf}

Note that the isomorphism \eqref{Rs} between Lie algebras induces an isomorphism between their derivation algebras, say,
\[
R^\rs: \mathrm{Der}(\fg^\rs)\longrightarrow\mathrm{Der}(\fg).
\]
In particular, we denote
\begin{equation}\label{dkrs}
d_k^\rs=R^\rs\left(-\frac{1}{r h^\rs} z^{r h^\rs k+1}\frac{\od}{\od z}\right), \quad
k\in\Z.
\end{equation}
The action of $d_k^\rs$ on an element $Z\in \fg$ is written as $[d_k^\rs, Z]$,
then these derivations satisfy the following relations:
\begin{align}\label{dskds0}
& [d_k^{\rs}, R^\rs\left(X(l)\right)+\xi c]=-\frac{l}{r h^\rs} R^\rs\left(X(r h^\rs k+l)\right), \\
& [d_k^{\rs}, d_l^{\rs}]=(k-l) d_{k+l}^{\rs}, \qquad k, l\in\Z. \label{dskdsl}
\end{align}
The relations \eqref{dskdsl} show that $\{d_k^\rs\}$ generate a Virasoro algebra (with
trivial center), which is denoted as $\mathfrak{d}^{\rs}$. So we obtain the Kac-Moody-Virasoro algebra $\mathfrak{d}^{\rs}\ltimes\fg$.

\begin{lem}
For any $Z\in\fg$ and $k\in\Z$,  the element $d^\rs$ defined in \eqref{dsal}  satisfies the following identities:
\begin{equation}\label{dsdkbf}
\left(d^\rs\mid [d_k^\rs, Z] \right)=0, \quad k\in \Z.
\end{equation}
\end{lem}
\begin{prf}
It follows from \eqref{blfdsX} and \eqref{dskds0} that
we only need to check $\big(d^\rs\mid R^\rs\left(H_i(0)\right) \big)=0$ for
$i=0,1,2,\dots, \ell$, which can be easily verified by using \eqref{dsal}, \eqref{dsc} and \eqref{weyls}. Thus the lemma is proved.
\end{prf}

In \cite{Wa}, Wakimoto studied the relations between the derivations $d_k^\rs$ with two different gradations. Let us review some results that will be applied in Section~\ref{sec-vir} below. Given two gradations $\rs, \rs'\in\mathrm{S}$, we introduce a series of elements
\begin{equation}\label{rhok}
\rho_0^\rs(r h^{\rs'}\! k; \rs')=\sum_{i=1}^\ell r_i R^{\rs'}\left( H_i(r
h^{\rs'}\! k ) \right)\in \fg_{r h^{\rs'}\! k \, [\rs']}, \quad k\in\Z,
\end{equation}
in which the coefficients $r_i$ are given by
\begin{equation}\label{ri}
(r_1, r_2, \dots, r_\ell)=\frac{1}{r h^\rs}(s_1, s_2,\dots,
s_\ell)\mathring{A}^{-1},
\end{equation}
with $\mathring{A}=(a_{i j})_{1\le i, j\le \ell}$ being the $\ell\times\ell$ submatrix of the affine Cartan matrix $A=(a_{i j})_{0\le i, j\le \ell}$.
It is easy to see that
\[
[d_l^{\rs'}, \rho_0^\rs(r h^{\rs'}\! k; \rs')]=-k \rho_0^\rs\left(r h^{\rs'}(k+l); \rs'\right), \quad k,l\in\Z.
\]
According to Lemma~\ref{thm-HEFind}, the elements $\rho_0^\rs(r h^{\rs'}\! k; \rs')\in\fg$ are independent of $\rs'$ whenever $k\ne0$. When $k=0$, by using \eqref{weyls}
we know that the difference between
the element $\rho_0^\rs(0; \rs')\in\fg$ and the following one
\begin{align}\label{rhos}
\rho^\rs:=\rho_0^\rs(0; \rs^0)=\sum_{i=1}^\ell r_i \al_i^{\vee}\in\mathfrak{h}
\end{align}
belongs to the center of $\fg$.

In terms of the above notations, let us present Wakimoto's Lemma~2.4 in \cite{Wa} in the following lemma.
Note that we put an additional central element term in the formula \eqref{dd0} to get the commutation relation \eqref{dskdsl}.
\begin{lem} \label{thm-dks}
For $\rs^0=(1,0,\dots,0)$ and any $\rs\in\mathrm{S}$, the following
equalities hold true:
\begin{equation}\label{dd0}
d_k^{\rs}=\begin{cases}
              d_0^{\rs^0}-\rho^\rs+\dfrac{r}{2}(\rho^\rs\mid\rho^\rs) c, & k=0; \\
              \\
              d_k^{\rs^0}-\rho_0^\rs(r k_0 k; \rs^0), & k\ne0.
            \end{cases}
\end{equation}
\end{lem}
\begin{prf}
To simplify the notations, let us denote by $d_k$
the right hand side of \eqref{dd0}. We need to show that $d_k$ with $k\in \Z$ also satisfy the relations \eqref{dskds0} and \eqref{dskdsl}.
By using the definition \eqref{ri} of $r_i$ and the properties \eqref{kkvee} of $k_i$, we obtain the equalities:
\begin{align*}
&\sum_{i=1}^\ell r_i a_{i j}=\frac{s_j}{r h^\rs}, \quad 1\le j\le\ell; \\
&\sum_{i=1}^\ell r_i a_{i
0}=-\frac{1}{k_0}\sum_{i=1}^\ell\sum_{j=1}^\ell r_i a_{i j}k_j =
-\frac{1}{k_0 r h^\rs}\sum_{j=1}^\ell s_j k_j=- \frac{h^\rs -k_0 s_0
}{k_0 r h^\rs}=-\frac{1}{r k_0}+\frac{s_0}{r h^\rs}.
\end{align*}
It is straight forward to verify, for  $0\le j\le\ell$,
\begin{align}\label{d0e}
[d_0, e_j]=& \left[
d_0^{\rs^0}-\rho^\rs+\dfrac{r}{2}(\rho^\rs\mid\rho^\rs) c,
e_j\right]\notag\\ =& -\left(\frac{\dt_{j 0}}{r k_0}+\sum_{i=1}^\ell r_i a_{i
j} \right)e_j=-\frac{s_j}{r h^\rs}e_j.
\end{align}
Similarly, for $k\ne0$ and $0\le j\le\ell$ we have
\begin{align*}
[d_k, R^{\rs^0}\left(E_j(\dt_{j 0})\right) ]&=-\left(\frac{\dt_{j 0}}{r k_0}+\sum_{i=1}^\ell r_i a_{i
j}\right)R^{\rs^0}\left(E_j(r k_0 k+\dt_{j 0})\right)\\
&=-\frac{s_j}{r h^\rs}R^{\rs^0}\left(E_j(r k_0
k+\dt_{j 0})\right).
\end{align*}
Thus by using Lemma~\ref{thm-HEFind} we arrive at the relations
\begin{align*}
[d_k, R^\rs\left(E_j(s_j)\right) ]=-\frac{s_j}{r h^\rs} R^\rs\left(E_j(r h^\rs k +s_j)\right).
\end{align*}
In the same way, we can prove the relations
\[[d_k,  R^\rs\left(F_j(-s_j)\right) ]=\frac{s_j}{r h^\rs} R^\rs\left(F_j(r h^\rs k -s_j )\right).\]
Now by using Leibniz's rule we arrive at
\[
[d_k,  R^\rs\left(X(l)\right) ]=-\frac{l}{r h^\rs}  R^\rs\left(X( r h^\rs k+l)\right), \quad k, l\in\Z.
\]
Finally, we check the commutation relation \eqref{dskdsl} for $d_k$
as follows:
\begin{align*}
[d_k, d_l]=&(k-l)d_{k+l}^{\rs^0}+\frac{r k_0 (l-k)}{r
k_0}\rho_0^\rs(r k_0 (k+l); \rs^0)\\
&\quad+\dt_{k,-l}\frac{ r k_0 k}{k_0}
\left( \rho_0^\rs(0; \rs^0)\mid \rho_0^\rs(0; \rs^0)\right) c
\\
=& (k-l)\left( d_{k+l}^{\rs^0}-\rho_0^\rs(r k_0 (k+l);
\rs^0)+\dt_{k,-l} \frac{r}{2} \left( \rho^\rs\mid
\rho^\rs\right) c\right)\\
=&(k-l)d_{k+l}, \quad k,l\in\Z.
\end{align*}
Thus the lemma is proved.
\end{prf}

The above lemma yields the following corollary (we repeat the fact that the element $\rho_0^{\rs}(r h^{\rs'} k; \rs')$ is independent of $\rs'$ whenever $k\neq0$).
\begin{cor}
Given any gradations $\rs, \rs'\in\mathrm{S}$ and  integers $k,l\in\Z$, the element $d_k^{\rs'}-d_k^{\rs}\in\fg$ is represented as follows:
\begin{equation}\label{dd}
d_k^{\rs'}-d_k^{\rs}=\begin{cases}
              \rho^\rs-\rho^{\rs'}-\dfrac{r}{2}\left((\rho^\rs\mid\rho^\rs)-(\rho^{\rs'}\mid\rho^{\rs'})  \right) c , & k=0; \\
              \\
             \rho_0^\rs(r h^\rs k; \rs)-\rho_0^{\rs'}(r h^\rs k; \rs), & k\ne0.
            \end{cases}
\end{equation}
\end{cor}

\section{Tau covers of Drinfeld-Sokolov hierarchies and their solutions}

In this section, we first recall the Drinfeld-Sokolov hierarchy associated to an affine Kac-Moody algebra and construct a tau cover of it,
then we reformulate this tau cover in terms of the dressing operator of the
Drinfeld-Sokolov hierarchy. This formulation of the tau cover
plays an important role in our study
of the Virasoro symmetries of the Drinfeld-Sokolov hierarchies,
which is done in the next section. Based on this tau cover, we also construct power series solutions of the
initial value problem of the Drinfeld-Sokolov hierarchy.

\subsection{Drinfeld-Sokolov hierarchies and their tau covers}
Let $\fg$ be an affine Kac-Moody algebra of rank $\ell$.
Apart from the principal gradation $\mathds{1}=(1,1,\dots,1)$, we also fix
a gradation $\rs=(s_0,s_1,s_2,\dots,s_\ell)\in\mathrm{S}$ with $s_i\le1$
($\rs\le\mathds{1}$ for short), and denote
\[
\fg^k=\fg_{k\,[\mathds{1}]}, \quad \fg_k=\fg_{k\,[\rs]}, \quad k\in\Z.
\]
In what follows, we will use notations like $\fg_{\ge l}=\bigoplus_{k\ge
l}\fg_k$, $\fg^{<l}=\bigoplus_{k< l}\fg^k$ etc.

Let us briefly review the construction of the generalized Drinfeld-Sokolov hierarchy associated to the triple
$(\fg, \rs, \mathds{1})$ mainly following the notations used in
\cite{LWZZ} (cf. the original definition given in \cite{DS,dGHM}).
We first introduce a Borel subalgebra
\begin{equation}\label{Borel}
\mathcal{B}=\left\{X\in\fg_{0}\cap\fg^{\le0}\mid (d^\rs\mid
X)=0\right\},
\end{equation}
and consider operators of the form
\begin{equation}\label{sL}
\sL=\frac{\p}{\p x}+\Ld_1+Q,\quad Q\in C^{\infty}(\mathbb{R},
\mathcal{B})
\end{equation}
with $x$ being the coordinate of $\R$. Note that the Lie bracket on $\fg$ can be extended naturally to
$\C\frac{\p}{\p x} \ltimes C^{\infty}(\mathbb{R}, \fg)$, then we
have the following {\em dressing lemma}.
\begin{lem}[\cite{DS,LWZZ}]\label{thm-dr}
For an operator $\sL$ of the form given in \eqref{sL}, there exists a unique
function $U(Q)\in C^\infty(\R, \fg^{<0})$ satisfying the following
two conditions:
\begin{align}
\rm{(i)}\quad   &e^{-\ad_{U(Q)}}\sL=\frac{\p}{\p x}+\Ld_1+H(Q), \quad
H(Q)\in
C^\infty(\R, \mathcal{H}\cap\fg^{<0}),\label{UL}\\
\rm{(ii)}\quad &\left(d^\rs\mid e^{\ad_{U(Q)}}\Ld_j\right)=0,\quad
\forall j\in J_{+}.\label{ULdc}
\end{align}
Moreover, both $U(Q)$ and $H(Q)$ are $x$-differential polynomials
with zero constant terms of the components of $Q$ w.r.t a basis of
$\mathcal{B}$ (differential polynomials of $Q$ for short).
\end{lem}

The Borel subalgebra $\mathcal{B}$
contains a nilpotent subalgebra
$\mathcal{N}:=\fg_{0}\cap\fg^{<0}$, that is, the subalgebra
generated by the elements $f_i$ with $s_i=0$.  It follows from \eqref{Ld1Ld}
and the Serre relations \eqref{efal} that
\begin{equation}\label{NB}
  \ad_{\Ld_1}\mathcal{N}=\ad_{I}\mathcal{N}\subset\mathcal{B} \quad \hbox{with} \quad  I:=\sum_{i\,\mid\, s_i=0}e_i.
\end{equation}
Since $\mathcal{N}\cap\mathcal{H}=\{0\}$ (see \eqref{zh-1} and \S\,14 of \cite{Kac}), the map
$\ad_{\Ld_1}:\mathcal{N}\to\mathcal{B}$ is an injection. Thus one can choose an $\ell$-dimensional
subspace $\mathcal{V}$ of $\mathcal{B}$ such that
\begin{equation}\label{BNV}
  \mathcal{B}=\ad_{\Ld_1}\mathcal{N}\oplus\mathcal{V}.
\end{equation}
Let us fix a complement subspace
$\mathcal{V}$ in \eqref{BNV} henceforth, and consider operators of the form
\begin{equation}\label{sLV}
\sL^\mV=\frac{\p}{\p x}+\Ld_1+Q^\mathcal{V}, \quad
Q^\mathcal{V}\in C^\infty(\R, \mathcal{V}).
\end{equation}
By using the method of \cite{DS,dGHM}, one can prove the following results.
\begin{lem}\label{thm-LV}
The following assertions hold true:
\begin{itemize}
\item[(i)]  For an operator $\sL$ of the form \eqref{sL},
there exists a unique function $N\in C^\infty(\R, \mathcal{N})$ such
that
\begin{equation}\label{LNLV}
\sL^\mV=e^{\ad_N}\sL,
\end{equation}
takes the form of \eqref{sLV}. Moreover,
both $N$ and $Q^\mathcal{V}$ are differential polynomials of $Q$
with zero constant terms.
\item[(ii)] For an operator $\sL^\mV$ of the form \eqref{sLV}, let $U(Q^\mV)$ be the function determined
by Lemma \ref{thm-dr}, then there is a unique
function $R(Q^\mV, \Ld_j)\in C^\infty(\R, \mathcal{N})$ for any fixed $j\in J_+$
such that the commutator
\[
\left[-(e^{\ad_{U(Q^\mV)} }\Ld_j)_{\ge 0}+R(Q^\mV, \Ld_j),
\sL^\mV\right]
\]
takes value in $\mV$. Moreover, the components of $R(Q^\mV, \Ld_j)$
are differential polynomials of $Q^\mV$ with zero constant terms.
\end{itemize}
\end{lem}

In the above lemma and in what follows, the subscripts ``$\ge0$'' and ``$<0$'' of
a $\fg$-valued function mean the projection to the corresponding component of the decomposition  $\fg=\fg_{\ge0}\otimes\fg_{<0}$.

Due to the second assertion of the above lemma, we can formulate the
Drinfeld-Sokolov hierarchy as follow.
\begin{defn}
The Drinfeld-Sokolov hierarchy associated to the triple $(\fg, \rs,
\mathds{1})$ is given by the following evolutionary
equations:
\begin{align}
\frac{\pd \sL^\mV}{\pd t_j}=\left[-(e^{\ad_{U(Q^\mV)} }\Ld_j)_{\ge
0}+R(Q^\mV, \Ld_j), \sL^\mV\right], \quad j\in J_{+}. \label{LVt}
\end{align}
\end{defn}

It can be verified that the flows \eqref{LVt} are compatible
with each other. In particular, one has $\p/\p t_1=\p/\p x$, so from now on we identify $t_1$ with $x$.
Let us choose a basis $\eta_1, \eta_2,
\dots, \eta_\ell$ of the subspace $\mathcal{V}$, and represent $Q^\mV$
in the form
\begin{equation}\label{QV}
Q^\mV=\sum_{i=1}^\ell u_i \eta_i.
\end{equation}
Then the Drinfeld-Sokolov hierarchy \eqref{LVt} can be represented
in terms of the unknown function $\mathbf{u}:=(u_1,u_2,\dots,u_\ell)$ as follows:
\begin{equation}\label{ut}
\frac{\pd u_i}{\pd t_j}=X_{j}^i(\mathbf{u}, \mathbf{u}',
\mathbf{u}'', \dots),\quad i=1,\dots, \ell;\, j\in J_+.
\end{equation}
Here $X_j^i$ are differential polynomials of
$\mathbf{u}$ (the prime means to take the
derivative with respect to $x$). In particular, one has $X^i_1=u_i'$.

Now let us define the differential polynomials
\begin{equation}\label{Omij}
\Om^\rs_{k j}(\mathbf{u}, \mathbf{u}', \mathbf{u}'',
\dots)=\frac{1}{h^\rs}\left(d^\rs\mid
\left[\left(e^{\ad_{U(Q^\mV)}}\Ld_k\right)_{\ge0},
e^{\ad_{U(Q^\mV)}}\Ld_j\right]\right), \quad k,j\in J_+.
\end{equation}
\begin{prp}[\cite{LWZZ}] \label{thm-Om}
The  differential polynomials $\Om^\rs_{k j}=\Om^\rs_{k
j}(\mathbf{u}, \mathbf{u}', \mathbf{u}'', \dots)$ satisfy the relations
\begin{equation}\label{Omsym}
\Om^\rs_{k j}=\Om^\rs_{j k}, \quad \frac{\pd\Om^\rs_{k j}}{\pd
t_l}=\frac{\pd\Om^\rs_{l j}}{\pd t_k},\quad j, k, l\in J_+.
\end{equation}
In particular,
\begin{equation}\label{Omhj}
\Om^\rs_{1 j}=\frac{j}{h}h_j, \quad j\in J_+,
\end{equation}
where $h$ is the Coxeter
number of $\fg$, $h_j=-(\Ld_j\mid H(Q^\mV))$ and $H(Q^\mV)$
is determined by Lemma~\ref{thm-dr}.
\end{prp}

We denote $\bt=\{t_j\mid j\in J_+\}$. Then the first assertion of the
proposition implies that, for a given solution $\bu(\bt)$ of the
Drinfeld-Sokolov hierarchy \eqref{LVt}, there locally exists a
function $\tau^\rs=\tau^\rs(\bf{t})$, called the tau function, such that
\begin{equation}\label{tauOm}
\frac{\pd^2\log\tau^\rs}{\pd t_k\pd t_j}=\Om^\rs_{k j}(\mathbf{u},
\mathbf{u}', \mathbf{u}'', \dots)|_{\mathbf{u}=\mathbf{u}(\bt)},
\quad  j,k\in J_+.
\end{equation}
Note that $\log\tau^\rs$ is determined by $\bu(\bt)$ up to the addition of a linear function of $\bt$.

\begin{defn}
The tau cover of the Drinfeld-Sokolov hierarchy associated to $(\fg, \rs, \one)$
is defined as the following hierarchies of the unknown functions $f$,
$\omega_j\ (j\in J_+)$, $u_i\ (i=1, 2, \dots, \ell)$:
\begin{equation}\label{tau-cover}
\frac{\p f}{\p t_k}=\omega_k, \quad \frac{\p \omega_j}{\p t_k}=\Om^\rs_{k j}(\mathbf{u},
\mathbf{u}', \mathbf{u}'', \dots),
\quad \frac{\p u_i}{\p t_k}=X^i_k(\mathbf{u},
\mathbf{u}', \mathbf{u}'', \dots), \quad k\in J_+.
\end{equation}
\end{defn}

\begin{rmk}\label{rmk-tauham}
When $\rs=\one$, the tau functions defined respectively in
\eqref{tauOm} and in \cite{EF,Mi} coincide. When $\rs=\rs^0$, the formulae \eqref{tauOm} are equivalent to the following ones given in \cite{Wu}:\begin{equation}\label{Htau}
\frac{\pd^2 \log\tau}{\pd t_k\,\pd
t_j}=\frac{j}{(\Ld_j\mid\Ld_{-j})}\left(-\Ld_j\mid
\pd_x^{-1}\frac{\pd H(Q^\mV)}{\pd t_k}\right), \quad j,k\in J_+.
\end{equation}
In the above formulae all components of ${\pd H(Q^\mV)}/{\pd t_k}$ are total
x-derivatives of differential polynomials of $\bu$ due to \eqref{Omhj}). Note that, when $\fg$ is of ADE or twisted type, the tau function defined in \eqref{Htau} coincides with the one for the Kac-Wakimoto hierarchy \cite{KW,HM}.
\end{rmk}

\begin{rmk}\label{rmk-uh}
It is known that the Drinfeld-Sokolov hierarchy \eqref{LVt} has a
Hamiltonian structure, and the functions $h_j$ given in Proposition~\ref{thm-Om} are
densities of the Hamiltonians \cite{BGHM,DS}. From Proposition~\ref{thm-Om} it follows that these densities of the
Hamiltonians satisfy the tau-symmetry condition \cite{DZ}. Moreover, if we denote $\mathbf{h}=(h_{m_1}, h_{m_2}, \dots, h_{m_\ell} )$ with $m_1, m_2, \dots, m_\ell$ being the first $\ell$ positive exponents of $\fg$ given in  \eqref{mirh}, then it is known that $\mathbf{u}\mapsto \mathbf{h}=\mathbf{h}(\mathbf{u}, \mathbf{u}', \dots)$ is a Miura-type transformation; conversely, one can represent $\mathbf{u}=\mathbf{u}(\mathbf{h}, \mathbf{h}', \dots)$ as differential polynomials of $\mathbf{h}$.
\end{rmk}

The tau covers of the integrable hierarchies play a crucial role in the application of Drinfeld-Sokolov hierarchies to the study of topological field theory. In fact, the tau functions correspond to the partition functions, and the functions $\omega_j$, $\Omega_{kj}$
correspond to the one-point and the two-point correlators respectively.

\subsection{A reformulation of the tau cover}

In this subsection, we are to show that the tau cover \eqref{tau-cover} of the Drinfeld-Sokolov hierarchy can be reformulated as the following hierarchy of differential equations for an unknown  function $V$, depending on the variables $\bt=\{t_j\mid j\in
J_+\}$  and taking value in $\fg_{<0}$:
\begin{equation}\label{nablaV}
\nabla_{t_j, V}V=\left(e^{\ad_V}\Ld_j\right)_{<0}, \quad j\in J_+,
\end{equation}
where
\begin{equation}\label{nablaUt}
\nabla_{t_j, V} V=\sum_{m\ge0}\frac1{(m+1)!}(\ad_V)^m\frac{\pd
V}{\pd t_j}.
\end{equation}

To prove the above assertion, let us first expand the function $V$ with respect to the decomposition \eqref{fgks} in the form
\[V=\sum_{k\le-1} V_k,\quad V_k=V|_{\fg_k}, \]
then the equations \eqref{nablaV} can be represented recursively as follows:
\begin{align}
\frac{\p V_{-1}}{\p t_j}=& \left.\left(e^{\ad_V}\Ld_j\right)\right|_{\fg_{-1}}, \label{Vm1t} \\
\frac{\pd V_k}{\pd
t_j}=&\left.\left(e^{\ad_V}\Ld_j\right)\right|_{\fg_k}-T_{j k},\quad k\le-2, \label{Vmkt}
\end{align}
where
\[T_{j k}=\sum_{m=1}^{-k-1}\frac{1}{(m+1)!}\sum_{ k_i\le-1,\,
k_1+\dots+k_{m+1}=k} \ad_{V_{k_1}}\dots \ad_{V_{k_m}}\frac{\pd
V_{k_{m+1}}}{\pd t_j}.\]

\begin{lem}\label{thm-Vt}
For any solution $V$ of \eqref{nablaV}, the following equalities hold true:
\begin{align}
\frac{\pd}{\pd
t_j}\left(e^{\ad_V}\Ld_i\right)&=\left[\left(e^{\ad_V}\Ld_j\right)_{<0},
e^{\ad_V}\Ld_i\right], \label{eVt} \\
\frac{\pd}{\pd
t_i}\frac{\pd}{\pd
t_j}V &=\frac{\pd}{\pd
t_j}\frac{\pd}{\pd
t_i}V\,.  \label{Vtij}
\end{align}
Here  $i, j\in J_+$.
\end{lem}
\begin{prf}
Denote $Y={\pd V}/{\pd t_j}$, then the equalities \eqref{eVt} can be verified as follows:
\begin{align}
\hbox{l.h.s.}&=\sum_{k\ge 0} \frac1{(k+1)!} \sum_{l=0}^k (\ad_V)^l \ad_{Y}(\ad_V)^{k-l} \Ld_i \notag\\
&=\sum_{k\ge 0} \frac1{(k+1)!} \sum_{l=0}^k \sum_{m=0}^l \binom{l}{m}\ad_{(\ad_V)^m Y}(\ad_V)^{l-m+k-l} \Ld_i \notag\\
&=\sum_{k\ge 0} \frac1{(k+1)!} \sum_{m=0}^k \binom{k+1}{m+1}\ad_{(\ad_V)^m Y}(\ad_V)^{k-m} \Ld_i \notag\\
&=\sum_{m\ge 0}\sum_{k\ge m} \frac1{(k-m)! (m+1)!} \ad_{(\ad_V)^m Y}(\ad_V)^{k-m} \Ld_i \notag\\
&=\sum_{m\ge 0} \frac1{(m+1)!} \ad_{(\ad_V)^m Y}e^{{\ad_V}}
\Ld_i=\left[\nabla_{t_j, V} V, e^{\ad_V}\Ld_i\right]=\hbox{r.h.s.}
\label{eVtprf}
\end{align}
To show the validity of the equalities \eqref{Vtij}, let us denote $A_j=e^{\ad_V}\Ld_j$. It is easy to see that $[A_i, A_j]=0$ for $i,j\in J_+$. Then,
for any $k\in J_+$, it follows from \eqref{eVt} that
\begin{align}
&\left[ \frac{\pd}{\pd
t_i}, \frac{\pd}{\pd
t_j} \right]  \left(e^{\ad_V}\Ld_k\right)  \nn \\
=&[ [ (A_i)_{<0}, A_j]_{<0}, A_k]+[ (A_j)_{<0}, [(A_i)_{<0}, A_k] ]\nn\\
&\quad-[ [ (A_j)_{<0}, A_i]_{<0}, A_k]- [ (A_i)_{<0}, [(A_j)_{<0}, A_k] ] \nn \\
=& ([ [ (A_i)_{<0}, A_j]_{<0}, A_k]+[ [ (A_j)_{\ge0}, (A_i)_{<0} ]_{<0}, A_k])\nn\\
&\quad+([ (A_j)_{<0}, [(A_i)_{<0}, A_k] ] + [[(A_j)_{<0}, A_k],  (A_i)_{<0} ])  \nn \\
=&[ [ (A_i)_{<0}, (A_j)_{<0}], A_k]+[ (A_j)_{<0}, (A_i)_{<0}], A_k] ]  =0 .  \label{eVtij}
\end{align}
On the other hand, the left hand side of \eqref{eVtij} can be expanded to
\begin{equation}\label{eVtij2}
\sum_{m\ge1}\frac{1}{m!}\sum_{p=0}^{m-1}(\ad_V)^p\ad_Z(\ad_V)^{m-1-p}\Ld_k=0,
\end{equation}
where $Z=\left[ {\pd}/{\pd t_i}, {\pd}/{\pd t_j} \right] V$.
Since $\rs\le\mathds{1}$, we have
$Z\in \fg_{<0}\subset\fg^{<0}$. Suppose that $Z\ne 0$, and let $l<0$ be the largest integer such that
$Z_l:=Z|_{\fg^l}\ne0$ with respect to the decomposition \eqref{prin}. We take $k=1$ in \eqref{eVtij2} and consider the highest degree term of its left hand side to obtain
\[
 [Z_l, \Ld_1]=0.
\]
It implies that $Z_l$ lies in
$\mathcal{H}\cap\fg^{<0}$, and that $l$ is in fact a negative exponent. Let us take $k=-l$ in \eqref{eVtij2} and consider the highest degree term, then we arrive at $Z_l=0$ due to \eqref{Ldij}, which contradicts our assumption that $Z_l\ne 0$.  So the equalities \eqref{Vtij} hold true, and the lemma is proved.
\end{prf}

It follows from the equalities \eqref{Vtij} that the flows
\eqref{nablaV} are compatible, so the systems of differential equations \eqref{nablaV} form an integrable hierarchy.
We proceed to introduce the tau function of the hierarchy
\eqref{nablaV}, and then establish its relation with the tau cover \eqref{tau-cover}
of the Drinfeld-Sokolov hierarchy.

Given a solution $V$ of the equations \eqref{nablaV}, we introduce a collection of functions $\{\om_j\mid j\in J_+\}$ as follows:
\begin{equation}\label{omV}
\om_j=-\frac{1}{h^\rs}\left(d^\rs\mid e^{\ad_V}\Ld_j\right), \quad
j\in J_+.
\end{equation}
Note that this notations (also $\tau^\rs$ below) have been used in the last subsection as part of the unknown functions of the tau cover \eqref{tau-cover}. We will show later that they actually coincide.
\begin{lem}
The functions $\om_j$ satisfy the following equations:
\begin{equation}\label{omij}
\frac{\pd\om_j}{\pd t_k}=\frac{\pd \om_k}{\pd t_j}, \quad j, k\in
J_+.
\end{equation}
\end{lem}
\begin{prf}
By using \eqref{blfdsX} and \eqref{eVt}, we have, for $j,
k\in J_+$,
\begin{align}\label{omij2}
\frac{\pd\om_j}{\pd t_k}=& -\frac{1}{h^\rs}\left(d^\rs\mid
\left[\left(e^{\ad_V}\Ld_k\right)_{<0},\left(e^{\ad_V}\Ld_j\right)_{\ge0}\right]
\right)\notag\\
=&\frac{1}{h^\rs}\left(d^\rs\mid
\left[\left(e^{\ad_V}\Ld_k\right)_{\ge0},\left(e^{\ad_V}\Ld_j\right)_{<0}\right]
\right)=\frac{\pd \om_k}{\pd t_j}.
\end{align}
The lemma is proved.
\end{prf}

From the above lemma it follows the existence of a function
$\tau^\rs=\tau^\rs(\bt)$ such that
\begin{equation}\label{tauom}
\frac{\p \log\tau^\rs}{\p t_j}=\omega_j, \quad j\in J_+.
\end{equation}
\begin{defn}
The function $\tau^\rs$ is called the tau function of the hierarchy of differential equations
\eqref{nablaV}.
\end{defn}

The following theorem is the main result of this section, which shows the equivalence between  the hierarchy of differential equations \eqref{nablaV} and the tau cover \eqref{tau-cover}
of the Drinfeld-Sokolov hierarchy \eqref{LVt}. A proof of the theorem will be given in the next subsection.
\begin{thm}\label{thm-taucover}
Let $\fg$ be an affine Kac-Moody algebra of rank $\ell$ with two
gradations $\rs\le\mathds{1}$, and a decomposition
\eqref{BNV} be fixed for the Borel subalgebra $\mathcal{B}$ defined in \eqref{Borel}. Then the following two assertions hold true:
\begin{itemize}
  \item[\rm{(i)}] For any solution $V$ of the equations \eqref{nablaV},  there is an operator (recall $x=t_1$)
\begin{equation}\label{LV2}
\sL^\mV=\frac{\p}{\p x}+\Ld_1+Q^\mathcal{V}, \quad
Q^\mV=\sum_{i=1}^\ell u_i \eta_i\in C^\infty(\R, \mathcal{V})
\end{equation}
of the form \eqref{sLV} and \eqref{QV}
such that the functions $u_1, \dots, u_\ell$ are differential polynomials of the components of $V$ and they, together with $\om_j\, (j\in J_+)$ and $f=\log\tau^\rs$ defined in \eqref{omV} and in \eqref{tauom} respectively, give a solution of the tau cover \eqref{tau-cover} of the Drinfeld-Sokolov hierarchy \eqref{LVt}. Moreover,  the components of the function $V$
can be represented uniquely as elements of the ring
\begin{equation}\label{Rsg}
\mathcal{R}=\C\left[u_i^{(k)}, \om_j \mid 1\le i\le\ell, \,
k\in\Z_{\ge0}, \, j\in J_+\right]
\end{equation}
with zero constant terms, and we denote $V$ as
\begin{equation}\label{Vuom}
V=\mathbf{V}(u_i^{(k)}, \om_j \mid 1\le i\le\ell, \,
k\in\Z_{\ge0}, \, j\in J_+).
\end{equation}
\item[\rm{(ii)}] If the functions $f=\log\tau^{\rs}$, $\omega_j \,(j\in J_+)$  and $u_1, \dots, u_{\ell}$ satisfy the tau cover \eqref{tau-cover} of the Drinfeld-Sokolov hierarchy \eqref{LVt}, then the function $V$ given by \eqref{Vuom} solves the equations \eqref{nablaV}.
\end{itemize}
\end{thm}

\subsection{Proof of Theorem~\ref{thm-taucover}}
Let $V$ be a solution of the hierarchy of differential equations \eqref{nablaV}. We consider an operator of the form
\begin{equation}\label{sL2}
\sL=\frac{\p}{\p t_1}+\Ld_1+Q \quad \hbox{with} \quad  Q=\left(e^{\ad_V}\Ld_1\right)_{\ge0}-\Ld_1+\om_1 c.
\end{equation}
Since $\Ld_1\in (\fg_0\cup\fg_1)\cap\fg^1$ and $V$ takes value in
$\fg_{<0}\subset\fg^{<0}$, we know that $Q\in \fg_0\cap\fg^{\le0}$;
on the other hand, by using \eqref{omV} we obtain
\[
(d^\rs\mid Q)=\left(d^\rs \mid e^{\ad_V}\Ld_1
\right)+\om_1(d^\rs\mid c)=-h^\rs \om_1+h^\rs\om_1=0.
\]
So the function $Q$ takes value in the Borel subalgebra
$\mathcal{B}$, and the operator $\sL$
is of the form \eqref{sL}. From the first assertion of
Lemma~\ref{thm-LV}, it follows the existence of a unique function $N$ taking value in the nilpotent subalgebra $\mathcal{N}$ such that
 \begin{equation}\label{sLV2}
\sL^\mV:=e^{\ad_N}\sL=\frac{\p}{\p t_1}+\Ld_1+Q^\mathcal{V}
\end{equation}
takes the form \eqref{LV2}. Note that
both $N$ and $Q^\mV$ can be represented as differential polynomials of $Q$, so all these three functions can be represented as differential polynomials of $V$.
\begin{lem}\label{thm-UHLV}
For the operator $\sL^\mV$ defined in \eqref{sLV2}, the functions $U(Q^\mV)$
and $H(Q^\mV)$ given via Lemma~\ref{thm-dr} are uniquely determined by the
following equations:
\begin{equation}\label{UHLV}
e^{\ad_{U(Q^\mV)}}=e^{\ad_N}e^{\ad_V}e^{-\ad_\Om}, \quad
H(Q^\mV)=-\frac{\p\Om}{\p t_1},
\end{equation}
where
\begin{equation}\label{Omom}
\Om=\sum_{j\in J_+}\frac{\om_j}{j}\Ld_{-j}.
\end{equation}
\end{lem}

\begin{prf}
By using the Baker-Campbell-Hausdorff formula (see \cite{Jac} for
example), there is a unique $\fg^{<0}$-valued function $W$ such that
\[
e^{\ad_{W}}=e^{\ad_N}e^{\ad_V}e^{-\ad_\Om}.
\]
Clearly, we have $[\Om,\Ld_j]=-\om_j c$ for any $j\in J_+$, so
\begin{align}
\sL^\mV=&e^{\ad_N}\left(\frac{\p}{\p t_1}+ e^{\ad_V}\Ld_1 -
\nabla_{t_1,V}V\right)+\om_1 c \notag\\
=&e^{\ad_N} e^{\ad_V}\left(\frac{\p}{\p t_1}+\Ld_1\right)+\om_1 c \notag\\
=&e^{\ad_N} e^{\ad_V}e^{-\ad_\Om}\left(\frac{\p}{\p t_1}+\Ld_1-\frac{\p\Om}{\p t_1}\right)
+[\Om,\Ld_1]+\om_1 c \notag\\
=&e^{\ad_W}\left(\frac{\p}{\p t_1}+\Ld_1-\frac{\p\Om}{\p
t_1}\right).\label{zh-2}
\end{align}
By using the facts that $[d^\rs, N]=0$ and the commutation relation \eqref{Ldij}, we obtain
\begin{align}
&\left(d^\rs\mid e^{\ad_W}\Ld_j\right)=\left(e^{-\ad_N}d^\rs\mid
e^{\ad_V}e^{-\ad_\Om}\Ld_j\right)\notag\\
=&\left(d^\rs\mid
e^{\ad_V}(\Ld_j-[\Om,\Ld_j])\right)=-h^\rs\om_j+\om_j(d^\rs\mid
c)=0.\label{zh-3}
\end{align}
Since ${\p\Om}/{\p t_1}\in\mH\cap\fg^{<0}$,  it follows from \eqref{zh-2}, \eqref{zh-3} and Lemma~\ref{thm-dr} that
\[
W=U(Q^\mV), \quad -\frac{\p\Om}{\p t_1}=H(Q^\mV).
\]
The lemma is proved.
\end{prf}

\begin{lem}
The functions $\om_j$ and the operator $\sL^\mV$ defined in \eqref{omV} and \eqref{sLV2} respectively satisfy the following equations:
\begin{align}
\frac{\p\om_j}{\p t_k}&=\frac{1}{h^\rs}\left(d^\rs \left|
\left[\left(e^{\ad_{U(Q^\mV)}}\Ld_k\right)_{\ge0},
e^{\ad_{U(Q^\mV)}}\Ld_j\right]\right.\right), \label{Omij2} \\
\frac{\pd \sL^\mV}{\pd t_j}&=\left[-(e^{\ad_{U(Q^\mV)} }\Ld_j)_{\ge
0}+R(Q^\mV, \Ld_j), \sL^\mV\right],  \label{LVt2}
\end{align}
where $j,k\in J_+$ and $R(Q^\mV, \Ld_j)$ are determined by
Lemma~\ref{thm-LV}.
\end{lem}
\begin{prf}
By using \eqref{eVt} and Lemma~\ref{thm-UHLV}, the equations
\eqref{Omij2} can be verified as follows:
\begin{align*}
\frac{\pd\om_j}{\pd t_k}=&-\frac{1}{h^\rs}\left(d^\rs\left|
\left[\left(e^{\ad_V}\Ld_k\right)_{<0},e^{\ad_V}\Ld_j\right] \right.\right)
\\
=&-\frac{1}{h^\rs}\left(d^\rs\left|
\left[\left(e^{\ad_V}e^{-\ad_\Om}\Ld_k\right)_{<0},e^{\ad_V}e^{-\ad_\Om}\Ld_j-\om_j c\right] \right.\right)
\\
=&-\frac{1}{h^\rs}\left(e^{\ad_N}d^\rs\left|
\left[\left(e^{\ad_N}e^{\ad_V}e^{-\ad_\Om}\Ld_k\right)_{<0},
e^{\ad_N}e^{\ad_V}e^{-\ad_\Om}\Ld_j\right] \right.\right)
\\
=&-\frac{1}{h^\rs}\left(d^\rs\left|
\left[\left(e^{\ad_{U(Q^\mV)}}\Ld_k\right)_{<0},e^{\ad_{U(Q^\mV)}}\Ld_j\right]\right.
\right)\\
=&\frac{1}{h^\rs}\left(d^\rs\left|
\left[\left(e^{\ad_{U(Q^\mV)}}\Ld_k\right)_{\ge0},e^{\ad_{U(Q^\mV)}}\Ld_j\right]\right.
\right).
\end{align*}
In order to prove the equations \eqref{LVt2}, by using \eqref{nablaV} we rewrite the operator $\sL$ defined in \eqref{sL2}
as follows:
\begin{equation}\label{L3}
\sL=e^{\ad_V}\left(\frac{\p}{\p t_1}+\Ld_1\right)+\om_1 c,
\end{equation}
hence we have
\begin{align}
\frac{\p\sL}{\p t_j} =&\left[\nabla_{t_j,V}V,
e^{\ad_V}\left(\frac{\p}{\p t_1}+\Ld_1\right)
\right]+\frac{\p\om_1}{\p t_j} c\\
=&\left[\left(e^{\ad_V}\Ld_j\right)_{<0}, e^{\ad_V}\left(\frac{\p}{\p
t_1}+\Ld_1\right)
\right]+\frac{\p\om_1}{\p t_j} c  \nn \\
=&\left[-\left(e^{\ad_V}\Ld_j\right)_{\ge0},
e^{\ad_V}\left(\frac{\p}{\p t_1}+\Ld_1\right)
\right]+\frac{\p\om_1}{\p t_j} c
\nn \\
=&\left[-\left(e^{\ad_V}e^{-\ad_\Om}\Ld_j\right)_{\ge0}+\om_j
c, \sL-\om_1 c \right]+\frac{\p\om_1}{\p t_j} c
\nn \\
=&\left[-\left(e^{\ad_V}e^{-\ad_\Om}\Ld_j\right)_{\ge0},
\sL\right]+\left(-\frac{\p\om_j}{\p t_1}+\frac{\p\om_1}{\p
t_j}\right) c
\nn \\
=&\left[-\left(e^{\ad_V}e^{-\ad_\Om}\Ld_j\right)_{\ge0}, \sL\right].
\end{align}
So it follows from \eqref{sLV2} that
\begin{align}
\frac{\p\sL^\mV}{\p t_j} =&\left[\nabla_{t_j,N}N,
\sL^\mV\right]+e^{\ad_N}
\left[-\left(e^{\ad_V}e^{-\ad_\Om}\Ld_j\right)_{\ge0}, \sL\right]
\nn\\
=&\left[-\left(e^{\ad_{U(Q^\mV)}}\Ld_j\right)_{\ge0}+\nabla_{t_j,N}N,
\sL^\mV\right]. \label{LVt3}
\end{align}
Note that both sides of \eqref{LVt3} take
value in the subspace $\mV$, and that the function $\nabla_{t_j,N}N$ takes value in the
nilpotent subalgebra $\mathcal{N}$. Hence, according to the second
assertion of Lemma~\ref{thm-LV}, we have
\[
\nabla_{t_j,N}N=R(Q^\mV, \Ld_j).
\]
Here the components of the right hand side of the above equation are differential polynomials of $Q^\mV$.
Therefore, the equation
\eqref{LVt2} holds true. The lemma is proved.
\end{prf}

Denote by $\Om^\rs_{k j}$ the right hand side of \eqref{Omij2}, then from the above lemma we know that the functions
$f=\log\tau^{\rs}$, $\omega_j \,(j\in J_+)$ defined by \eqref{omV}, \eqref{tauom} and the functions $u_1, \dots, u_{\ell}$ given in \eqref{LV2} satisfy the
tau cover \eqref{tau-cover} of the Drinfeld-Sokolov hierarchy \eqref{LVt}.

To finish the proof of the first assertion of Theorem \ref{thm-taucover}, we need to show that the function $V$ can be represented via $u_i$ with $i=1,2,\dots,\ell$ and $\om_j$ with $j\in J_+$. For this purpose let us first prove the following lemma.
\begin{lem}\label{thm-YNX}
Given any $X\in\fg^{<0}$, there exists a unique elment $(M, Y)$ of $\mathcal{N}\times \fg_{<0}$ such that
\begin{equation}\label{YNX}
e^{\ad_Y}=e^{\ad_M}e^{\ad_X}.
\end{equation}
Moreover, both $M$ and $Y$ can be represented as polynomials of the components of $X$.
\end{lem}
\begin{prf}
According to the Baker-Campbell-Hausdorff formula and properties of the adjoint representation,  equation \eqref{YNX} is equivalent to
\begin{equation}\label{YNX2}
Y=M+X+\frac{1}{2}[M,X]+\frac{1}{12}([M,[M,X] ]+[X,[X,M]
])-\frac{1}{24}[X,[M,[M,X] ] ]+\dots.
\end{equation}
Let us represent $Z=M, X, Y$ in the form $Z=\sum_{k\le-1}Z_k$ with
$Z_k\in\fg^k$. By using the
fact that $\fg^{<0}=\mathcal{N}\oplus\fg_{<0}$, we can solves $M_k$ and
$Y_k$ uniquely in a recursive way from \eqref{YNX2}, and they are
clearly polynomials of $X$. The lemma is proved.
\end{prf}

Recall that the $\fg^{<0}$-valued functions $U(Q^\mV)$ and $\Om$ are defined by \eqref{UHLV} and \eqref{Omom} respectively, so they determine a $\fg^{<0}$-valued
function $\tilde{V}$ by
\[
e^{\ad_{\tilde{V}}}=e^{\ad_{U(Q^\mV)}}e^{\ad_\Om},
\]
which implies
\[
e^{\ad_V}=e^{-\ad_N}e^{\ad_{\tilde{V}}}.
\]
By using Lemma~\ref{thm-YNX}, we know that the $\mathcal{N}$-valued function
$N$ and the $\fg_{<0}$-valued function $V$ must be polynomial of
$\tilde{V}$, so they are polynomials of
$U(Q^\mV)$ and $\Om$. Thus, we arrive at the fact that
the components of $V$ are elements of the ring $\mathcal{R}$ (define by \eqref{Rsg})
with zero constant terms. So the first assertion of Theorem \ref{thm-taucover} is proved.

In order to prove the second assertion of the theorem, let us assume that
\[
\{f=\log\tau^\rs, \,\om_j, \, u_i\mid j\in J_+, \, i=1,2,\dots,\ell\}
\]
is a solution of the tau cover \eqref{tau-cover} of the Drinfeld-Sokolov hierarchy, with  $Q^\mV=\sum_{i=1}^\ell u_i\eta_i$ taking value in a fixed subspace $\mV$ of $\mathcal{B}$. For the operator
$\sL^\mV=\p/\p x+\Ld_1+Q^\mV$, let $U(Q^\mV)$ and $H(Q^\mV)$ be the functions determined via Lemma~\ref{thm-dr}. Denote $\Om=\sum_{j\in J_+}\frac{\om_j}{j}\Ld_{-j}$, then it follows from \eqref{Omhj} that
\begin{equation}\label{HOm}
H(Q^\mV)=\sum_{j\in J_+}\frac{(\Ld_j\mid H(Q^\mV) )}{h}\Ld_{-j}=-\sum_{j\in J_+}\frac{\Om_{1 j}^\rs}{j}\Ld_{-j}=-\frac{\p\Om}{\p x}.
\end{equation}
For any $j\in J_+$, the above operator $\sL^\mV$ satisfies \eqref{LVt}, which can be recast to
\begin{equation}\label{LVt4}
\frac{\p\sL^\mV}{\p t_j}=\left[ \left(e^{\ad_{U(Q^\mV)}}\Ld_j\right)_{<0}+R(Q^\mV,\Ld_j), \sL^\mV\right]-e^{\ad_{U(Q^\mV)}}[\Ld_j, H(Q^\mV)].
\end{equation}
Note that the second term on the right hand side of the above equation is equal to $\Om_{1 j}^\rs c$. On the other hand, by using the dressing formula \eqref{UL} for $\sL^\mV$ again we have
\begin{equation}\label{LVt5}
\frac{\p\sL^\mV}{\p t_j}=\left[ \nabla_{t_j, U(Q^\mV)}U(Q^\mV), \sL^\mV\right]+e^{\ad_{U(Q^\mV)}}\frac{\p H(Q^\mV)}{\p t_j}.
\end{equation}
Denote
\begin{equation}\label{Gj}
G_j=e^{-\ad_{U(Q^\mV)}}\left(\nabla_{t_j, U(Q^\mV)}U(Q^\mV)-\left(e^{\ad_{U(Q^\mV)}}\Ld_j\right)_{<0}-R(Q^\mV,\Ld_j) \right),
\end{equation}
then it satisfies, by using the equations \eqref{LVt4} and \eqref{LVt5}, that
\begin{equation}\label{}
\left[ G_j, \frac{\p}{\p x}+\Ld_1+H(Q^\mV)\right]+\frac{\p H(Q^\mV)}{\p t_j}-\Om_{1 j}^\rs c=0.
\end{equation}
By using the same argument that is used in the proof of Lemma~3.4 of \cite{Wu}, we know that $G_j$ takes values in $\mathcal{H}\cap\fg^{<0}$ and that
\begin{equation}\label{GHLd}
-\frac{\p G_j}{\p x}+\frac{\p H(Q^\mV)}{\p t_j}=0, \quad [G_j, \Ld_1]-\Om_{1 j}^\rs c=0.
\end{equation}
Since both $G_j$ and $\p\Om/\p t_j$ are differential polynomials of $u_1, u_2, \dots, u_\ell$ with zero constant terms, the first equality of \eqref{GHLd} together with \eqref{HOm} leads to
\[
G_j=-\frac{\p\Om}{\p t_j},
\]
which clearly satisfies the second equality given in \eqref{GHLd}.

It follows from the Baker-Campbell-Hausdorff formula and Lemma~\ref{thm-YNX} that there is a unique function $(N, V)$ taking value in $\mathcal{N}\times\fg_{<0}$ such that
\begin{equation}\label{}
e^{\ad_V}=e^{-\ad_N}e^{\ad_{U(Q^\mV)} }e^{\ad_\Om}.
\end{equation}
Moreover, both $V$ and $N$ are differential polynomials in the ring $\mathcal{R}$ with zero constant terms. Then we have
\begin{align*}
\nabla_{t_j, V}V=&\frac{\p}{\p t_j}-e^{\ad_V}\frac{\p}{\p t_j}=\frac{\p}{\p t_j}-e^{-\ad_N}e^{\ad_{U(Q^\mV)}}\left(\frac{\p}{\p t_j}-\frac{\p\Om}{\p t_j}\right) \\
=&\frac{\p}{\p t_j}-e^{-\ad_N}\left(\frac{\p}{\p t_j}-\nabla_{t_j, U(Q^\mV)}U(Q^\mV)+e^{\ad_{U(Q^\mV)}}G_j\right) \\
=& \nabla_{t_j, -N}(-N) - e^{-\ad_N} \left( -\left(e^{\ad_{U(Q^\mV)}}\Ld_j\right)_{<0}-R(Q^\mV,\Ld_j)  \right) \\
=&\left(e^{-\ad_N}e^{\ad_{U(Q^\mV)}}\Ld_j\right)_{<0} + \nabla_{t_j, -N}(-N)+  e^{-\ad_N} R(Q^\mV,\Ld_j),
\end{align*}
where we used \eqref{Gj} to derive the third equality. Observe that $\nabla_{t_j, -N}(-N)+  e^{-\ad_N} R(Q^\mV,\Ld_j)$ takes value in $\mathcal{N}\subset\fg_0$, hence it must vanish. So we arrive at the equation
\[
\nabla_{t_j, V}V=\left(e^{-\ad_N}e^{\ad_{U(Q^\mV)}}\Ld_j\right)_{<0} =\left(e^{\ad_V}\Ld_j\right)_{<0}.
\]
Therefore, we complete the proof of Theorem \ref{thm-taucover}.

\subsection{Formal power series solutions of the tau cover}\label{sec-sol}
Let us consider the Cauchy problem of the equations
\eqref{tau-cover} with initial values:
\begin{equation}\label{initial}
\left.u_i^{(k)}\right|_{\bt=0}=U_i^k, \quad  \left.\om_j\right|_{\bt=0}=W_j, \quad 1\le
i\le\ell, ~ k\in\Z_{\ge0}, ~ j\in J_+,
\end{equation}
where $U_i^k$ and $W_j$ are arbitrary constants. To solve this problem, we introduce the following notations:
\begin{equation}\label{}
\om_{j_1 j_2\dots j_m}=\frac{\pd^m\log\tau^\rs}{\pd t_{j_1}\pd t_{j_2}\dots\pd t_{j_m}}, \quad j_1, j_2, \dots, j_m\in J_+
\end{equation}
for $m\ge2$, then their initial values are given by
\begin{equation}\label{Wj1jm}
W_{j_1 j_2\dots j_m}:=\left.\om_{j_1 j_2\dots j_m} \right|_{\bt=0}=\left.\frac{\p^{m-2} \Om_{j_1
j_2}^\rs(\bu, \bu',\dots)}{ \pd {t_{j_3}}\dots\pd {t_{j_m}} }\right|_{ u_i^{(k)}\mapsto U_i^k}.
\end{equation}
Thus, we can write down the formal power series solution to the equations \eqref{tau-cover} with the above initial data via the tau function given by
\begin{equation}\label{tauW}
\log\tau^\rs(\bt)=\sum_{m\ge1}\sum_{j_1, j_2, \dots, j_m\in J_+}\frac{W_{j_1 j_2\dots j_m}}{m!}t_{j_1}t_{j_2}\dots t_{j_m}+\mathrm{const}.
\end{equation}
Without loss of generality, the constant term will be omitted below.

The solution \eqref{tauW} can be represented alternately as follows.
We note that the initial conditions \eqref{initial} are
equivalent to the following data:
\begin{equation}\label{initial2}
\mu_i(x):=u_i|_{t_p=x \dt_{p 1}}, \quad \om_j|_{\bt=0}=W_j, \quad
1\le i\le\ell, ~ j\in J_+.
\end{equation}
With the help of these data, the functions $w_j(x):=\om_j|_{t_p=x\dt_{p 1}}$ can be solved
from the equations:
\begin{equation}\label{initialw}
{w_j}'(x)=\left.\Om^\rs_{1 j}(\bu, \bu', \dots)\right|_{u_i\mapsto
\mu_i(x)}, \quad w_j(0)=W_j;
\end{equation}
moreover, similar to \eqref{Wj1jm}, the following functions can be calculated:
\begin{equation}\label{wj1jm}
w_{j_1 j_2\dots j_m}(x):=\left.\om_{j_1 j_2\dots j_m} \right|_{t_p=x\dt_{p 1}}=\left.\frac{\p^{m-2} \Om_{j_1
j_2}^\rs(\bu, \bu',\dots)}{ \pd {t_{j_3}}\dots\pd {t_{j_m}} }\right|_{u_i\mapsto\mu_i(x)}
\end{equation}
for $ j_1, j_2, \dots, j_m\in J_+$ and $m\ge2$. Thus, the solution \eqref{tauW} can also be represented as follows:
\begin{equation}\label{tauw}
\log\tau^\rs(\bt)=\int^{t_1}w_1(x)\od
x+
\sum_{m\ge1}\sum_{j_1,\dots,j_m\in J_{>1}}\frac{w_{j_1
\dots j_m}(t_1)}{m!}t_{j_1}\dots t_{j_m}.
\end{equation}

On the other hand, since the unknown functions $u_1, \dots, u_\ell$ of the tau cover
\eqref{tau-cover} of the Drinfeld-Sokolov hierarchy can be represented, via Miura-type transformations, by
${\om'_{m_1}}, \dots {\om'_{m_\ell}}$ (see the equalities \eqref{Omhj} and
Remark~\ref{rmk-uh}), hence the initial data \eqref{initial2} can be
replaced by
\begin{equation}\label{initial3}
w_{m_i}(x):=\om_{m_i}|_{t_p=x\dt_{p 1}}, \quad \om_j|_{\bt=0}=W_j,
\quad 1\le i\le\ell, ~ j\in J_{>m_\ell}.
\end{equation}
Therefore, we conclude the following result.
\begin{prp}\label{thm-cauchy}
For the system of equations \eqref{tau-cover} with initial data given by any of
\eqref{initial}, \eqref{initial2} or \eqref{initial3}, there
exists a unique formal power series solution, up to the addition of a constant to $\log\tau^\rs$, given by
\eqref{tauW} or \eqref{tauw}.
\end{prp}

\begin{rmk}
According to Theorem~\ref{thm-taucover}, the initial value $V(0):=V|_{\bt=0}$ is determined by \eqref{initial}, which provides an alternative way to compute the initial values \eqref{Wj1jm} as follows. Denote $A_j(0)=e^{\ad_{V(0)}}\Ld_j$, then by using \eqref{omV} and \eqref{eVt} we have
\begin{align*}
W_{j k}=&-\frac{1}{h^\rs}\left(d^\rs\mid \left[A_j(0)_{<0},
A_k(0)\right] \right).
\end{align*}
By using Leibniz's rule, one has
\[
W_{j k l}=-\frac{1}{h^\rs}\left(d^\rs\mid \left[\left[ A_j(0)_{<0}, A_k(0)\right]_{<0},
A_l(0)\right]  + \left[A_k(0)_{<0},
\left[A_j(0)_{<0}, A_l(0) \right] \right] \right),
\]
and the values $W_{j_1 j_2\dots j_m}$ for $m\ge4$ can be computed recursively.
\end{rmk}

\subsection{Examples}
At the end of this section, let us illustrate the system of equations \eqref{nablaV} and its formal solutions with some examples.
\begin{exa}\label{exa-A1KdV}
Let $\fg$ be of type $A_1^{(1)}$, for which the Coxeter number is $h=2$ and the exponents are given by all odd integers. Let the elements $\Ld_j$ be chosen as in \cite{DS}. We consider $\rs=\rs^0=(1,0)$. In this case let us choose the subspace $\mV=\C f_1$ (recall the Chevalley generators in Subsection~\ref{sec-g}), and represent the function $Q^\mV$ in the form  $Q^\mV=-u f_1$. Denote ($\ell=1$ in the present case)
\begin{equation}\label{fi1im}
f_{i_1 i_2 \dots i_m}=\left[ f_{i_1}, [\dots, [f_{i_{m-1} } ,f_{i_m}]\dots] \right], \quad 0\le i_1, i_2, \dots, i_m\le\ell, ~ m\ge2,
\end{equation}
then the function $V$ solving the equations \eqref{nablaV} is represented in terms of $u$ and $\omega_1, \omega_3, \dots$ as follows:
\[
V=\gm_1 f_0+\gm_2 f_{0 1}+ \gm_3 f_{0  0 1} +\ka_3 f_{1 0 1}+\gm_4 f_{0 1 0 1}+
\dots,
\]
where
\begin{align*}
&\gm_1=\om_1, \quad \gm_2= \frac{1}{4}\left(2 {\om_1}^2+u\right), \\
&\gm_3=-\frac{1}{24} \left(2 {\om_{1}}^3+4 {\om_{3}}+3
   {\om_{1}} u+u'\right),
\quad \ka_3=-\frac{1}{12} \left(2 {\om_{1}}^3-2 {\om_{3}}+3
   {\om_{1}} u+u'\right), \\
&\gm_4=\frac{1}{96} \left(4 {\om_{1}}^4+8 \om_1 \om_3+12 {\om_{1}}^2 u +8 {\om_{1}}
   u'+6 u^2 +3 u''\right).
\end{align*}
In particular, the functions $\om_1$, $\om_3$ and $u$ are related by
\begin{equation}\label{om13}
\frac{\p\om_1}{\p t_1}=\Om_{1 1}^{\rs^0}=\frac{1}{2} u, \quad \frac{\p\om_3}{\p t_1}=\Om_{1 3}^{\rs^0}=\frac{1}{8} \left(3u^2+u''\right).
\end{equation}
By using the relation $\p\om_3/\p t_1=\p\om_1/\p t_3$ we can derive from \eqref{om13}
the KdV equation
\[\frac{\p u}{\p t_3}=\frac32 u u'+\frac14 u'''.\]
In this case, the Drinfeld-Sokolov hierarchy \eqref{LVt2} is just the KdV hierarchy.

Applying the approach used in the previous subsection, we obtain the following tau function of the KdV hierarchy with the initial data $u|_{t_{j>1}=0}=t_1$:
\begin{align}\label{KdVtau1}
\log\tau^{\rs^0}=& \frac{{t_1}^3}{12}+  W_1{t_1} +
\left(\frac{{t_1}^3}{8}+W_3\right) {t_3}+
   \left(\frac{5 {t_1}^4}{64}+\frac{5 {t_1}}{32}+W_5\right){t_5}+
   \left(\frac{7 {t_1}^5}{128}+\frac{35
   {t_1}^2}{128}+W_7\right){t_7} \nn\\
   & +\left(\frac{3
   {t_1}^3}{16}+\frac{3}{64}\right) {t_3}^2+\left(\frac{45
   {t_1}^4}{128}+\frac{15 {t_1}}{32}\right) {t_3} {t_5} + \dots,
\end{align}
where $W_k$ are constants. In particular, if we take $W_j=\dt_{j 3}/16$, then this tau function corresponds to the well-known {\em Witten-Kontsevich tau function} \cite{Ko,Wi} of the KdV hierarchy.
\end{exa}

\begin{exa}\label{exa-A1mKdV}
Let $\fg$ be of type $A_1^{(1)}$ and $\rs=\mathds{1}=(1,1)$. In
this case the nilpotent subalgebra $\mathcal{N}$ is trivial, and the
subspace $\mV=\mathcal{B}=\C (\al_1^\vee-\al_0^\vee)$. Let
$Q^\mV=\frac{v}{2}(\al_1^\vee-\al_0^\vee)$, then by using the relation
\[e^{\ad_V}=e^{\ad_{U(Q^\mV)}}e^{\ad_\Om}\]
we can write down the unknown function $V$ in \eqref{nablaV} as follows:
\[
V=\gm_1 f_0+\ka_1 f_1+\gm_2 f_{0 1}+ \gm_3 f_{0  0 1} +\ka_3 f_{1 0 1}+\gm_4 f_{0 1 0 1}+
\dots,
\]
where
\begin{align*}
&\gm_1=\frac{1}{2} (2 {\om_1}-v), \quad \ka_1=\frac{1}{2} (2
   {\om_1}+v), \quad  \gm_2= \frac{1}{4} \left(2 {\om_1}
   v+v'\right), \\
&\gm_3=\frac{1}{48} \left(2 {\om_1} v^2-6
   {\om_1} v'-4 {\om_1}^2 v-8 {\om_3}+4 v^3-3 v''+v
   v'\right), \\
&\ka_3=\frac{1}{48} \left(-2 {\om_1} v^2-6
   {\om_1} v'-4 {\om_1}^2 v+8 {\om_3}+4 v^3-3 v''-v
   v'\right), \\
&\gm_4=\frac{1}{96} \left(-8 {\om_1} v^3+6
   {\om_1} v''+4 {\om_1}^2 v'+8 {\om_3} v+3
   v^{(3)}-17 v^2 v'\right).
\end{align*}
The functions $\om_1$, $\om_3$ and $v$ satisfy the equations
\begin{equation}\label{om13A1p}
\frac{\p\om_1}{\p t_1}=\Om_{1 1}^\one=-\frac{1}{2}v^2, \quad \frac{\p\om_3}{\p t_1}=\Om_{1 3}^\one=\frac{1}{8} \left(3 v^4+(v')^2-2v v''\right).
\end{equation}
Then the relation $\p\om_3/\p t_1=\p\om_1/\p t_3$ leads to
the modified KdV equation
\[
\frac{\p v}{\p t_3}=-\frac{3 }{2}v^2 v'+\frac{1}{4}v'''.
\]
The Drinfeld-Sokolov hierarchy in this case is the modified KdV hierarchy, which is related to the KdV hierarchy via the Miura transformation $u=-v^2+v'$. It is easy to see that the tau functions $\tau^{\mathds{1}}$ and $\tau^{\rs^0}$ of these two integrable hierarchies are related by the formula
\[
v=2\frac{\p}{\p t_1}\log\frac{\tau^{\rs^0}}{\tau^{\mathds{1}}}.\
\]
Similar to the above example, one can write down the formal power series expression of the tau function $\log\tau^\mathds{1}$ with any given initial data.
\end{exa}

\begin{exa}\label{exa-A2tau}
Let $\fg$ be of type $A_2^{(1)}$, of which the
Coexter number is $h=3$, the set of exponents is $J=3\Z\pm1$, and
the basis elements $\Ld_j$ of the principal Heisenberg subalgebra are chosen as in \cite{DS}. Let us take $\rs=\rs^0=(1,0,0)$, and fix a
subspace $\mV=\C(f_1+f_2)\oplus\C f_{1 2}$. We write
\[
Q^\mV=-u f_2-v f_{1 2},
\]
where $f_{i_1\dots i_m}$ are defined as in  \eqref{fi1im}.
Then the function $V$ solving \eqref{nablaV} can be represented as
\begin{equation}\label{}
V=\gamma_1 f_0+\gamma_2 f_{0 1} + \ka_2 f_{0 2}  + \gamma_3 f_{1 0 2} + \ka_3 f_{2 0 1}+\dots,
\end{equation}
where
\begin{align*}
&\gamma_{1}={\om_{1}}, \quad \gamma_{2}= \frac{1}{6} \left(3{\om_{1}}^2+3
   \om_{2}+u\right), \quad \gamma_{3}=\frac{1}{18} \left(-3{\om_{1}} u-3
  {\om_{1}}^3+9 \om_{2}{\om_{1}}-4 u'+6 v\right), \\
& \ka_{2}= \frac{1}{6} \left(3{\om_{1}}^2-3
   \om_{2}+u\right),
\quad
\ka_{3}=
   \frac{1}{18} \left(-3{\om_{1}} u-3{\om_{1}}^3-9 \om_{2}
  {\om_{1}}+2 u'-6 v\right).
\end{align*}
The second-order derivatives of $\log\tau^\rs$ with respect to $t_1$ and $t_2$ are given by
\begin{equation}\label{OmA2}
\Om_{1 1}^\rs=\frac{u}{3}, \quad \Om_{1 2}^\rs=\frac{2 v }{3}-\frac{u' }{3}, \quad \Om_{2 2}^\rs=-\frac{2}{9} u ^2-\frac{u'' }{9}.
\end{equation}
In particular, the relation $\pd^2 \Om_{1 1}^\rs/\pd {t_2}^2=\pd^2 \Om_{2 2}^\rs/\pd {t_1}^2$ gives the Boussinesq equation
\[
\frac{\p^2 u}{\p {t_2}^2}=-\frac{2}{3} \left(u ^2\right)''-\frac{u^{(4)} }{3}.
\]
\end{exa}

\begin{exa}\label{exa-A22tau}
Let  $\fg$ be of type $A_2^{(2)}$ and $\rs=\rs^0=(1,0)$. The Coxeter number is $h=3$, the set of exponents is $J=6\Z\pm1$, and the generators  $\Ld_j$ are normalized as in \cite{Wu} with a constant $\nu=\sqrt{2}$ that appears in \eqref{Ld1Ld}. We fix a subspace $\mV=\C f_1$ of $\mathcal{B}$, and write the function $Q^\mV=-\sqrt{2}\,u f_1$. Then, the function $V$ solving the equation \eqref{nablaV} has the expression
\[
V=\gm_1 f_0+\gm_2 f_{0 1}+ \gm_3 f_{0 0 1} +\gm_4 f_{0 0 0 1}+ \gm_5 f_{0 0 0
0 1} + \ka_5 f_{1 0 0 0 1} \dots,
\]
where
\begin{align*}
&\gm_1=\frac{{\om_1}}{\sqrt{2}}, \quad \gm_2= \frac{1}{6}
   \left(3 {\om_1}^2+u\right), \quad
\gm_3= -\frac{1}{36 \sqrt{2}} \left(3
   {\om_1} u+ 3  {\om_1}^3+
   u'\right), \\
&\gm_4=\frac{1}{432} \left(3 {\om_1} u'+3
   {\om_1}^2 u+u''+u^2\right),\\
&\gm_5=\frac{1}{ 8640\sqrt{2}}\left(6   {\om_1}^5 + 36
   {\om_5}- 10
   {\om_1} u''-10 {\om_1} u^2-10
   {\om_1}^2 u'-3   u^{(3)}-10   u u'\right),  \\
&\ka_5=\frac{1}{2160\sqrt{2}}\left(9   {\om_1}^5-36
     {\om_5}- 10   {\om_1} u''-5   {\om_1}
   u^2-15   {\om_1}^2 u'-2   u^{(3)}-5   u
   u'\right).
\end{align*}
The functions $\om_1$, $\om_5$ and $u$ satisfy the equations
\begin{equation}\label{om15}
\frac{\p\om_1}{\p t_1}=\Om_{1 1}^{\rs^0}=\frac{1}{3} u, \quad \frac{\p\om_5}{\p t_1}=\Om_{1 5}^{\rs^0}=-\frac{1}{324}\left(20 u^3+30 u u''+3
u^{(4)}\right).
\end{equation}
By using $\p\om_5/\p t_1=\p\om_1/\p t_5$ we arrive at
the Sawada-Kotera equation
\[
\frac{\p u}{\p t_5}=-\frac{1}{108}\left(20 u^3+30 u u''+3
u^{(4)}\right)'.
\]
Suppose that $u|_{t_{j>1}=0}=3 t_1$, then the tau function is given by
\begin{align}\label{SKtau1}
\log\tau^{\rs^0}=& \frac{{t_1}^3}{6}+{t_1} W_1+ \left(W_5-\frac{5
{t_1}^4}{12}\right){t_5}+
   \left(-\frac{7 {t_1}^5}{15}-\frac{7
   {t_1}^2}{12}+W_7\right){t_7} \nn\\
   &\quad+\left(\frac{5
   {t_1}^5}{2}+\frac{25 {t_1}^2}{12}\right) {t_5}^2 + \dots,
\end{align}
where $W_j$ are constants.
\end{exa}

\section{Virasoro constraints for Drinfeld-Sokolov hierarchies}\label{sec-vir}

We consider in this section the Virasoro constraints for the tau cover \eqref{tau-cover} of
the Drinfeld-Sokolov hierarchy.

\subsection{Virasoro symmetries}
Let us first present the Virasoro symmetries of the Drinfeld-Sokolov hierarchy in terms of the $\fg_{<0}$-value function $V$.
Observe that the commutation relations \eqref{Ldij} between the generators for the principal
Heisenberg subalgebra $\mathcal{H}$ are preserved under the scaling
transformations $\Ld_{\pm j}\mapsto\ld_j^{\pm1}\Ld_{\pm j}$ for
arbitrary nonzero constants $\ld_j$ with $j\in J_+$, so we can adjust these generators such that they also satisfy the following commutation relations:
\begin{equation}\label{dLd}
[d_k^\mathds{1}, \Ld_j]=-\frac{j}{r h}\Ld_{j+r h k}, \quad k\in\Z, ~
j\in J_+.
\end{equation}
Suppose that a function $V$ of $\bt=\{t_j\mid j\in J_+\}$ taking
value in $\fg_{<0}$ solves the system of equations \eqref{nablaV}.
We denote $\Xi=\sum_{j\in J_+}t_j\Ld_{j}$ and define
\begin{equation}\label{Bk}
B_k=e^{\ad_V}e^{-\ad_\Xi} d_k^\mathds{1}-d_k^\rs, \quad k\in \Z.
\end{equation}
From
\eqref{dd} and \eqref{dskdsl} it follows that $B_k$ takes value in $\fg$ and that they satisfy the commutation relations
\begin{equation}\label{Bkl}
[B_k+d_k^\rs, B_l+d_l^\rs]=(k-l)(B_{k+l}+d_{k+l}^\rs), \quad k,l\in\Z.
\end{equation}
Similar to the system of equations \eqref{nablaV}, we introduce the following
evolutionary differential equations of $V$:
\begin{equation}\label{nablabtV}
\nabla_{\beta_k, V} V=-\left(B_k\right)_{<0},
\end{equation}
where
\begin{enumerate}
\item[(I)] $k\ge-1$ when $r  h^\rs=1$, namely, $\fg$ is of untwisted type $X_\ell^{(1)}$
and $\rs$ is equivalent to $\rs^0=(1,0,0,\dots,0)$ via a diagram automorphism of $\fg$;
\item[(II)] $k\ge0$ when $r h^\rs>1$, namely, all cases except those in class (I). In
particular, it includes all cases that correspond to twisted affine Kac-Moody algebras.
\end{enumerate}
The reason that we take such values of the indices $k$ will be explained  in the proof of Lemma~\ref{thm-betat}.
\begin{lem}\label{zh-4}
Any solution $V$ of the evolutionary equations \eqref{nablabtV}
also satisfies the following equations:
\begin{align}
&\frac{\pd B_l}{\pd
\beta_k}=\left[-\left(B_k\right)_{<0},
B_l+d_l^\rs\right] , \quad j\in J_+, \label{Bbt} \\
& \frac{\pd}{\pd
\beta_k}\left(e^{\ad_V}\Ld_j\right)=\left[-\left(B_k\right)_{<0},
e^{\ad_V}\Ld_j\right] , \quad j\in J_+,  \label{eVbt}
\end{align}
where the range of $k$ is specified in the above cases (I) or (II).
\end{lem}

\begin{lem}\label{thm-betat}
Let $V$ be a solution of the systems \eqref{nablaV} and \eqref{nablabtV},
then it also satisfies the following equations:
\begin{align}\label{Bt}
&\frac{\p B_k}{\p t_j}=\left[-\left(e^{\ad_V}\Ld_j\right)_{\ge0}, B_k+d_k^\rs\right], \\
&  \frac{\p}{\p t_j}\frac{\p}{\p \beta_k}V=\frac{\p}{\p \beta_k}\frac{\p}{\p t_j}V.  \label{Vtbt}
\end{align}
Here $j\in J_+$ and $k$ are given in the above cases (I) or (II)..
\end{lem}
\begin{prf}
By using \eqref{nablaV} we can verify the validity of \eqref{Bt} as follows:
\begin{align*}
\frac{\p B_k}{\p t_j}=&\left[\nabla_{t_j, V}V,
B_k+d_k^\rs\right]+e^{\ad_V}\left[\nabla_{t_j, -\Xi}(-\Xi),
e^{-\ad_\Xi}d_k^\mathds{1}\right] \\
=&\left[\left(e^{\ad_V}\Ld_j\right)_{<0},
B_k+d_k^\rs\right]+e^{\ad_V}e^{-\ad_\Xi}\left[-\Ld_j,
d_k^\mathds{1}\right] \\
=&\left[\left(e^{\ad_V}\Ld_j\right)_{<0},
B_k+d_k^\rs\right]-\left[e^{\ad_V}\Ld_j,
B_k+d_k^\rs\right]\\
=&\left[-\left(e^{\ad_V}\Ld_j\right)_{\ge0},
B_k+d_k^\rs\right].
\end{align*}
For the same reason as given in the proof of \eqref{Vtij}, in order to show \eqref{Vtbt} it suffices to verify
\[
\left[ \frac{\pd}{\pd
t_j}, \frac{\pd}{\pd
\beta_k} \right]  \left(e^{\ad_V}\Ld_i\right)
=0, \quad \forall i\in J_+.
\]
Let us write $A_j=e^{\ad_V}\Ld_j$ for short, then the left hand side of the above equation is equal to
\begin{align}
\mathrm{l.h.s.}=&[ [ -(A_j)_{\ge0}, -B_k-d_k^\rs]_{<0}, A_i]+[- (B_k)_{<0}, [(A_j)_{<0}, A_i] ] \nn\\
&\quad-[ [ -(B_k)_{<0}, A_j]_{<0}, A_i]- [ (A_j)_{<0}, [-(B_k)_{<0}, A_i] ] \nn \\
=& (-[ [(B_k)_{<0},   (A_j)_{\ge0}]_{<0}, A_i]+[ [ (B_k)_{<0}, A_j]_{<0}, A_i] \nn\\
&\quad- ([(B_k)_{<0}, [(A_j)_{<0}, A_i] ]  - [ (A_j)_{<0}, [(B_k)_{<0}, A_i] ] )  \nn \\
=&[ [ (B_k)_{<0}, (A_j)_{<0}], A_i] -[ [ (B_k)_{<0}, (A_j)_{<0}], A_i]=0 .  \label{eVtbt}
\end{align}
Here in the derivation of the second equality we used the fact that $[(A_j)_{\ge0},d_k^\rs]_{<0}=0$ when $k$ takes values specified in Cases (I) and (II). The lemma is proved.
\end{prf}

It follows from the above lemma that the flows \eqref{nablabtV} are symmetries of the system of equations \eqref{nablaV}.

\begin{prp}
The symmetries \eqref{nablabtV}  of the system of equations \eqref{nablaV} yield the following symmetries of the tau cover \eqref{tau-cover} of the Drinfeld-Sokolov hierarchy:
\begin{align}
\frac{\pd\log\tau^\rs}{\pd\beta_k} &= \frac{1}{h^\rs}\left(d^\rs\mid B_k\right), \label{logtaubeta}  \\
\frac{\pd\omega_j}{\pd\beta_k} &=\frac{1}{h^\rs}\left(d^\rs\mid \left[\left(B_k\right)_{<0}, e^{\ad_V}\Ld_j\right]\right),  \label{ombeta} \\
\frac{\pd\sL^\mV}{\pd \beta_k} &= \left[\nabla_{\beta_k, N}N -(e^{\ad_N}B_k)_{<0}, \sL^\mV\right]+\frac{1}{h^\rs}\left(d^\rs\mid \left[\left(B_k\right)_{<0}, e^{\ad_V}\Ld_1\right]  \right) c \label{Lbt}
\end{align}
with $k$ being given as in Cases (I) and (II).
Here $N$ is the $\mathcal{N}$-valued function determined by \eqref{sLV2}, whose components are differential polynomials of the components of $V$.
\end{prp}
\begin{prf}
We first note that the equations \eqref{ombeta} follow directly from \eqref{eVbt} and \eqref{omV}.
By using \eqref{sLV2}, \eqref{UHLV} and \eqref{L3} we obtain
\begin{align*}
\frac{\pd\sL^\mV}{\pd \beta_k} =& \left[\nabla_{\beta_k, N}N, e^{\ad_N}e^{\ad_V}\left(\frac{\p}{\p t_1}+\Ld_1\right)\right]+e^{\ad_N}\left[-(B_k)_{<0}, e^{\ad_V}\left(\frac{\p}{\p t_1}+\Ld_1\right)\right]+ \frac{\p\om_1}{\p\beta_k} c \\
=&\left[\nabla_{\beta_k, N}N -(e^{\ad_N}B_k)_{<0}, \sL^\mV-\om_1 c\right]+ \frac{\p\om_1}{\p\beta_k} c  \\
=&\left[\nabla_{\beta_k, N}N -(e^{\ad_N}B_k)_{<0}, \sL^\mV \right]+ \frac{\p\om_1}{\p\beta_k} c,
\end{align*}
so from \eqref{ombeta} we also have \eqref{Lbt}. On the other hand, from \eqref{Bt} it follows that
\[
\frac{\p}{\p t_j}\frac{\pd\log\tau^\rs}{\pd\beta_k}= \frac{1}{h^\rs}
\left(d^\rs\mid \left[-\left(e^{\ad_V}\Ld_j\right)_{\ge0},
B_k\right]\right) =-\frac{1}{h^\rs}\left(d^\rs\mid
\left[e^{\ad_V}\Ld_j,
(B_k)_{<0}\right]\right)=\frac{\pd\om_j}{\pd\beta_k}.
\]
Therefore the proposition is proved.
\end{prf}

\begin{thm}\label{thm-vir}
The symmetries $\p/\p\beta_k$ of the tau cover of the Drinfeld-Sokolov hierarchy satisfy the following Virasoro
commutation relations:
\begin{equation}\label{taubtkl}
\left[\frac{\p}{\p \beta_l}, \frac{\p}{\p \beta_k}\right]\log\tau^\rs=(k-l)  \frac{\p\log\tau^\rs}{\p \beta_{k+l}}.
\end{equation}
Moreover, these symmetries can be represented in the form
\begin{equation}
\frac{\pd\log\tau^\rs}{\pd\beta_k} =
\begin{cases}
\dfrac{1}{r h}\displaystyle\sum_{j\in J_{+} } (j+r h)
t_{j+r h}\omega_{j}+\dfrac{1}{2 r
h}\sum_{j\in J_+\mid i+j=r h}i j t_i t_j,   & k=-1\ \mbox{for Case (I)}; \\
\dfrac{1}{r h}\displaystyle\sum_{j\in J_+} j
t_j\omega_{j}+C^\rs, & k=0 \ \mbox{for Case (I, II)};\\
\dfrac{1}{h^\rs}\left(d^\rs\mid
e^{\ad_{U(Q^\mV)} }d_k^\mathds{1}-d_k^\rs\right)+\dfrac{1}{2 r h}\displaystyle\sum_{j\in J_+\mid i+j=r h k}\omega_i\omega_j \\
\quad +\dfrac{1}{r h}\sum_{j\in J_+} j
t_j\omega_{j+r h},  & k\ge1 \ \mbox{for Case (I, II)}.
\end{cases}
\label{logtaubeta2}
\end{equation}
Here $C^\rs$ is a constant given by
\begin{equation}\label{Cs}
C^\rs=\left(d^\rs\mid
\rho^\rs-\rho^{\mathds{1}} \right)-\dfrac{r h^\rs}{2}\left((\rho^\rs\mid\rho^\rs)-(\rho^{\mathds{1}}\mid\rho^{\mathds{1}})\right).
\end{equation}
with $\rho^\rs$ defined by \eqref{rhos}.
\end{thm}
\begin{prf}
By using \eqref{Bbt} and \eqref{logtaubeta} we obtain
\begin{align*}
\left[\frac{\p}{\p \beta_l}, \frac{\p}{\p \beta_k}\right]\log\tau^\rs =& \frac{1}{h^\rs}\left(d^\rs\mid
 [-(B_l)_{<0},B_k+d_k^\rs]-[-(B_k)_{<0},B_l+d_l^\rs]\right) \\
 =& \frac{1}{h^\rs}\left(d^\rs\mid
 [(B_k)_{\ge0},(B_l)_{<0}]+[(B_k)_{<0},(B_l)_{\ge0}]\right) \\
=& \frac{1}{h^\rs}\left(d^\rs\mid
 [B_k,B_l]\right) \\
 =& \frac{1}{h^\rs}\left(d^\rs\mid
 [B_k+d_k^\rs,B_l+d_l^\rs]-(k-l)d_{k+l}^\rs\right) \\
 =& \frac{1}{h^\rs}\left(d^\rs\mid
 (k-l)B_{k+l}\right) =(k-l)  \frac{\p\log\tau^\rs}{\p \beta_{k+l}}.
\end{align*}
Here to derive the second and the fourth equalities we have used \eqref{dsdkbf}, and the fifth equality is due to \eqref{Bkl}. So the first assertion of the theorem holds true.
By using \eqref{UHLV} and the fact that $[d^\rs, N]=0$ we have
\begin{align}
&\left(d^\rs\mid B_k\right)\nn\\
=&\left(d^\rs\mid e^{-\ad_N}e^{\ad_{U(Q^\mV)}}e^{\ad_\Om}e^{-\ad_{\sum_{i\in
J_+}t_i\Ld_i}}d_k^\mathds{1}-d_k^\rs\right) \nn\\
=&\Biggl(e^{\ad_N}d^\rs\mid e^{\ad_{U(Q^\mV)}}e^{\ad_{\sum_{j\in
J_+}{j}^{-1}\om_j\Ld_{-j}}}\biggl(d_k^\mathds{1}-\sum_{i\in
J_+}\frac{i t_i}{r h}\Ld_{i+r h k}\nn \\
&\quad+\frac{\dt_{k,-1}}{2 r h}
\sum_{0<i<r h}i (r h-i) t_i t_{r h-i}  c\biggr) -d_k^\rs\Biggr) \nn\\
=&\Biggl(d^\rs\mid e^{\ad_{U(Q^\mV)}}\biggl(d_k^\mathds{1}-\sum_{j\in
J_+}\frac{\om_j}{r h}\Ld_{-j+r h k}+\frac{1}{2 r h} \sum_{0<j<r h
k}\om_j \om_{r h k-j}  c  -\sum_{i\in J_+}\frac{i
t_i}{r h}\Ld_{i+r h k} \nn\\
 &\quad+\sum_{i>\max(0, -r h k)}\frac{i t_i}{r h}\om_{i+r h
k} c+\frac{\dt_{k,-1}}{2 r h} \sum_{0<i<r h}i (r h-i) t_i t_{r h-i}
 c\biggr) -d_k^\rs\Biggr) \nn\\
=&\left(d^\rs\mid
e^{\ad_{U(Q^\mV)}}d_k^\mathds{1}-d_k^\rs\right)+\frac{h^\rs}{2 r h}
\sum_{0<j<r h
k}\om_j \om_{r h k-j} \nn\\
&\quad+\frac{h^\rs}{r h}\sum_{i>\max(0, -r h k)}{i t_i}\om_{i+r h k}+
\frac{\dt_{k,-1}h^\rs}{2 r h} \sum_{0<i<r h}i (r h-i) t_i t_{r h-i}.\label{zh-5}
\end{align}
Here to derive the second and the third equalities we have used the normalization equations \eqref{dLd} and \eqref{Ldij}, and to derive the last equality we have used the condition \eqref{ULdc}.
Note that the first term on the right hand side of \eqref{zh-5} vanishes when $k=-1$ for Case (I), and it is a constant when $k=0$, i.e.
\begin{equation}\label{Cs2}
C^\rs=\left(d^\rs\mid
d_0^\mathds{1}-d_0^\rs\right)=\left(d^\rs\mid
\rho^\rs-\rho^{\mathds{1}} \right)-\dfrac{r h^\rs}{2}\left((\rho^\rs\mid\rho^\rs)-(\rho^{\mathds{1}}\mid\rho^{\mathds{1}})\right)
\end{equation}
due to \eqref{dd}. Thus the theorem is proved.
\end{prf}

\begin{exa}
When $\rs$ is the principal gradation $\one$ we have $C^\mathds{1}=0$, and when $\rs$ is the homogeneous gradation $\rs^0$ we have
\begin{equation}\label{C1C0}
C^{\rs^0}=\frac{r k_0}{2}(\rho^{\mathds{1}}\mid\rho^{\mathds{1}})
=\frac{k_0}{2 r h^2}(1,1,\dots,1)\mathring{A}^{-1}D(1,1,\dots,1)^T
\end{equation}
with $D=\mathrm{diag}\left(k_1/k_1^{\vee}, k_2/k_2^{\vee}, \dots, k_\ell/k_\ell^{\vee}\right)$. Note that in the derivation of the second equality we have used \eqref{blf} and the facts $\rho^{\rs^0}=0$, $(d^{\rs^0}\mid \rho^\one)=0$.
\end{exa}

Due to the commutation relation \eqref{taubtkl},  we give the following definition.
\begin{dfn}
The flows defined by \eqref{logtaubeta}--\eqref{Lbt} are
called the \emph{Virasoro symmetries} (of the tau cover) of the
Drinfeld-Sokolov hierarchy \eqref{LVt}  associated to $(\fg, \rs, \mathds{1})$.
\end{dfn}

\subsection{Virasoro constraints}

Let us consider solutions of the Drinfeld-Sokolov hierarchy associated to $(\fg, \rs, \mathds{1})$ that satisfy either of the following equations:
\begin{align}\label{str}
&\sum_{p\in J_+}\left(\frac{p+h}{h}t_{p+h}-a_{p}\right)\frac{\pd
\log\tau^{\rs}}{\pd t_p}+\dfrac{1}{2 h}\sum_{i,j\in J_+, \, i+j=h}i j t_i t_j=0, \quad \hbox{for Case (I)}, \\
&\sum_{p\in J_+}\left(\frac{p}{r h}t_{p}-b_p\right)\frac{\pd
\log\tau^{\rs}}{\pd t_p}+C^\rs=0, \quad \hbox{for Cases (I) and (II)}. \label{sim}
\end{align}
Here $a_{p}$ and $b_p$ are constants that vanish except for finitely many of exponents $p\in J_+$.
We call the equation \eqref{str} the {\em (generalized) string equation}, and the equation \eqref{sim} the {\em similarity equation} for it is related to the so-called similarity reductions of the Drinfeld-Sokolov hierarchy.

\begin{thm}\label{thm-Virconstraint}
The equations \eqref{str} and \eqref{sim} lead respectively to the following Virasoro constraints for the tau function of the Drinfeld-Sokolov hierarchy \eqref{LVt}:
\begin{align}\label{strVir}
&S_{k}(\log\tau^\rs)-\sum_{p\in J_+}a_{p}\frac{\pd
\log\tau^{\rs}}{\pd t_{p+h (k+1) } }=0, \quad k=-1,0,1,2,\dots  \hbox{ for Case (I)}, \\
&S_k(\log\tau^\rs)-\sum_{p\in J_+}b_p\frac{\pd \log\tau^{\rs}}{\pd
t_{p+r h k} }=0, \quad k=0,1,2,\dots  \hbox{ for Cases (I) and (II),}
\label{simVir}
\end{align}
where $S_k(\log\tau^\rs)$ denote the right hand side of \eqref{logtaubeta2}.
\end{thm}
\begin{rmk}\label{rmk-virtau}
From \eqref{Omhj} and Remark~\ref{rmk-uh} it follows that $Q^\mV$ can be represented by the second order derivatives of $\log\tau^\rs$ with respect to the time variables. Hence the right hand side of \eqref{logtaubeta2} can always be represented in terms of the functions $\om_j=\p\log\tau^\rs/\p t_j$ together with the time variables.
\end{rmk}

\begin{prf}
The constraints \eqref{strVir} are proved in \cite{Wu} for $\rs=\rs^0$ and
$a_{p}=\dt_{p 1}$. For the general case the proof is almost the
same, so we omit it here. Let us check the validity of the constraint \eqref{simVir}. Note that the similarity equation \eqref{sim} is just the equation
\begin{align}
&\frac{\pd\log\tau^{\rs}}{\pd\beta_{0}}=\sum_{p\in J_+}b_p\frac{\pd
\log\tau^{\rs}}{\pd t_p}.
\label{sim0}
\end{align}
Since $\p/\p\beta_{0}$ is a symmetry of the system of equations  \eqref{nablaV}, we have $\pd V/\pd\beta_0=\sum_{p\in J_+}b_p\pd V/\pd t_p$, from which it follows that
\[
\nabla_{\beta_{0},V}V-\sum_{p\in J_+}b_p \nabla_{t_p,V}V=0.
\]
This equation together with \eqref{nablabtV} and \eqref{nablaV} leads to
the equation
\begin{equation}\label{str2}
-\left(e^{\ad_V}e^{-\ad_\Xi}d^\mathds{1}_{0}-d_{0}^\rs+\sum_{p\in
J_+} b_p e^{\ad_V}\Ld_p\right)_{<0}=0.
\end{equation}
To simplify the notations in this proof, let us identify $\fg$ with its realization \eqref{gAs}. By using \eqref{dkrs}, \eqref{dd} and \eqref{dLd} we have
\[
d_k^\rs=z^{r h^\rs}d_{k-1}^\rs, \quad d_k^\mathds{1}=z^{r
h^\rs}d_{k-1}^\mathds{1}, \quad \Ld_{p+r h}=z^{r h^\rs} \Ld_p.
\]
Then the subscript ``$<0$'' in the equation \eqref{str2} can be understood as to take the negative part of the
Laurent series in $z$. For any $k\ge1$, by multiplying $z^{r h^\rs k}$ to the equation \eqref{str2} we obtain
\begin{equation}\label{}
-\left(e^{\ad_V}e^{-\ad_\Xi}d^\mathds{1}_{k}-d_{k}^\rs+\sum_{p\in
J_+}b_p e^{\ad_V}\Ld_{p+r h k}\right)_{\le0}=0.
\end{equation}
Thus, from \eqref{logtaubeta} and  \eqref{omV}, it follows that
\[
-\frac{\pd\log\tau^{\rs}}{\pd\beta_{k}}+\sum_{p\in J_+}b_p\frac{\pd
\log\tau^{\rs}}{\pd t_{p+r h k} }=0, \quad k=1,2,\dots.
\]
Thus the theorem is proved.
\end{prf}

\begin{dfn}
We call the series of equations given in \eqref{strVir} the Virasoro constraints of the first type, and the ones given in \eqref{simVir} the Virasoro constraints of the second type.
\end{dfn}

\subsection{Solutions of Witten-Kontsevich and of Brezin-Gross-Witten types}

In this subsection, we illustrate the solutions of the Drinfeld-Sokolov hierarchy that satisfy the Virasoro constraints \eqref{str} or \eqref{sim}.

We first consider the Virasoro constraint \eqref{str} with $\rs=\rs^0$ and $a_{p}=\dt_{p 1}$, which is just the string equation in the literature.  In this case, by taking the derivatives of the string equation with respect to $t_{m_i}$ and $t_1$, we obtain
\begin{align}\label{ommi0}
&\left.\frac{\p^{k} \om_{m_i}}{\p {t_1}^{k}}\right|_{\bt=0}=\dt_{i,
\ell}\dt_{k,2}\frac{h-1}{h}, \quad 1\le i\le\ell, ~ k\ge 1.
\end{align}
Here we used \eqref{mirh} and the fact that $m_\ell=h-1$ in this special  case.
As mentioned in Remark~\ref{rmk-uh}, the unknown functions $u_1, \dots,
u_\ell$ of the system of equations \eqref{tau-cover} can be represented by $\om'_{m_1}, \dots \om'_{m_\ell}$ via a Miura-type
transformation, hence the
initial values $U_i^k=\left.u_i^{(k)}\right|_{\bt=0}$ are determined
by \eqref{ommi0}. Furthermore, for any $j\in J_+$ we take the derivative of the string equation with respect to $t_{j+h}$, then we obtain
\[
W_j:=\om_j|_{\bt=0}=\left.\frac{h}{j+h}\Om_{1,
j+h}^\rs\right|_{u_i^{(k)}\mapsto U_i^k}.
\]
From Proposition~\ref{thm-cauchy} it follows that
$\log\tau^{\rs^0}$ is determined up to a constant term. Such a tau
function, which satisfies the Virasoro constraints of the first type with $\rs=\rs^0$ and $a_{p}=\dt_{p 1}$,
is called the {\em topological solution} of the
Drinfeld-Sokolov hierarchy. In particular, the topological solution
for $A_1^{(1)}$ is the well-known Witten-Kontsevich tau
function given in \eqref{KdVtau1}. We remark that another algebraic procedure was proposed in
\cite{CW2} to compute explicitly the topological solution via Hankel
determinants.

In contrast to the string equation \eqref{str}, the similarity
equation \eqref{sim} is weaker. Let us proceed to consider solutions of the Drinfeld-Sokolov hierarchy satisfying this constraint.

\begin{prp}\label{thm-simred}
We impose the Virasoro constraint \eqref{sim} with $b_p=\dt_{ p 1}$ on the tau function of the Drinfeld-Sokolov hierarchy, i.e.
\begin{equation}\label{sim2}
\sum_{p\in J_+}\left(\frac{p}{r h}t_{p}-\dt_{p 1}\right)\frac{\pd
\log\tau^{\rs}}{\pd t_p}+C^\rs=0.
\end{equation}
Then its solution $\log\tau^\rs$ is determined by the following $\ell-1$ parameters:
\begin{equation}\label{ommi2}
W_{m_i}=\om_{m_i}|_{\bt=0}, \quad i=2,3,\dots,\ell.
\end{equation}
\end{prp}
\begin{prf}
Let us consider the initial data
$w_j(x)=\om_j|_{t_p=x \dt_{p 1}}$ with $j\in J_+$ for any solution of the system \eqref{tau-cover} satisfying the constraint \eqref{sim2}. Firstly, we take
$t_{p}=x \dt_{p 1}$ in \eqref{sim2} to arrive at
\begin{equation}\label{w1BGW}
w_1(x)= C^\rs\frac{r h}{r h-x}.
\end{equation}
For any $j\in J_{>1}$, by taking the derivative of \eqref{sim2} with
respect to $t_j$ we obtain
\begin{equation*}
\left.\left(\left(\frac{t_1}{r h}-1\right)\Om_{1 j}^\rs+\frac{j}{r
h}\om_j\right)\right|_{t_p=x \dt_{p 1}}=0,
\end{equation*}
which is equivalent to
\begin{equation}\label{prfBGW}
\left(\frac{x}{r h}-1\right){w_j}'(x)+\frac{j}{r h}w_j(x)=0.
\end{equation}
In particular, it follows from \eqref{ommi2} that
\begin{equation}\label{wjBGW}
w_j(x)=W_j\left(\frac{r h}{r h-x}\right)^j, \quad j=m_2, m_3, \dots,
m_\ell.
\end{equation}
Here $W_j$ are arbitrary constants.
Then the functions $\mu_i(x)=u_i|_{t_p=x\dt_{p 1}}$ are determined by \eqref{w1BGW} and \eqref{wjBGW} due to Remark~\ref{rmk-uh}.
By using \eqref{prfBGW} again and the fact that $\Om_{j k}^\rs$ are differential polynomials of $\mathbf{u}=(u_1, \dots, u_\ell)$, we have
\begin{equation}\label{}
w_j(x)=\left.\frac{r h-x}{j}\Om_{1
j}^\rs\right|_{u_i\mapsto\mu_i(x)}, \quad j\in J_{>m_\ell}.
\end{equation}
The above constructed functions $w_j$ and $\mu_i$ gives a collection of initial data for the tau cover \eqref{tau-cover}, so they lead to a unique solution $(f, \omega_j, u_i)$. Since the left hand side of \eqref{sim2} is a symmetry of the tau cover \eqref{tau-cover}, the solution $(f, \omega_j, u_i)$ also satisfy \eqref{sim2}.
The proposition is proved.
\end{prf}

\begin{exa}
Let $\fg$ be of type $A_1^{(1)}$ and $\rs=\rs^0$. By using \eqref{C1C0}, we have $C^{\rs^0}=1/16$. Then the equation \eqref{w1BGW} gives
\[
w_1(x)=\frac{1}{8(2-x)}, \quad \mu_1(x)=2 w_1'(x)=\frac{1}{4(2-x)^2}.
\]
By using Proposition~\ref{thm-cauchy}, we obtain the following solution
\begin{align}
\log\tau^{\rs^0}=&-\frac{1}{8} \log
\left(1-\frac{t_1}{2}\right)+\frac{9 {t_3}}{128
   (2-{t_1})^3}+\frac{225 {t_5}}{1024 (2-{t_1})^5}+\frac{55125 {t_7}}{32768 (2-{t_1})^7} \nn\\
   &\quad+\frac{567 {t_3}^2}{1024 (2-{t_1})^6}+\frac{388125 {t_3}{t_5}}{32768 (2-{t_1})^8}+\dots \nn\\
=&\frac{{t_1}}{16}+\frac{{t_1}^2}{64}+\frac{{t_1}^3}{192}+\frac{9
{t_3}}{1024}+\frac{27 {t_3} {t_1}}{2048}+\frac{27 {t_3}
   {t_1}^2}{2048}+\frac{45 {t_3} {t_1}^3}{4096}+\frac{567 {t_3}^2}{65536} +\frac{1701 {t_3}^2 {t_1}}{65536}\nn\\
   &\quad+\frac{11907 {t_3}^2 {t_1}^2}{262144}+ \frac{3969 {t_3}^2 {t_1}^3}{65536}+\frac{225 {t_5}}{32768}+\frac{1125 {t_5} {t_1}}{65536}+\frac{3375 {t_5} {t_1}^2}{131072}+\frac{7875 {t_5} {t_1}^3}{262144}\nn\\
   &\quad+\frac{388125 {t_3}
   {t_5}}{8388608}+\frac{388125
   {t_3} {t_5} {t_1}}{2097152}+\frac{3493125 {t_3} {t_5} {t_1}^2}{8388608}+\frac{5821875 {t_3} {t_5}
   {t_1}^3}{8388608}+\frac{55125 {t_7}}{4194304}\nn\\
   &\quad+\frac{385875 {t_7}
   {t_1}}{8388608}+\frac{385875 {t_7}
   {t_1}^2}{4194304}+\frac{1157625 {t_7}
   {t_1}^3}{8388608}+\dots. \nn
\end{align}
The tau function $\tau^{\rs^0}$ is the so-called Brezin-Gross-Witten tau
function of the KdV hierarchy \cite{BG,GW}, and it gives a
generating function for certain intersection numbers on the moduli
space of stable curves \cite{Nor}.
\end{exa}

\begin{exa}
Let $\fg$ be of type $A_2^{(2)}$ and $\rs=\rs^0$. In this case,
we have $C^{\rs^0}=1/36$ and $\ell=1$. The unique tau function determined by \eqref{sim2} is given by
\begin{align}
\log\tau^{\rs^0}=&-\frac{1}{6} \log \left(1-\frac{1}{6} {t_1}\right)
-\frac{91{t_5}}{648 (6-t_1)^5} -\frac{2821{t_7}}{3888
(6-t_1)^7}+\frac{54145
   {t_5}^2}{1728 (6-t_1)^{10}}+\dots \nn \\
=& \frac{1}{36}t_1+\frac{{t_1}^2}{432}+\frac{{t_1}^3}{3888}
-\frac{91 }{5038848}){t_5}-\frac{455 {t_1}
{t_5}}{30233088}-\frac{455 {t_1}^2 {t_5} }{60466176}-\frac{3185
{t_1}^3 {t_5}
   }{1088391168} \nn\\
&\quad -\frac{2821}{1088391168}{t_7}-\frac{19747 {t_1}
{t_7}}{6530347008}-\frac{19747 {t_1}^2 {t_7}
}{9795520512}-\frac{19747 {t_1}^3 {t_7} }{19591041024}
\nn\\
&\quad+\frac{54145   {t_5}^2}{104485552128}+\frac{270725 {t_1}{t_5}^2
}{313456656384}+\frac{2977975 {t_1}^2 {t_5}^2
   }{3761479876608}+\frac{2977975 {t_1}^3  {t_5}^2
   }{5642219814912}+\dots. \nn
\end{align}
\end{exa}

\begin{rmk}
Suppose that the flows \eqref{logtaubeta} are replaced by
\begin{equation}\label{logtaubt0C}
\frac{\pd\log\tau^\rs}{\pd\beta_k} = \frac{1}{h^\rs}\left(d^\rs\mid
B_k\right)+\dt_{k 0}C
\end{equation}
with an arbitrary constant $C$, then together with
\eqref{ombeta}--\eqref{Lbt} they still give a series of symmetries of the tau cover \eqref{tau-cover} of the Drinfeld-Sokolov hierarchy. It is easy to see that the constant $C$ does not
change the communication relations \eqref{taubtkl} for $k, l\ge0$,
and that the similarity equation \eqref{sim0} also induces the
Virasoro constraints of the form \eqref{simVir}. Under this setting, the conclusion
of Proposition~\ref{thm-simred} should be modified as follows: the
solution $\log\tau^\rs$ of the Drinfeld-Sokolov hierarchy
\eqref{LVt} satisfying the similarity equation $\pd\log\tau^\rs/\pd\beta_0=\pd\log\tau^\rs/\pd t_1$ is characterized by $\ell$ parameters
\[
W_{m_i}:=\om_{m_i}|_{\bt=0}, \quad i=1, 2, \dots, \ell.
\]
In particular, one sees $W_1=C^\rs+C$. As an example, such a tau
function of the KdV hierarchy that depends on one parameter was derived by Alexandrov,
Bertola and Ruzza \cite{Al,BR} (cf. \cite{DYZ}), and they showed that this tau function
satisfies the Virasoro constraints \eqref{simVir}.
\end{rmk}

\section{From Drinfeld-Sokolov hierarchies to equations of Painlev\'e type}

We recall that in Proposition~\ref{thm-simred} the similarity equation \eqref{sim} with parameters $b_p=\delta_{p1}$ leads to a system of linear ODEs \eqref{prfBGW}, which can be solved whenever the initial values \eqref{ommi2} are given. In what follows, we will choose $b_p$ in other ways, say $b_p=\delta_{p j}$ for an exponent $j\in J_{>1}$, then the above mentioned ODEs are nonlinear and of Painlev\'e type. We show in this section that there exist affine Weyl group actions on the solution spaces of these Painlev\'e type ODEs. In the particular cases when the affine Kac-Moody algebra $\fg$ is of type $A_\ell^{(1)}$, $C_\ell^{(1)}$ and $D_{2n+2}^{(1)}$, such kind of affine Weyl group actions were given in \cite{FS-A,FS-D,Na,NY-A}.

\subsection{Similarity reductions of Drinfeld-Sokolov hierarchies}

Let us study solutions of the Drinfeld-Sokolov hierarchies that are constrained by the similarity equations.
\begin{thm}\label{thm-Lax}
Given a solution of the tau cover of the Drinfeld-Sokolov hierarchy associated to $(\fg,\rs,\one)$ that satisfies the similarity equation \eqref{sim}, then the following equation holds true:
\begin{equation}\label{simred}
\left[\frac{\pd}{\pd x}+L, r h^\rs d_0^\rs+M\right]=0,
\end{equation}
where
\begin{align}\label{simL}
&L=\left.(\Ld_1+Q^\mV)\right|_{t_p=x\dt_{p 1}},\\
&M={r
h^\rs}\left.\Biggl(\sum_{p\in J_+}\left(b_p-\frac{\dt_{p 1}}{r
h}x\right)
\left(e^{\ad_{U(Q^\mV)}}\Ld_p\right)_{\ge0} +e^{\ad_N}(\rho^\rs-\rho^\mathds{1})\Biggr)\right|_{t_p=x\dt_{p
1}}  \label{simM}
\end{align}
with $Q^\mV$ and $N$ given in \eqref{sLV} and \eqref{LNLV}.
\end{thm}
\begin{prf}
Recall that the similarity equation \eqref{sim} leads to
\eqref{str2}, from which it follows that
\begin{align*}
&\Biggl[ e^{\ad_V}\left(\frac{\p}{\p t_1}+\Ld_1\right),
\Biggl(e^{\ad_V}e^{-\ad_\Xi}d^\mathds{1}_{0}-d_{0}^\rs+\sum_{p\in
J_+}b_p e^{\ad_V}\Ld_p\Biggr)_{\ge0}+d_{0}^\rs \Biggr] \\
=&\Biggl[ e^{\ad_V}\left(\frac{\p}{\p t_1}+\Ld_1\right),
e^{\ad_V}e^{-\ad_\Xi}d^\mathds{1}_{0} +\sum_{p\in
J_+}b_p e^{\ad_V}\Ld_p  \Biggr] \\
=&e^{\ad_V}e^{-\ad_\Xi}\Biggl[e^{\ad_\Xi}\left( \frac{\p}{\p t_1}+\Ld_1\right),
d^\mathds{1}_{0}+\sum_{p\in
J_+}b_p e^{\ad_\Xi}\Ld_p \Biggr] \\
=&e^{\ad_V}e^{-\ad_\Xi}\Biggl[ \frac{\p}{\p t_1},
d^\mathds{1}_{0}+\sum_{p\in J_+}b_p\Ld_p \Biggr] =0.
\end{align*}
With the help of \eqref{dd},  \eqref{L3} and \eqref{dLd}, we can rewrite the above equation as
\begin{equation}\label{simredV}
\Biggl[\frac{\p}{\p t_1}+\Ld_1+Q, \sum_{p\in J_+}\left(b_p-\frac{p}{r
h}t_p\right)
\left(e^{\ad_V}\Ld_p\right)_{\ge0}+\rho^\rs-\rho^\mathds{1}+d_{0}^\rs\Biggr]=0.
\end{equation}
On the other hand, by using \eqref{sLV2} and \eqref{sim}
we have
\begin{align*}
&e^{\ad_N}\left( \frac{\p}{\p t_1}+\Ld_1+Q\right)=\frac{\p}{\p t_1}+\Ld_1+Q^\mathcal{V}, \nn \\
&\sum_{p\in J_+}\left(b_p-\frac{p}{r h}t_p\right)
e^{\ad_N}\left(e^{\ad_V}\Ld_p\right)_{\ge0}\\
=&\sum_{p\in J_+}\left(b_p
-\frac{p}{r h}t_p\right) e^{\ad_N}\left(\left(e^{\ad_V}\Ld_p\right)_{\ge0}+\om_p\cdot
c\right)-C^\rs c \nn\\
=&\sum_{p\in J_+}\left(b_p -\frac{p}{r h}t_p\right)
\left(e^{\ad_N}e^{\ad_V}e^{-\ad_\Om}\Ld_p\right)_{\ge0}-C^\rs c \nn\\
=&\sum_{p\in J_+}\left(b_p -\frac{p}{r h}t_p\right)
\left(e^{\ad_{U(Q^\mathcal{V})} }\Ld_p\right)_{\ge0}-C^\rs c, \nn
\end{align*}
where the last equality is due to \eqref{UHLV}, so by using \eqref{simredV} we arrive at \eqref{simred}. The theorem is proved.
\end{prf}

From any solution $(L, M)$ of equation \eqref{simred} we can obtain a solution of the Drinfeld-Sokolov hierarchy. In fact, from $L$ we obtain the initial data $\mu_i(x)=\left.u_i\right|_{t_p=x\dt_{p 1}}$ with $i=1,2,\dots,\ell$, then we can solve $w_j(x)=\left.\om_j\right|_{t_p=x\dt_{p 1}}$ for all $j\in J_+$ from the following equations:
\begin{align*}
&\frac{j}{r
h}w_j(x) + \left.\Biggl(\frac{x}{r h}\Om_{1 j}^\rs -  \sum_{p\in J_+}b_p\Om_{j p}^\rs\Biggr)\right|_{u_i\mapsto \mu_i(x)}=0, \quad j\in J_{>1},
\end{align*}
which are  derived from \eqref{sim} by taking the derivative with respect to $t_j$ and by letting  $t_m=x \dt_{m 1}$. Then the solution $\log\tau^\rs$ of the Drinfeld-Sokolov hierarchy is determined via Proposition~\ref{thm-cauchy}.

We call the equation \eqref{simred} a similarity reduction of the Drinfeld-Sokolov hierarchy. It is a system of ODEs of the unknown functions
$u_i|_{t_p=x\dt_{p 1}}$ and $\om_j|_{t_p=x\dt_{p 1}}$. If we take a matrix realization of
$\mathcal{G}$ as in \cite{DS,Kac} and identify $\fg$ with its realization \eqref{gAs}, then equation \eqref{simred} is just the compatibility condition of the following Lax pair of an unknown vector function $\Psi=\Psi(x;z)$:
\begin{equation}\label{Laxpair}
z\frac{\pd\Psi}{\pd z}=M\Psi, \quad \frac{\pd\Psi}{\pd x}=-L\Psi.
\end{equation}

\subsection{Equations of Painlev\'e type}
Let us give some examples of the similarity reduction \eqref{simred} with
$b_p=\dt_{p k}$ for some $k\in J_{>1}$.

\begin{exa}\label{exa-A1sim3}
Let $\fg$ be of type $A_1^{(1)}$, with gradations $\rs\le\one$. We consider the similarity equation \eqref{sim} with
$b_p=\dt_{p 3}$, that is,
\begin{equation}\label{A1sim3}
\sum_{p\in\Zop}\left(\frac{p}{2}t_p-\dt_{p 3}\right)\frac{\pd\log\tau^\rs}{\pd
t_p}+C^\rs=0.
\end{equation}
By taking its second order derivative with respect to
$t_1$ we arrive at the equation
\begin{equation}\label{ODEA1sim3}
\left.\left(-\frac{\p\Om^\rs_{ 1 3}}{\p t_1}+\frac{1}{2}t_1 \frac{\p\Om^\rs_{1 1
}}{\p t_1}+\Om^\rs_{1 1}\right)\right|_{t_p=x \dt_{p 1} }=0.
\end{equation}
We want to write down this equation more explicitly for the cases $\rs=\one$ and $\rs=\rs^0$.

\noindent
(i)\, When $\rs=\one$, let us follow the notations used in  Example~\ref{exa-A1mKdV} and denote $\ld(x):=v|_{t_p=x\dt_{p 1}}$. By using \eqref{om13A1p} we can represent the equation \eqref{ODEA1sim3} in the form
\[
-\left( \dfrac{3 \ld^4}{8}-\dfrac{\ld \ld''
}{4}+\dfrac{{(\ld')}^2}{8}\right)'-\frac{1}{2}x\left(\frac{\ld^2}{2}\right)'-\frac{\ld^2}{2}=0.
\]
It leads to the second Painlev\'{e} equation (P2)
\begin{equation}\label{P2}
\ld''=2 \ld^3+2 x \ld+\mathrm{const}.
\end{equation}
The formal power series solution of this equation has the form
\[
\ld(x)=a + b x+ c x^2+ p_3(a,b,c)x^3+p_4(a,b,c)x^4+\dots,
\]
where $a$, $b$ and $c$ are arbitrary parameters, and $p_i$ are certain
polynomials of these parameters. If a matrix realization of
$\fg$ is taken as in \cite{DS}, then we have the Lax equation for P2 given by the similarity reduction \eqref{simred} with $r h^\rs d_0^\rs$ replaced by $-z\frac{\p}{\p z}$ and
\[L=\left(
\begin{array}{cc}
 -\ld & z \\
 z & \ld
\end{array}
\right),\
M=\left(
\begin{array}{cc}
  -2 z^2 \ld +\ld^3+ x \ld-\frac{1}{2}\ld'' & 2 z^3-z(\ld^2+x+\ld') \\
 2 z^3-z(\ld^2+x-\ld') & 2 z^2 \ld- \ld^3- x \ld+\frac{1}{2}\ld''
\end{array}
\right).
\]

\noindent
(ii)\, When $\rs=\rs^0$, let us follow the notations used in Example~\ref{exa-A1KdV} and denote $\mu(x):=u|_{t_p=x\dt_{p 1}}$. By using \eqref{om13} we can rewrite the equation \eqref{ODEA1sim3} in the form
\[
-\left(\frac{3}{8}\mu^2+\frac{1}{8}\mu''\right)'+\frac{1}{2} x
\frac{\mu'}{2}+\frac{\mu}{2}=0,
\]
which is just the Painlev\'{e} equation (P34; see, e.g. \cite{BR})
\begin{equation}\label{P34}
\mu'''=-6 \mu \mu'+ 2 x \mu' +4 \mu.
\end{equation}
Its formal power series solution can be represented as
\[
\mu(x)=a + b x+ c x^2+ q_3(a,b,c)x^3+q_4(a,b,c)x^4+\dots
\]
with arbitrary parameters $a$, $b$ and $c$,  and $q_i$ are certain polynomials of these parameters. It leads to solutions of the KdV hierarchy
that satisfies the similarity equation. In particular, if we take
$(a,b,c)=(0,1,0)$, then we obtain the solution of the KdV hierarchy corresponding to the Witten-Kontsevich
tau function  given by \eqref{KdVtau1} with $W_j=\dt_{j 3}/16$.

The Lax equation for the Painlev\'e equation P34 is given by \eqref{simred}
with
\[
L=\left(
\begin{array}{cc}
 0 & z-\mu \\
 1 & 0
\end{array}
\right),\ M=\left(
\begin{array}{cc}
 -\frac{1}{2 }(\mu'-1) & 2 z^2- z(\mu+ x)-\frac{1}{2 }(2 \mu^2-2 x \mu+\mu''+4 w_1) \\
 2 z+\mu-x & \frac{1}{2}(\mu'-1)
\end{array}
\right).
\]
Here in the matrix $M$ the function $w_1$ satisfies $w'_1(x)=\frac12 \mu(x)$. We observe that when $\tilde{\mu}(x):=\mu(x)-x\ne0$, the equation \eqref{P34} can be rewritten as
\begin{equation}\label{ODEKdV2}
\tilde{\mu}''=\frac{ {(\tilde{\mu}')}^2}{2 \tilde{\mu}}-2 \tilde{\mu}^2- 2 x
\tilde{\mu}+\mathrm{const}\cdot\frac{1}{\tilde{\mu}}
\end{equation}
which is the Painlev\'{e} equation P4$'$ given in Appendix B of \cite{CM}. It is easy to see that the equation \eqref{P34} is related to the second Painlev\'{e} equation \eqref{P2} via the Miura transformation $\mu=-\ld^2+\ld'$.
\end{exa}

\begin{exa}\label{exa-A2ODE}
Let $\fg$ be of type $A_2^{(1)}$ and $\rs=\rs^0$. We follow the notations used in Example~\ref{exa-A2tau} and take $b_p=\delta_{p2}$. Then the  similarity equation \eqref{sim} has the expression
\begin{equation}
\sum_{p\in
J_+}\left(\frac{p}{3}t_p-\dt_{p 2}\right)\frac{\pd\log\tau^{\rs^0}}{\pd
t_p}+C^{\rs^0}=0.
\end{equation}
and the similarity reduction \eqref{simred} is given by
\begin{align*}
L=\left(
\begin{array}{ccc}
 0 & 0 & z-\ld \\
 1 & 0 & -\mu \\
 0 & 1 & 0 \\
\end{array}
\right),
\quad M=& \left(
\begin{array}{ccc}
 2 \mu+1 & M_{1 2}  & M_{1 3} \\
 -x & -\mu & -\ld+\frac{2 }{3}x \mu+3 z-{w_1} \\
 3 & -x & -\mu-1 \\
\end{array}
\right),
\end{align*}
where
\[\mu(x)=u|_{t_p=x \dt_{p 1}},\quad \ld(x)=v|_{t_p=x \dt_{p 1}},\]
and
\[
M_{1 2}=\mu'-\frac{1}{3}x \mu-{w_1}-\ld+3 z, \quad M_{1 3}= \frac{1}{3}\mu''-\frac{1}{3}x \mu'-
   \mu-\frac{1}{3} \mu^2+\frac{2}{3}x \ld- {w_2}- x z.
\]
The similarity reduction can be represented as the following system of ODEs:
\begin{equation}\label{fgode}
\frac{\mu^{(4)}}{4}+\frac{x^2 \mu''}{12}+\mu \mu''+\frac{7 x
   \mu'}{12}+\left(\mu'\right)^2+\frac{2 \mu}{3}=0, \quad \ld'=\frac{1}{6} \left(2 \mu+3 \mu''+x \mu' \right).
\end{equation}
Via the following replacements of variables:
\[
x \mapsto \bigg(\frac{3}{4}\bigg)^{\frac{1}{4}}z, \quad \mu(x)\mapsto \frac{1}{2\sqrt{3}}g(z),
\]
the first equation in \eqref{fgode} is recast to the equation (2.9) given in \cite{CK}, which can be solved in terms of solutions of the Painlev\'{e} equation P4.
\end{exa}

\begin{exa} \label{exa-A22Vir}
Let $\fg$ be of type $A_2^{(2)}$ and $\rs=\rs^0=(1,0)$. Let us take $b_p=\delta_{p7}$, then the similarity equation \eqref{sim} has the expression
\begin{equation}\label{V0A22}
\sum_{p\in
J_+}\left(\frac{p}{6}t_p-\dt_{p 7}\right)\frac{\pd\log\tau^{\rs^0}}{\pd
t_p}+C^{\rs^0}=0.
\end{equation}
Following the notations of
Example~\ref{exa-A22tau} we have $\Om_{1 1}^\rs=u/3$ and
\begin{align}
&\Om_{1 7}^\rs=-\frac{1}{1944}\left(56 u^4+84 u
\left(u'\right)^2+168 u^2 u''+42 \left(u''\right)^2+42 u' u''' +42 u
u^{(4)} +3 u^{(6)}\right).
\end{align}
Denote $\mu(x)=u|_{t_j=\dt_{j 1}x}$, then the similarity reduction of the Drinfeld-Sokolov hierarchy can be rewritten as the following nonlinear ODE:
\[
\left(56 \mu^4+84 \mu \left(\mu'\right)^2+168 \mu^2 \mu''+42
\left(\mu''\right)^2+42 \mu' \mu''' +42 \mu \mu^{(4)} +3 \mu^{(6)}\right)'+108 x
 \mu' +216 \mu=0.
\]
This ODE can be verified to pass the Painlev\'{e}
test (see, e.g., Chapter~2 of \cite{CM}). More precisely, the leading behavior of its solution at movable singularities is
$-3(x-x_0)^{-2}$, and the Fuchs indices are $j=-1, 2, 3, 4, 7, 8,
12$ that correspond to seven arbitrary parameters.
We also know that the formal power series solution of this ODE has the form
\[
\mu(x)=\sum_{i=0}^6 a_i x^i+\sum_{i\ge7}p_i(a_0,a_1,\dots,a_6)x^i
\]
with arbitrary parameters $a_0$, $a_1$, $\dots$, $a_6$ and certain
polynomials $p_i$ of these parameters.
\end{exa}

\subsection{Affine Weyl group actions}

In this subsection we consider the similarity reduction \eqref{simred} with $\rs=\one$ (hence $\mV=\mathcal{B}\subset\fg^0$), i.e.
\begin{equation}\label{LaxLM}
\left[\frac{\pd}{\pd x}+\Ld_1+Q, -d+M\right]=0.
\end{equation}
Here we denote $Q|_{t_p=x \dt_{p 1}}$ by $Q$, and we use the
fact that $-d=-d^\one=r h d_0^\one$.
The functions $Q$ and $M$ now take values in $\mathcal{B}$ and $\fg^{\ge0}$ respectively,
and the similarity reduction \eqref{LaxLM} gives a system of ODEs of the unknown functions $u_i|_{t_p=\dt_{p 1}x}$ with $i=1, 2,\dots, \ell$.
Motivated by a series of work of Noumi, Yamada et al (see \cite{NY} and references therein), let us study the affine Weyl group actions on the space of solutions of these ODEs. For this purpose, we introduce the following scalar functions (see the notations in Subsection~\ref{sec-g}):
\begin{equation}\label{taphi}
\ta_i=\frac{k_i^\vee}{k_i}(\al_i^\vee\mid Q), \quad
\chi_i=\frac{k_i}{k_i^\vee}-(\al_i^\vee\mid M), \quad
\vp_i=\frac{1}{\nu}(f_i\mid M),\quad
i=0,1,\dots,\ell.
\end{equation}
Then we have
\begin{equation}\label{tachi}
\sum_{i=0}^\ell k_i\ta_i=0, \quad \sum_{i=0}^\ell k_i^\vee \chi_i=h.
\end{equation}
Note that
$\boldsymbol{\ta}:=(\ta_1,\ta_2,\dots,\ta_\ell)$ gives a coordinate system
for the space $\mV=\mathcal{B}$, and the functions $\chi_i$ and
$\vp_i$ are elements of the ring $\C\left[x, \boldsymbol{\ta},
\boldsymbol{\ta}', \boldsymbol{\ta}'', \dots\right]$.

{\begin{thm}\label{thm-weylaction}
The following assertions hold true:
\begin{itemize}
\item[\rm (i)] The equation \eqref{LaxLM} implies that $\chi_i$ are constant functions.
\item[\rm (ii)] For a fixed set of constants $\left\{\chi_0, \dots, \chi_\ell\mid \sum_{i=0}^\ell k_i^\vee\chi_i=h\right\}$, the equation \eqref{LaxLM} can be represented as a system of ODEs of the unknown functions $\theta_1, \dots, \theta_\ell$ in the form
\begin{equation}\label{vpta}
\vp_i'+\ta_i\vp_i+\chi_i=0, \quad i=0,1,2,\dots,\ell.
\end{equation}
\item[\rm (iii)] Denote $\xi_j=\chi_j/(\nu \vp_j)$ for $j=0,1,\dots,\ell$.
Then the equation \eqref{LaxLM} has the B\"acklund
transformations
\begin{equation}\label{Rj}
\sR_j: (Q, M) \mapsto (\tilde{Q}, \tilde{M}), \quad j=0,1,\dots,\ell
\end{equation}
defined by the following relations:
\begin{align}
\frac{\p}{\p
x}+\tilde{Q}+\Ld_1&=e^{-\xi_j\ad_{f_j}}\left(\frac{\p}{\p
x}+Q+\Ld_1\right)-\frac{\nu k_j}{h k_j^\vee}\xi_j c, \label{Qtilde} \\
-d+\tilde{M}&=e^{-\xi_j\ad_{f_j}}\left(-d+M\right)-\frac{ \chi_j}{h}
 c.\label{Mtilde}
\end{align}
More explicitly, for $i, j=0,1,2,\dots,\ell$, we have
\begin{equation}\label{}
\sR_j(\chi_i)=\chi_i-a_{i j}\chi_j, \quad \sR_j(\ta_i)=\ta_i+a_{j
i}\frac{\chi_j}{\vp_j}, \quad
\sR_j(\vp_i)=\frac{1}{\nu}\left(e^{\xi_j\ad_{f_j}}f_i\mid M\right).
\end{equation}
Here $A=(a_{i j})_{0\le i, j\le \ell}$ is the Cartan matrix of $\fg$.
\end{itemize}
\end{thm}}
\begin{prf}
We expand $M=\sum_{k=0}^m M_k$ with $M_k$ taking value in $\fg^k$, and restrict the equation \eqref{LaxLM} to each component of the decomposition $\fg=\bigoplus_{k\in\Z}\fg^k$.

Firstly, the $\fg^0$-component of the equation \eqref{LaxLM} can be represented as
\begin{equation}\label{M0eq}
M_0'+[Q,M_0]+[Q, -d]=0.
\end{equation}
Since $Q$ and $M_0$ take value in $\fg^0\subset\mathfrak{h}$, we have $[Q,M_0]=[Q,-d]=0$. Hence  $M_0'=0$, and we obtain the first assertion of the theorem from the definition \eqref{taphi}.

In order to prove the second assertion of the theorem, we only need to substitute the constants $\chi_i$ into the $\fg^1$-component of the equation \eqref{LaxLM}, namely
\begin{align}
M_1'+[Q,M_1]+[\Ld_1,M_0]+[\Ld_1,-d]=0.
\end{align}
Note that $(e_i\mid f_j)=\dt_{i j}k_i/k_i^\vee$,
hence the above equation is equivalent to the following ones:
\begin{align}\label{}
\frac{1}{\nu}(f_i\mid M_1')
=- \frac{1}{\nu}(f_i\mid [Q,M_1])- \frac{1}{\nu}(f_i\mid [\Ld_1,M_0])- \frac{1}{\nu}(f_i\mid [\Ld_1,-d]), \quad i=0,1,\dots, \ell. \label{MLdeg1}
\end{align}
The left hand side is just $\vp_i'$ where $\vp_i$ is defined in \eqref{taphi}.
On the other hand, by using the definition of $\vp_i$ again,
we have
\begin{align}\label{M1}
M_1=\sum_{i=0}^\ell\frac{\nu k_i^\vee}{ k_i}\vp_i e_i,
\end{align}
so the right hand side of \eqref{MLdeg1} can be written as
\begin{align*}
\mathrm{r.h.s.}
  &=-\frac{1}{\nu}([M_1,f_i]\mid Q )+\frac{1}{\nu}([\Ld_1,f_i]\mid M_0 )- \frac{1}{\nu}(f_i\mid \Ld_1) \\
  &=-\frac{ k_i^\vee}{ k_i}\vp_i(\al_i^\vee\mid Q)+ (\al_i^\vee\mid M_0)- \frac{k_i}{k_i^\vee}\\
  &=- \ta_i\vp_i-\chi_i.
\end{align*}
Thus the second assertion is proved.

From \eqref{Qtilde} and \eqref{Mtilde} it is easy to see
\[
\left[\frac{\p}{\p x}+\tilde{Q}+\Ld_1, -d+\tilde{M}\right]=0.
\]
To prove the third assertion of the theorem, we need to show that $\tilde{Q}$ takes value in $\mathcal{B}$ and that $\tilde{M}$ has an expression as \eqref{simM}. Firstly, note that the action $\sR_j$ yields
\begin{align}
 \tilde{Q} &=\xi_j' f_j+Q+[Q,\xi_j f_j]+[\Ld_1, \xi_j f_j]+\frac{1}{2}[[\Ld_1, \xi_j f_j],\xi_j f_j]-\frac{\nu k_j}{h k_j^\vee}\xi_j c \nn\\
 &= Q+\nu \xi_j\left( \al_j^\vee-\frac{ k_j}{h k_j^\vee} c\right)+\varrho_j f_j, \label{Qtilde1}
\end{align}
where
\begin{align*}
\varrho_j&=\frac{k_j^\vee}{ k_j}\left( e_j\mid \xi_j' f_j+[Q,\xi_j f_j]+\frac{1}{2}[[\Ld_1, \xi_j f_j],\xi_j f_j]\right) \\
&=\xi_j'- \frac{k_j^\vee}{ k_j}\xi_j\left([e_j, f_j] \mid Q\right)-\frac{1}{2}\xi_j^2\left([e_j, f_j] \mid[\Ld_1, f_j]\right)\frac{k_j^\vee}{ k_j} \\
&=\xi_j'- \frac{k_j^\vee}{ k_j}\xi_j\left(\al_j^\vee\mid Q\right)-\frac{1}{2}\xi_j^2\left(\al_j^\vee \mid \nu \al_j^\vee\right)\frac{k_j^\vee}{ k_j} \\
&=\xi_j'-\ta_j \xi_j-\nu\xi_j^2\\
&=-\nu\xi_j^2\left( \left(\frac{1}{\nu\xi_j}\right)'+\ta_j\frac{1}{\nu\xi_j}+1\right) \\
&=-\nu\xi_j^2\left( \frac{\vp_j'}{\chi_j}+\ta_j \frac{\vp_j}{\chi_j}+1  \right).
\end{align*}
From \eqref{vpta} we obtain $\varrho_j=0$. Moreover, by using \eqref{blf} we have
\[
(d\mid \tilde{Q})=\left(d\mid Q +\nu \xi_j\bigg( \al_j^\vee-\frac{ k_j}{h k_j^\vee} c\bigg)\right)=(d\mid Q)+0=0.
\]
Hence $\tilde{Q}$ is a function taking values in $\mathcal{B}$.

Secondly, we proceed to show that the function $\tilde{M}$ can be represented in the form \eqref{simM}. To this end, we define
\[
K_j=\sum_{k\in J_+ }\frac{1}{k}\zeta_{j k}\Ld_{-k},\quad j=0,1,\dots,\ell,
\]
where
\begin{equation}\label{zetajk}
\zeta_{j k}=\frac{\xi_j}{h}\left(f_j\mid e^{\ad_{U(Q)}}\Ld_k\right), \quad k\in J_+
\end{equation}
with $U(Q)$ being the $\fg^{<0}$-valued function defined by Lemma~\ref{thm-dr} for the operator $\p/\p x+\Ld_1+Q$.
Since the function $K_j$ takes value in $\mathcal{H}\cap\fg^{<0}$, a $\fg^{<0}$-valued function $X_j$ can be defined by
\begin{equation}
e^{\ad_{X_j}}=e^{-\xi_j\ad_{ f_j}}e^{\ad_{U(Q)}}e^{\ad_{K_j}}.
\end{equation}
By using the fact that
\[
\zeta_{j 1}=\frac{\xi_j}{h}\left(f_j\mid e^{\ad_{U(Q)}}\Ld_1\right)=\frac{\xi_j}{h}\left(f_j\mid\Ld_1\right)=\frac{\nu k_j}{h k_j^\vee}\xi_j,
\]
we can rewrite \eqref{Qtilde} as follows:
\begin{align*}
\frac{\p}{\p
x}+\tilde{Q}+\Ld_1=&e^{-\xi_j\ad_{f_j}}e^{\ad_{U(Q)}}\left(\frac{\p}{\p
x}+\Ld_1+H(Q)\right)-\zeta_{j 1}c \\
=&e^{-\xi_j\ad_{f_j}}e^{\ad_{U(Q)}}e^{\ad_{K_j}}\left(\frac{\p}{\p
x}+\Ld_1+H(Q)+K_j'\right)\\
=&e^{\ad_{X_j}}\left(\frac{\p}{\p
x}+\Ld_1+\tilde{H}\right).
\end{align*}
Here $H(Q)$ is given by Lemma~\ref{thm-dr}, and the function $\tilde{H}=H(Q)+K_j'$ takes values in $\mathcal{H}\cap\fg^{<0}$. Moreover, for any $k\in J_+ $, we have
\begin{align*}
\left(d\mid e^{\ad_{X_j}}\Ld_k\right)=& \left(e^{\xi_j\ad_{f_j}} d\mid e^{\ad_{U(Q)}}e^{\ad{K_j}}\Ld_k \right) \\
=& \left(d+\xi_j f_j\mid e^{\ad_{U(Q)}}\Ld_k-\zeta_{j k}c\right) \\
=& \left(d \mid e^{\ad_{U(Q)}}\Ld_k \right) -\zeta_{j k} \left(d \mid c\right)+\xi_j\left( f_j\mid e^{\ad_{U(Q)}}\Ld_k\right) \\
=&0- \zeta_{j k}h +h \zeta_{j k} =0.
\end{align*}
Thus we arrive at $X_j=U(\tilde{Q})$.
On the other hand, since
\begin{align}
  \ad_{f_j} [e_k, e_l]&=-\dt_{j k}[\al_j^\vee, e_l]-\dt_{j l}[e_k, \al_j^\vee]=-\dt_{j k}a_{j l}e_l+\dt_{j l}a_{j k}e_k, \label{adfee} \\
 (\ad_{f_j})^2 [e_k, e_l]&=\dt_{j k}a_{j l}\dt_{j l}\al_j^\vee-\dt_{j l}a_{j k}\dt_{j k}\al_j^\vee=0,  \label{adffee}
\end{align}
the restriction of the function $\tilde{M}$
to $\fg^{<0}$ reads
\begin{align}\label{Mm1tilde}
\tilde{M}^{<0}&=[-d, \xi_j f_j]+[M_0, \xi_j f_j]+\frac{1}{2}[[M_1,
\xi_j f_j], \xi_j f_j],
\end{align}
which in fact takes values in $\fg^{-1}$. So we have, for any $i=0,1,2,\dots,\ell$,
\begin{align*}
(e_i\mid \tilde{M}^{<0}) &= \xi_j([d,e_i]\mid f_j)-\xi_j( [e_i, f_j]\mid M_0)+\frac{1}{2}\xi_j^2([ [e_i, f_j],f_j]\mid M_1) \\
&=\dt_{i j}\xi_j\left(\frac{k_i}{k_i^\vee} - ( \al_j^\vee\mid M_0)+\frac{1}{2}\xi_j(-2 f_j\mid M_1)\right) \\
&=\dt_{i j}\xi_j\left( \frac{k_i}{k_i^\vee}
+ (\chi_j-\frac{k_i}{k_i^\vee} )-\xi_j \nu \vp_j\right)
\\
&=\dt_{i j}\xi_j\left( \chi_j  -\xi_j  \frac{\chi_j}{\xi_j}\right)=0,
\end{align*}
which implies that $\tilde{M}^{<0}$ vanishes.
Hence, from the definitions of $\tilde{M}$ and $M$ it follows that
\begin{align}
\tilde{M}=&\left(d+e^{-\xi_j\ad_{f_j}}(-d+M)\right)_{\ge0}-\frac{ \chi_j}{h}c \nn\\
=&\left(e^{-\xi_j\ad_{f_j}}M\right)_{\ge0}-\frac{ \chi_j}{h}c \nn\\
=& rh\sum_{p\in J_+ }\left(b_p-\frac{\dt_{p 1}}{r h}x\right)\left(e^{-\xi_j\ad_{f_j}}e^{\ad_{U(Q)}}\Ld_p\right)_{\ge0}-\frac{ \chi_j}{h}c \nn\\
=& rh\sum_{p\in J_+ }\left(b_p-\frac{\dt_{p 1}}{r h}x\right)\left(e^{-\xi_j\ad_{f_j}}e^{\ad_{U(Q)}}e^{\ad_{K_j}}\Ld_p+\zeta_{j p}c\right)_{\ge0}-\frac{ \chi_j}{h}
c \nn\\
=& rh\sum_{p\in J_+ }\left(b_p-\frac{\dt_{p 1}}{r h}x\right)\left(e^{\ad_{U(\tilde{Q})}}\Ld_p\right)_{\ge0}+\ve_j
c,
\end{align}
where the center term $\ve_j
c$ vanishes due to the definitions of $\zeta_{j k}$ and $\xi_j$, namely,
\begin{align*}
\ve_j=&rh\sum_{p\in J_+ }\left(b_p-\frac{\dt_{p 1}}{r h}x\right) \zeta_{j p}-\frac{\chi_j}{h} \\
=&rh\sum_{p\in J_+ }\left(b_p-\frac{\dt_{p 1}}{r h}x\right)\frac{\xi_j}{h}\left(f_j\mid e^{\ad_{U(Q)}}\Ld_p\right) -\frac{\chi_j}{h} \\
=& \frac{\xi_j}{h}(f_j\mid M) -\frac{\chi_j}{h} \\
=& \frac{\chi_j}{h\nu\vp_j}\nu\vp_j -\frac{\chi_j}{h}=0.
\end{align*}
Thus $\sR_j$ is a B\"acklund transformation of the equation
\eqref{LaxLM}.

Finally,  by using \eqref{M1} and \eqref{adffee}, we see that the restriction of $\tilde{M}$ to $\fg^0$ is given by
\begin{align*}
\tilde{M}_{0}&=M_0+[M_1, \xi_j f_j]-\frac{\chi_j}{h} c=M_0+
\frac{k_j^\vee}{k_j}\nu\vp_j\xi_j\al_j^\vee-\frac{\chi_j}{h}\cdot
c=M_0+\chi_j\left(\frac{k_j^\vee}{k_j}\al_j^\vee-\frac{1}{h}\cdot
c\right),
\end{align*}
hence
\begin{equation}
\sR_j(\chi_i)=\frac{k_i}{k_i^\vee}-(\al_i^\vee\mid \tilde{M}_0)=\frac{k_i}{k_i^\vee}-(\al_i^\vee\mid M_0)-\chi_j\frac{k_j^\vee}{k_j}(\al_i^\vee\mid\al_j^\vee) =\chi_i-a_{i j}{\chi_j}.
\end{equation}
By using \eqref{Qtilde1} and \eqref{Mtilde}, it is straight forward to verify
\begin{align}\label{}
 \sR_j(\ta_i)&=\frac{k_i^\vee}{k_i}(\al_i^\vee\mid \tilde{Q}) =\frac{k_i^\vee}{k_i}(\al_i^\vee\mid Q)+\frac{k_i^\vee}{k_i}\nu\xi_j(\al_i^\vee\mid \al_j^\vee)=\ta_i+a_{j i}\frac{\chi_j}{\vp_j}, \\
\sR_j(\vp_i)&=\frac{1}{\nu}(f_i\mid \tilde{M})=\frac{1}{\nu}(f_i\mid
e^{-\xi_j\ad_{f_j}}M)=\frac{1}{\nu}( e^{\xi_j\ad_{f_j}}f_i\mid M).
\end{align}
Therefore the theorem is proved.
\end{prf}

The above theorem shows that the actions of the B\"acklund transformations $\sR_j$ on $\chi_0, \chi_1,\dots, \chi_\ell$
generate an affine Weyl group associated to the Cartan matrix $A=(a_{i j})_{0\le i,j\le\ell}$.
Moreover, by using a general result of \cite{NY}, we have the
following proposition.
\begin{prp}
The actions of the B\"acklund transformations $\sR_j$, with $j=0, 1,\dots,\ell$, on the space of solutions of the equation
\eqref{LaxLM} satisfy the following relations:
\begin{equation}\label{WeylR}
{\sR_j}^2=\mathrm{Id}, \quad (\sR_i \sR_j)^{m_{i j}}=\mathrm{Id}\quad \hbox{for} \quad i\ne j,
\end{equation}
where $m_{i j}=2, 3, 4, 6$ or $\infty$ when $a_{i j}a_{j i}=0, 1,
2, 3$ or $\ge4$ respectively.
\end{prp}
\begin{prf}
For the Cartan matrix $A=(a_{i j})_{0\le i, j\le\ell}$ of the affine Kac-Moody algebra $\fg$, a certain nilpotent Poisson algebra $\mathcal{K}$ was constructed by Noumi and Yamada in \cite{NY}
(in fact, an even more general setting has been considered there,
but here only the case of affine type is concerned). The
Poisson algebra $\mathcal{K}$ is generated by $\phi_i\in\fg^*$
together with a set of parameters $\ld_i$ with $i=0,1,2,\dots,\ell$,
say,
\begin{equation}\label{}
\mathcal{K}=\C(\ld_i, \phi_i, \{\phi_i, \phi_j\}, \{\phi_i,
\{\phi_j, \phi_k\}\}, \dots ).
\end{equation}
The Poisson bracket satisfies $\{\ld_i, \phi_j\}=0$
and the following locally nilpotent conditions
\begin{equation}\label{nPoiss}
(\ad_{\{\,,\,\}\, \phi_j})^{1-a_{i j} }\phi_i=0, \quad i\ne j.
\end{equation}
For any $j=0,1,2,\dots,\ell$, let $\sg_j$ be an automorphism of
$\mathcal{K}$ such that
\begin{equation}\label{}
\sg_j(\ld_i)=\ld_i-a_{i j}\ld_j, \quad \sg_j(\phi_i)=\phi_i.
\end{equation}
Then, on $\mathcal{K}$ there is a class of automorphisms given by
\begin{equation}\label{RjNY}
\sR_j=\exp\left(\frac{\ld_j}{\phi_j}\ad_{\{\,,\,\}\,
\phi_j}\right)\circ\sg_j, \quad j=0,1,\dots,\ell.
\end{equation}
It is shown in \cite{NY} that such automorphisms $\sR_j$
satisfy the relations \eqref{WeylR}, namely, they give a realization
of the affine Weyl group for the Cartan matrix $A=(a_{i j})_{0\le i, j\le\ell}$.

Noumi and Yamada also explained a Lie theoretic background for the
above nilpotent Poisson algebra. More exactly, one can choose (see
\S\,4.1 in \cite{NY})
\begin{equation}\label{}
\phi_i(X)=(f_i\mid X), \quad \{\phi_j, \phi_i\}(X)=-\left([f_j,
f_i]\mid X\right), \qquad X\in\fg,
\end{equation}
such that the nilpotent conditions \eqref{nPoiss} are satisfied due to the Serre relations \eqref{adee}.  In terms of our
notations, if we take
\begin{equation}\label{}
\phi_i\left(\frac{1}{\nu}M\right)=\vp_i, \quad
\ld_i=-\frac{\chi_i}{\nu}, \quad  i,j=0,1,\dots,\ell,
\end{equation}
then the isomorphisms \eqref{RjNY} coincide with those given in
Theorem~\ref{thm-weylaction} (note that the equation \eqref{LaxLM}
can be represented in the variables $\vp_i$ and parameters $\chi_i$ due
to \eqref{vpta}). Thus the relations \eqref{WeylR} for $\sR_j$
defined by \eqref{Rj} are verified, and the proposition is proved.
\end{prf}

\subsection{Examples}
Let us give more details of the system of ODEs \eqref{vpta} and its discrete symmetries for
some examples.

\begin{exa}
Let $\fg$ be of type $A_\ell^{(1)}$ with $\ell\ge2$, then its Cartan matrix is given by
\[
a_{i j}=a_{j i}=\begin{cases}
                 2, & i=j; \\
                 -1, & i-j=\pm1;\\
                 0, & \hbox{ else}.
               \end{cases}
               \]
Here and throughout the present example, the indices $i,j,k\in\Z/(\ell+1)\Z$. Note that the Kac labels and their duals are given by $k_i=k_i^\vee=1$, the Coxeter number is $h=\ell+1$ and the constant in \eqref{Ld1Ld} reads  $\nu=1$.
Let us consider the equation \eqref{LaxLM} induced by
the similarity equation \eqref{sim} with $b_p=\dt_{p 2}$, namely,
\begin{equation}\label{A2M}
M=h\left( e^{\ad_{U(Q)}}\Ld_2\right)_{\ge0}-x \left( e^{\ad_{U(Q)}}\Ld_1\right)_{\ge0}=
M_0+M_1+h \Ld_2
\end{equation}
with $\Ld_2=\sum_{k\in\Z/(\ell+1)\Z} [e_{k+1}, e_k]$ (see, e.g.
\cite{DS}).
According to Theorem~\ref{thm-weylaction}, we have
\begin{align}\label{RjvpA}
\sR_j(\vp_i)&=\left( e^{\xi_j\ad_{f_j}}f_i\mid M_1+h\Ld_2\right)
=(f_i\mid M_1)+h\xi_j\left( [f_j, f_i]\mid \Ld_2\right) =\vp_i+h
b_{i j}\xi_j,
\end{align}
where
\begin{align*}
b_{i j}=&\sum_{k\in\Z/(\ell+1)\Z}\left( [f_j, f_i]\mid [e_{k+1}, e_k]\right) \\
=&\sum_{k\in\Z/(\ell+1)\Z}\left( f_i\mid
[[e_{k+1}, e_k], f_j] \right) \\
=&\sum_{k\in\Z/(\ell+1)\Z}\left( f_i\mid \dt_{j, k+1}a_{j k}e_k-\dt_{j
k}a_{j,k+1}e_{k+1}\right) \\
=&\sum_{k\in\Z/(\ell+1)\Z}(\dt_{i k}\dt_{j, k+1}-\dt_{i, k+1}\dt_{j k})a_{j i}.
\end{align*}
Indeed, the numbers $b_{i j}$ indicate a direction on the Dynkin diagram of type $A_\ell^{(1)}$:
\begin{equation}\label{}
b_{i j} =\begin{cases}
          1, & j=i-1; \\
          -1, & j=i+1; \\
          0, & \hbox{ else}.
        \end{cases}
\end{equation}
By using \eqref{RjvpA} we also have
{\renewcommand\arraystretch{2}
\begin{align}\label{RjxiA}
\sR_j(\xi_i)&=\frac{\chi_i-a_{i j} \chi_j}{\vp_i+h b_{i j} \xi_j}
=\begin{cases}
          -\xi_i, & j=i; \\
          \dfrac{\chi_i+\chi_{i+1}}{\vp_i-h \xi_{i+1}}, & j=i+1; \\
          \dfrac{\chi_{i}+\chi_{i-1}}{\vp_{i}+h \xi_{i-1}}, & j=i-1; \\
          \xi_i, & \hbox{for other cases}. \\
        \end{cases}
\end{align}
}
From \eqref{RjvpA}--\eqref{RjxiA} it follows that
\[
\sR_k\sR_j(\vp_i)=\vp_i+h b_{i k}\xi_k+h b_{i j} \sR_k(\xi_j).
\]
It is straight forward to verify the following assertions:
\begin{itemize}
  \item[$\bullet$] If $k=j$, then $\sR_j(\xi_j)=-\xi_j$, and hence
  ${\sR_j}^2(\vp_i)=\vp_i$;
  \item[$\bullet$] If $a_{j k}=0$, then $\sR_k(\xi_j)=\xi_j$, and hence
  $\sR_k\sR_j(\vp_i)=\sR_j\sR_k(\vp_i)$;
  \item[$\bullet$] If $a_{j k}=-1$, namely $k-j=\pm1$, then
\begin{align}\label{}
\sR_j\sR_{j+1}\sR_j(\vp_i)&=\sR_j(\vp_i+h b_{i,j+1}\xi_{j+1}+h b_{i
j} \sR_{j+1}(\xi_j) ) \nn\\
&=\vp_i+h b_{i j}(\xi_{j}+ \sR_j\sR_{j+1}(\xi_j))+h
b_{i,j+1}\sR_j(\xi_{j+1}) \nn\\
&=\vp_i+h b_{i j}\sR_{j+1}(\xi_{j})+h
b_{i,j+1}\sR_j(\xi_{j+1})\nn\\
&=\sR_{j+1}\sR_j\sR_{j+1}(\vp_i),
\end{align}
in which the third equality holds true since
\begin{align*}
\xi_{j}+\sR_j\sR_{j+1}(\xi_j)&=\xi_{j}+\sR_j\left(\frac{\chi_j+\chi_{j+1}}{\vp_j-h\xi_{j+1}}\right)
=\xi_{j}+\frac{-\chi_j+\chi_{j+1}+\chi_j}{\vp_j-h\frac{\chi_{j+1}+\chi_{j}}{\vp_{j+1}+h\xi_j}}
\\
&=\xi_{j}+\frac{\chi_{j+1}(\vp_{j+1}+h\xi_j)}{\vp_j\vp_{j+1}-h \chi_{j+1}}
=\xi_{j}+\frac{\xi_{j+1}(\vp_{j+1}+h\xi_j)}{\vp_j-h \xi_{j+1}} \\
&=\frac{\xi_j \vp_j+\xi_{j+1}\vp_{j+1} }{\vp_j-h
\xi_{j+1}}=\frac{\chi_j +\chi_{j+1} }{\vp_j-h \xi_{j+1}}=\sR_{j+1}(\xi_j).
\end{align*}
\end{itemize}
Thus we arrive at the relations
\eqref{WeylR} based on the explicit representation
\eqref{RjvpA} of $\sR_j$. The result agrees with the one obtained in
\cite{NY-A} (see also \cite{SHC}), where a matrix realization of
$\fg$ was used.

In the current case the system of ODEs \eqref{vpta} can be  represented in an alternative form as follows. Let us expand $U(Q)=\sum_{m<0}U_{m}$ with $U_m$ taking value in $\fg^m$, then by using Lemma~\ref{thm-dr} we have
\[
[U_{-1}, \Ld_1]=Q, \quad (d\mid [U_{-1}, \Ld_1)=0.
\]
Let us introduce the notations:
\[
\tilde{\psi}_i=(e_i\mid U_{-1}), \quad \psi_i=\tilde{\psi}_{i}-\tilde{\psi}_{i-1}-\frac{x}{2(\ell+1)}.
\]
Then we have $\sum_{i=0}^\ell\psi_i=-x/2$, and by using the expansion of $M$ given in \eqref{A2M} we can represent $\ta_i$ and $\vp_i$ as follows:
\begin{align*}
\ta_i&=(\al_i^\vee\mid [U_{-1}, \Ld_1])=-([\al_i^\vee,\Ld_1]\mid U_{-1})=-\sum_{j=0}^\ell a_{i j}(e_j\mid  U_{-1})\\
&=\tilde{\psi}_{i-1}-2 \tilde{\psi}_i+\tilde{\psi}_{i+1}=\psi_{i+1}-\psi_{i}, \\
\vp_i&=(f_i\mid M)=(f_i\mid h [U_{-1},\Ld_2]-x\Ld_1)=h([\Ld_2,f_i]\mid U_{-1})-x (f_i\mid  \Ld_1) \\
&=h([\al_i^\vee,e_{i-1}]+[e_{i+1},\al_i^\vee]\mid U_{-1})-x  =h(-\tilde{\psi}_{i-1}+\tilde{\psi}_{i+1})-x\\
&=(\ell+1)(\psi_{i+1}+\psi_{i}).
\end{align*}
So the system of ODEs \eqref{vpta} can be represented as
\begin{equation}\label{chain}
\psi_{i+1}'+\psi_i'+\psi_{i+1}^2-\psi_{i}^2+\frac{\chi_i}{\ell+1}=0, \quad i\in\Z/(\ell+1)\Z.
\end{equation}
Note that the system \eqref{chain} is the nonlinear chain studied in \cite{Ad,VS}, which is related to the forth and the fifth Painlev\'e equations (P4 and P5) when $\ell=2$ and $3$ respectively.
\end{exa}

In contrast to the above example, the system of ODEs \eqref{vpta} may be complicated in general for the reason that $\vp_i$ are no longer linear functions of $\ta_i$. Let us illustrate this fact by the following examples.
We can obtain a system of ODEs of $\ta_i$ by substituting the expressions of $\vp_i$ into \eqref{vpta} and taking the conditions \eqref{tachi} into account.
\begin{exa}\label{exa-ODEA3}
Let $\fg$ be of type $A_3^{(1)}$, then its Kac labels and dual Kac labels are equal to $1$. For $b_p=\dt_{p 3}$, we have
\[
\ta_0+\ta_1+\ta_2+\ta_3=0, \quad \chi_0+\chi_1+\chi_2+\chi_3=4,
\]
and
\begin{align}\label{vpA3}
\vp_i&=\frac{1}{2}
   \ta_{i-1}'+\frac{1}{2} \ta_{i+1}'+2 \ta_{i+2}'+\frac{3}{4} {\ta_{i-1}}^2+\frac{3}{4} {\ta_{i+1}}^2-\frac{1}{8}\left(\ta_{i-1}+\ta_{i+1}-2\ta_{i+2}\right)^2-x
\end{align}
with $i\in\Z/4\Z$.
Observe that the system of ODEs \eqref{vpta} is invariant with respect to the rotation $\pi: i\mapsto i+1$ or the reflection $\sg: (0, 1, 2, 3)\mapsto(0, 3, 2, 1)$ of the indices.
\end{exa}

\begin{exa}
Let $\fg$ be of type $C_2^{(1)}$, then its Kac labels are given by $(k_0,k_1,k_2)=(1,2,1)$, its dual Kac labels are given by $(k_0^\vee,k_1^\vee,k_2^\vee)=(1,1,1)$, and the elements $\Ld_j$ are chosen as in \cite{DS}. Taking $b_p=\dt_{p 3}$, we have
\[
\ta_0+2\ta_1+\ta_2=0, \quad \chi_0+\chi_1+\chi_2=4,
\]
and
\[
\vp_i=\begin{cases}
                \ta_1'+2\ta_{2-i}'+{\ta_1}^2+\ta_1\ta_{2-i}-\dfrac{1}{2}{\ta_{2-i}}^2-x, & i=0,2; \\
                -\ta_0'-\ta_{2}'+\dfrac{1}{2}{\ta_0}^2-2\ta_0\ta_{2}+\dfrac{1}{2}{\ta_{2}}^2-2 x, & i=1.
              \end{cases}
\]
Observe that the system ODEs \eqref{taphi} can be obtained from the one in the previous example via the constraints $\ta_3=\ta_1$ and $\chi_3=\chi_1$, as well as the replacements  $\vp_1+\vp_3\mapsto\vp_1$ and $\chi_1+\chi_3\mapsto\chi_1$.
\end{exa}

\begin{exa}\label{exa-ODED4}
Let $\fg$ be of type $D_4^{(1)}$, then we have $k_i=k_i^\vee=2$ for $i=2$ and $k_i=k_i^\vee=1$ otherwise. Let the elements $\Ld_j$ be  normalized as in \cite{Wu}. Taking $b_p=\dt_{p 3}$, we have
\[
\ta_0+\ta_1+2\ta_2+\ta_3+\ta_4=0, \quad \chi_0+\chi_1+2\chi_2+\chi_3+\chi_4=6,
\]
and
\begin{equation}\label{ODED4}
\vp_i=\begin{cases}
  -3 \ta_{1-i}'+\dfrac{3}{2} \ta_{3}'+\dfrac{3}{2}
   \ta_{4}'+\dfrac{3}{2} {\ta_{1-i}}^2-\dfrac{3}{4}
   {\ta_{3}}^2-\dfrac{3}{4} {\ta_{4}}^2-x,  & i=0,1; \\
  \dfrac{3}{2}(\ta_0\ta_1+\ta_3\ta_4)-\dfrac{3}{4}(\ta_0+\ta_1)(\ta_3+\ta_4)-x, & i=2; \\
-3 \ta_{7-i}'+\dfrac{3}{2} \ta_{0}'+\dfrac{3}{2}
   \ta_{1}'+\dfrac{3}{2} {\ta_{7-i}}^2-\dfrac{3}{4}
   {\ta_{0}}^2-\dfrac{3}{4} {\ta_{1}}^2-x, & i=3,4.
\end{cases}
\end{equation}
The system of ODEs \eqref{taphi} is invariant with respect to the following reflections of indices:
\[
 \sg_1: (0,1,2,3,4)\mapsto(0,1,2,4,3), \quad  \sg_2: (0,1,2,3,4)\mapsto(1,0,2,3,4).
\]
Similar to the previous example, the reductions of the system of ODEs \eqref{taphi} with respect to the symmetries $\sg_1$ and $\sg_2$ give rise to the similarity reductions of the Drinfeld-Sokolov hierarchy associated to the affine Kac-Moody
algebras of type $B_3^{(1)}$ and $D_3^{(2)}$ respectively.
\end{exa}

\section{Concluding remarks }
In this paper we present a tau cover for
the Drinfeld-Sokolov hierarchy associated to any affine Kac-Moody
algebra $\fg$ with gradations $\rs\le\one$, and construct its Virasoro symmetries. This tau cover leads to an algorithm to construct
formal power series solution of the Cauchy problem of
the Drinfeld-Sokolov hierarchy with an arbitrary initial data. By
using this algorithm, we compute the formal solutions of the Drinfeld-Sokolov
hierarchy that satisfy two types of Virasoro constraints which are induced by the string equation and the similarity equation respectively.
In particular, the Virasoro constraints induced by the similarity equation lead to a system of ODEs of Painlev\'e type. When $\rs=\one$, the solution space of such ODEs admit an affine Weyl group actions, which generalizes the theory of Noumi, Yamada {\em et al} on the affine Weyl group symmetries for the equations of  Painlev\'e type.

In \cite{LWZZ} we proved a $\Gm$-reduction theorem for
the Drinfeld-Sokolov hierarchies. To explain this result, let $(\fg, \rs, \one)$ be a triple such that the affine Kac-Moody algebra $\fg$ possesses a diagram automorphism $\sg$ given in Tables~1--3 of \cite{LWZZ} and the gradation $\rs$ is consistent with $\sg$, then we can choose a basis $\Ld_j\, (j\in J)$ of the principal Heisenberg subalgebra $\mathcal{H}$ to be eigenvectors of $\sg$ with eigenvalues $\zeta_j$.  The
$\Gm$-reduction theorem asserts that the diagram automorphism $\sg$ induces an action on the flows $\p/\p t_j\, (j\in J_+)$ of the Drinfeld-Sokolov hierarchy, and the flow $\p/\p t_j$ is
invariant under the action of $\sg$ if and only if $\zeta_j=1$. Note that the folded Dynkin
diagram of $\fg$ with respect to $\sg$ corresponds to another affine Kac-Moody algebra, denoted by $\bar\fg$, on which there are two gradations $\bar{\rs}\le\one$ induced by the gradation $\rs\le\one$ of $\fg$ respectively. From the reduction procedure given in \cite{LWZZ}, we conclude:
\begin{itemize}
  \item[$\bullet$] If $b_p=0$ for any $p\in J_+$ with $\zeta_p\ne1$,
then $\sg$ induces an action on the space of solutions of the similarity reduction \eqref{simred};
  \item[$\bullet$] If $b_p=0$ unless $p$ is a positive exponent of $\bar\fg$, then any $\sg$-invariant solution of the similarity reduction \eqref{simred} also solves the corresponding similarity reduction for the Drinfeld-Sokolov hierarchy associated to $(\bar{\fg}, \bar{\rs}, \one)$.
\end{itemize}
These conclusions were illustrated by Examples~\ref{exa-ODEA3}--\ref{exa-ODED4}. We hope that such  results would help us to have a better understanding of properties of the higher order Painlev\'e-type equations related to Drinfeld-Sokolov hierarchies.

The Drinfeld-Sokolov hierarchies we consider in this paper are associated to the principal Heisenberg subalgebra of $\fg$. There are generalized Drinfeld-Sokolov hierarchies that are associated to other Heisenberg subalgebras of $\fg$, see for example \cite{FHM,dGHM,Kac}, and their similarity reductions also yield some ODEs of Painlev\'e type (see e.g. \cite{FS-P6,FS-D,FS-E,FS-A,KK}). For instance, it was derived by Fuji and Suzuki \cite{FS-P6} the sixth Painlev\'e equation from the similarity reduction of the generalized Drinfeld-Sokolov hierarchy associated to $\fg=D_4^{(1)}$ with a certain Heisenberg subalgebra different from the principal one, whose relation to the system \eqref{vpta} given by \eqref{ODED4} is unknown yet. It is natural to ask how the similarity reductions of the generalized Drinfeld-Sokolov hierarchies corresponding to different Heisenberg subalgebras are related to each other. We will study this question elsewhere.

\vskip 0.5truecm \noindent{\bf Acknowledgments.}
The authors thank Mattia Cafasso, Robert Conte and Yongbin Ruan for useful discussions, and they also thank Maxim Pavlov for his helpful comments.
The work is partially supported by NSFC  No.\,12071451, 11771238
and the NSFC for Distinguished Young Scholars No.\,11725104, and it is also partially supported by NSFC No.\,11831017, 11771461.


\begin{thebibliography}{99}

\bibitem{Ad} { V. E. Adler}. `Nonlinear chains and Painlev\'e equations', {\em Phys. D }73 (1994) 335--351.

\bibitem{Al} { A. Alexandrov}.
`Cut-and-join description of generalized Brezin-Gross-Witten model',
{\em Adv. Theor. Math. Phys. }22 (2018)  1347--1399
.
\bibitem{BR} { M. Bertola \& G. Ruzza}. `Brezin-Gross-Witten tau function and isomonodromic deformations', {\em Commun. Number Theory Phys. }13 (2019)  827--883.

\bibitem{BG}  { E. Brezin \& D. J. Gross}. `The external field problem in the large $N$ limit of QCD', {\em Physics
Letters B }97 (1980) 120--124.

\bibitem{BGHM} { N. J. Burroughs, M. F. de Groot, T. J. Hollowood \& J. L. Miramontes}. `Generalized Drinfeld-Sokolov hierarchies. II. The Hamiltonian
structures', {\em Comm. Math. Phys. }153 (1993)  187--215.

\bibitem{CW2}
{ M. Cafasso \& C.-Z. Wu}. `Borodin-Okounkov formula, string equation and
topological solutions of Drinfeld-Sokolov hierarchies', {\em Lett. Math. Phys. }109 (2019)  2681--2722.


\bibitem{CK} { P. A. Clarkson \& M. D. Kruskal}. `New similarity reductions of the Boussinesq equation', {\em  J. Math. Phys. }30 (1989)  2201--2213.

\bibitem{CM} { R. Conte \& M. Musette}. {\em The Painlev\'{e} Handbook} (Springer, Dordrecht, 2008).


\bibitem{DS}
{ V. G. Drinfeld \& V. V. Sokolov}. `Lie algebras and equations of {K}orteweg-de {V}ries type', {\em Current problems in mathematics }24, Itogi Nauki i
  Tekhniki, pages 81--180. Akad. Nauk SSSR, Vsesoyuz. Inst. Nauchn. i Tekhn.
  Inform., Moscow, 1984.

\bibitem{DYZ} { B. Dubrovin, D. Yang \& D. Zagier}. `On tau-functions for the KdV hierarchy',  Preprint arXiv: 1812.08488.

\bibitem{DZ}
{ B. Dubrovin \& Y. Zhang}.
`Normal forms of hierarchies of integrable PDEs, Frobenius manifolds
  and Gromov-Witten invariants',
{Preprint arXiv: 0108160}.

\bibitem{EF} { B. Enriquez \& E. Frenkel}. `Equivalence of two approaches to integrable
hierarchies of KdV type', {\em Comm. Math. Phys. }185 (1997)   211--230.

\bibitem{FJR}
{ H. Fan, T. Jarvis \& Y. Ruan}. `The Witten equation, mirror symmetry, and quantum singularity
  theory',
{\em Ann. of Math. }2 (2013)  1--106.

\bibitem{FHM} { L. Feher, J. Harnad \& I. Marshall}. `Generalized Drinfeld-Sokolov reductions and KdV type hierarchies',
{\em Comm. Math. Phys. }154 (1993) 181--214.

\bibitem{FF2} { B. Feigin \& E. Frenkel}. `Quantization of the Drinfeld-Sokolov reduction',
{\em Phys. Lett. B }246 (1990)  75--81.

\bibitem{FF1}  { B. Feigin \& E. Frenkel}. `Affine Kac-Moody algebras at the critical level
and Gelfand-Dikii algebras', {\em Infinite analysis, Part A, B} (Kyoto, 1991), 197--215, Adv.
Ser. Math. Phys., 16, World Sci. Publ., River Edge, NJ, 1992.

\bibitem{FBZ} { E. Frenkel \& D. Ben-Zvi}. {\em Vertex algebras and algebraic curves. Second edition. Mathematical Surveys and Monographs }88 (American Mathematical Society, Providence, RI, 2004).

\bibitem{FS-P6} { K. Fuji \& T. Suzuki}. `The sixth Painlev\'{e} equation arising from $D^{(1)}_4$ hierarchy', {\em J. Phys. A }39 (2006)  12073--12082.

\bibitem{FS-D}  { K. Fuji \& T. Suzuki}. `Higher order Painlev\'e system of type
$D^{(1)}_{2n+2}$ arising from integrable hierarchy', {\em Int. Math. Res.
Not. IMRN }2008, no. 1, Art. ID rnm 129, 21 pp.

\bibitem{FS-E}  { K. Fuji \& T. Suzuki}. `Coupled Painlev\'e VI system with $E^{(1)}_6$-symmetry', {\em J. Phys. A }42 (2009) 145205, 11 pp.

\bibitem{FS-A}  { K. Fuji \& T. Suzuki}. `Drinfeld-Sokolov hierarchies of type A and fourth order Painlev\'{e}
systems', {\em Funkcial. Ekvac. }53 (2010)
143--167.

\bibitem{GO} { P. G. Grinevich \& A. Y. Orlov}. `Virasoro Action on Riemann Surfaces, Grassmannians, det ${\overline\partial _J}$ and Segal-Wilson $\tau$-Function', {\em Problems of Modern Quantum Field Theory } (Springer, Berlin, Heidelberg, 1989) 86-106.

\bibitem{dGHM} { M. F. de Groot, T. J. Hollowood \& J. L. Miramontes}. `Generalized Drinfeld-Sokolov hierarchies', {\em Comm. Math. Phys. }145
(1992)  57--84.

\bibitem{GW} { D. J. Gross \& E. Witten}. `Possible third order phase transition in the large $N$
lattice gauge theory', {\em Phys. Rev., D }21 (1980) 446--453.

\bibitem{HM}
Hollowood, {  T. J. Hollowood \& J. L. Miramontes}. `Tau-functions and generalized integrable hierarchies',
{\em Comm. Math. Phys. }157 (1993) 99--117.

\bibitem{HMG} { T. Hollowood, J. L.  Miramontes \& J. S.  Guill{{\'e}}n}. `Additional symmetries of generalized integrable hierarchies', {\em J. Phys. A }27 (1994)  4629--4644.

\bibitem{Jac} { N. Jacobson}. {\em Lie Algebras. Republication of the 1962 original} (Dover Publications, Inc., New York, 1979).

\bibitem{Kac} { V. G. Kac}. {\em Infinite-dimensional Lie Algebras. Third edition} (Cambridge University Press, Cambridge, 1990, RI, 1989).

\bibitem{KW} { V. G. Kac \& M. Wakimoto}. `Exceptional hierarchies of soliton equations',
{\em Theta functions--Bowdoin 1987, Part 1} (Brunswick, ME, 1987),
191--237, Proc. Sympos. Pure Math., 49, Part 1, Amer. Math. Soc.,
Providence, RI, 1989.

\bibitem{KK} { S. Kakei \& T. Kikuchi}. `Affine Lie group approach to a derivative nonlinear
Schr\"odinger equation and its similarity reduction', {\em Int. Math. Res.
Not. }2004 (2004) 4181--4209.

\bibitem{Ko} { M. Kontsevich}. `Intersection theory on the moduli space of curves and the matrix Airy function',
{\em Comm. Math. Phys. }147 (1992) 1--23.

\bibitem{Kumar} { S. Kumar}. `Kac-Moody groups, their flag varieties and representation theory',
{\em Progress in Mathematics }204 (Birkh\"auser Boston, Inc., Boston, MA, 2002).

\bibitem{LRZ}
{ S.-Q. Liu, Y. Ruan \& Y. Zhang}. `BCFG Drinfeld-Sokolov hierarchies and FJRW-theory', {\em Invent. Math. }201 (2015) 711--772.

\bibitem{LWZZ}
{ S.-Q. Liu, C.-Z. Wu, Y. Zhang \& X. Zhou}. `Drinfeld-Sokolov hierarchies and diagram automorphisms of affine Kac-Moody algebras',
{\em Comm. Math. Phys. }375 (2020) 785--832.

\bibitem{Mi}
{ J. L. Miramontes}. `Tau-functions generating the conservation laws for generalized
  integrable hierarchies of {K}d{V} and affine {T}oda type', {\em Nuclear Phys. B }547 (1999) 623--663.

\bibitem{Na} { H. Nagoya}. `Quantization of differential systems with the affine Weyl group
symmetries of type $C^{(1)}_N$', {\em J. Math. Sci. Univ. Tokyo }15 (2008) 493--519.

\bibitem{Nor}  {  P. Norbury}. `A new cohomology class on the moduli space of curves', Preprint arXiv: 1712.03662.

\bibitem{Nou} {  M. Noumi}. `Affine Weyl group approach to Painlev\'{e} equations', {\em Proceedings of
the International Congress of Mathematicians }Vol. III Beijing  (2002), 497--509.

\bibitem{NY-A} {  M. Noumi \& Y. Yamada}. `Affine Weyl group symmetries in Painlev\'e type equations', {\em Toward the exact WKB analysis of differential equations, linear or non-linear} (Kyoto, 1998),
204, 245--259, Kyoto Univ. Press, Kyoto, 2000.

\bibitem{NY-W}{  M. Noumi \& Y. Yamada}. `Affine Weyl groups, discrete dynamical systems and Painlev\'e equations', {\em Comm. Math. Phys. }199 (1998) 281--295.

\bibitem{NY} {  M. Noumi \& Y. Yamada}. `Birational Weyl group action arising from a nilpotent Poisson algebra', {\em Physics and combinatorics }1999 (Nagoya) 287--319 (World Sci. Publ., River Edge, NJ, 2001).

\bibitem{SK} { K. Sawada \& T. Kotera}. `A method for finding $N$-soliton solutions of the
K.d.V. equation and K.d.V.-like equation', {\em Progr. Theoret. Phys. }51
(1974) 1355--1367.

\bibitem{SHC} { A. Sen, A. N. W. Hone \& P. A. Clarkson}.
`On the Lax pairs of the symmetric Painlev\'e equations', {\em  Stud. Appl.
Math. }117 (2006) 299--319.

\bibitem{VS} { A. P. Veselov \& A. B. Shabat}. `A dressing chain and the spectral theory of
Schr\"{o}dinger operator', {\em Funct. Anal. Appl. }27 (1993) 81--96.

\bibitem{Wa} {  M. Wakimoto}.
`Affine Lie algebras and the Virasoro algebra. I', {\em Japan. J. Math.
(N.S.) }12 (1986) 379--400.

\bibitem{Wi} { E. Witten}. `Two-dimensional gauge theories revisited', {\em J. Geom. Phys. }9 (1992) 303--368.

\bibitem{Wu}
{ C.-Z. Wu}. `Tau functions and Virasoro symmetries for Drinfeld-Sokolov
  hierarchies', {\em Adv.  Math. }306 (2017) 603--652.

\end{thebibliography}
\end{document}